\documentclass[12pt]{article}
\usepackage{amssymb}
\usepackage{bm}
\usepackage{a4}
\usepackage{amsmath}
\usepackage{graphicx}
\usepackage{cancel}
\usepackage{comment}
\usepackage{xcolor}
\usepackage{hyperref}
\usepackage{enumitem}

\evensidemargin \oddsidemargin
\def\1{{\bf 1}} 

\newcommand{\CC}{\mathbb{C}}
\newcommand{\RR}{\mathbb{R}}
\newcommand{\ZZ}{\mathbb{Z}}
\newcommand{\NN}{\mathbb{N}}

\newcommand{\vin}{\rotatebox[origin=c]{90}{$\in$}}
\newcommand{\la}{\langle}
\newcommand{\ra}{\rangle}

\newcommand{\Hi}{{\cal H}}

\newcommand{\A}{{\cal A}}
\newcommand{\B}{{\cal B}}
\newcommand{\C}{{\cal C}}
\newcommand{\E}{{\cal E}}
\newcommand{\F}{{\cal F}}
\newcommand{\I}{{\cal I}}
\newcommand{\J}{{\cal J}}
\newcommand{\T}{{\cal T}}

\newcommand{\Ps}{{\cal P}}
\newcommand{\Pt}{{\cal P}^t}
\newcommand{\Pe}{{\sf P}}

\newcommand{\Ss}{{\cal S}}

\newcommand{\bx}{{\bm x}}
\newcommand{\bD}{{\bf D}}
\newcommand{\bL}{{\bm L}}
\newcommand{\btheta}{{\bm\theta}}
\newcommand{\blambda}{{\bm \lambda}}
\newcommand{\bLambda}{{\bm \Lambda}}
\newcommand{\bpi}{{\bm\pi}}
\newcommand{\bphi}{{\bm\phi}}
\newcommand{\bpsi}{{\bm\psi}}

\newcommand{\be}{\begin{equation}}
\newcommand{\ee}{\end{equation}}
\newcommand{\bea}{\begin{eqnarray}}
\newcommand{\eea}{\end{eqnarray}}
\newcommand{\ba}{\begin{array}}
\newcommand{\ea}{\end{array}}
\def\nn{\nonumber \\}

\newtheorem{prop}{Proposition}[section]
\newtheorem{theorem}[prop]{Theorem}
\newtheorem{corollary}[prop]{Corollary}

\def\sq{\mbox{\rlap{$\sqcap$}$\sqcup$}}
\newenvironment{proof}[1]{\vspace{5pt}\noindent{\bf Proof #1}\hspace{6pt}}%
{\hfill\sq}
\newcommand{\bp}{\begin{proof}}
\newcommand{\ep}{\end{proof}\par\vspace{10pt}\noindent}

\begin{document}
\title{Fuzzy hyperspheres via confining potentials and energy cutoffs}
\date{}

\author{  Gaetano Fiore 
   \\   \\  
Dip. di Matematica e applicazioni, Universit\`a di Napoli ``Federico II'',\\
\& INFN, Sezione di Napoli, \\
Complesso Universitario  M. S. Angelo, Via Cintia, 80126 Napoli, Italy}

\maketitle

\begin{abstract} 
We simplify and complete the construction  of fully  $O(D)$-equivariant fuzzy  spheres $S^d_\Lambda$, for all dimensions $d\equiv D-1$, initiated in  [G. Fiore, F. Pisacane, 
J. Geom. Phys. 132 (2018), 423-451]. This is based on imposing a suitable energy cutoff on a quantum particle in $\RR^D$ subject to a confining potential well $V(r)$  
with a very sharp minimum on the sphere of radius $r=1$; the cutoff and the depth of the well diverge with $\Lambda\in\NN$.  As a result, the noncommutative Cartesian coordinates $\overline{x}^i$ generate the whole algebra of observables  $\A_{\Lambda}$ on the Hilbert space $\Hi_{\Lambda}$;  applying 
polynomials in the $\overline{x}^i$ to any $\bpsi\in\Hi_{\Lambda}$ we recover
the whole $\Hi_{\Lambda}$.
The commutators of   the $\overline{x}^i$ are proportional to the angular momentum components, as in  {\it Snyder} noncommutative spaces. 
$\mathcal{H}_{\Lambda}$, as 
carrier space of a reducible representation of $O(D)$, is isomorphic  to  the space  
of harmonic homogeneous polynomials of degree $\Lambda$ in the Cartesian coordinates of (commutative) $\RR^{D+1}$, which carries an irreducible representation $\bpi_\Lambda$  of  $O(D\!+\!1)\supset O(D)$.
Moreover, $\A_{\Lambda}$  is isomorphic to $\bpi_\Lambda\left(Uso(D\!+\!1)\right)$. 
We resp. interpret $\{\Hi_\Lambda\}_{\Lambda\in\NN}$, $\{\A_\Lambda\}_{\Lambda\in\NN}$ as fuzzy deformations of
the space  $\Hi_s:={\cal L}^2(S^d)$ of (square integrable) functions on $S^d$ and of 
 the associated algebra $\A_s$ of observables, because they resp. go to  $\Hi_s,\A_s$
 as $\Lambda$ diverges (with $\hbar$ fixed).
With suitable $\hbar=\hbar(\Lambda)\stackrel{\Lambda\to\infty}{\longrightarrow} 0$,
in the same limit $\A_\Lambda$ goes to the (algebra of functions on the) Poisson manifold $T^*S^d$; 
 more formally,  $\{\A_\Lambda\}_{\Lambda\in\NN}$ yields a fuzzy quantization of a coadjoint orbit of $O(D\!+\!1)$ that goes to the classical phase space $T^*S^d$. 
\end{abstract}


\section{Introduction}
\label{intro}

Noncommutative space(time) algebras are introduced and studied 
with various motivations, notably  to provide  an arena for \  regularizing ultraviolet (UV) divergences in quantum field theory (QFT) (see e.g. \cite{Snyder}), 
\ reconciling Quantum Mechanics and General Relativity in a satisfactory Quantum Gravity (QG) theory   (see e.g. \cite{DopFreRob95}), 
\ unifying fundamental interactions  (see e.g. \cite{ConLot90,ChaCon10}).
Noncommutative Geometry  (NCG) \cite{Connes,Madore99,GraFigVar00,Lan97} 
has become a sophisticated framework that  develops the whole machinery of differential geometry  on noncommutative spaces. 
Fuzzy spaces are particularly appealing noncommutative spaces: a fuzzy space is a sequence $\{\mathcal{A}\}_{n\in\NN}$
of {\it finite-dimensional} algebras such that
 $\mathcal{A}_n\overset{n\rightarrow\infty}\longrightarrow\mathcal{A}\!\equiv$algebra 
of regular functions on an ordinary manifold, with \ $dim(\mathcal{A}_n)\overset{n\rightarrow\infty}\longrightarrow\infty$. 
They have raised a big interest in the high energy physics community as a non-perturbative technique in  QFT  based on a finite discretization of space(time)  alternative to the lattice one: the main advantage is that the algebras 
$\A_n$ can carry representations of Lie groups (not only of discrete ones). They can be used also 
 for  internal (e.g. gauge) degrees of freedom (see e.g. \cite{AscMadManSteZou}),
or  as a new tool in string and $D$-brane theories (see e.g. \cite{AleRecSch99,HikNozSug01}).
The first and seminal fuzzy space is the 2-dimensional Fuzzy Sphere (FS) of Madore and Hoppe \cite{Mad92,HopdeWNic}, where ${\cal A}_n\simeq M_n(\CC)$, 
which is generated by coordinates $x^i$ ($i=1,2,3$) fulfilling
\be
[x^i,x^j]=\frac {2i}{\sqrt{n^2\!-\!1}}\varepsilon^{ijk}x^k, \quad
r^2:=x^ix^i=1,\qquad n\in\NN\setminus \{1\}                      \label{FS}
\ee
(sum over repeated indices is understood);
they are obtained by the rescaling $x^i=2L_i/{\sqrt{n^2\!-\!1}}$ of the elements
$L_i$ of  the standard basis of $so(3)$ in the unitary irreducible representation (irrep)
 $(\pi^l,V^l)$ of dimension   $n=2l\!+\!1$, i.e. where $V^l$ is the eigenspace of the Casimir 
$\bL^2=L_iL_i$ with eigenvalue $l(l+1)$. Ref. \cite{GroMad92,GroKliPre96'} first proposed a QFT based on it.
Each matrix in $M_n$ can be expressed as a polynomial in the $x^i$ that can be rearranged 
as  the expansion  in spherical harmonics of an element of $C(S^2)$
truncated at level $n$. Unfortunately, such a nice feature is  not shared by the fuzzy spheres
of dimension $d>2$: the product of two spherical harmonics is not a combination
of spherical harmonics, but an element in a larger algebra $\A_n$.
Fuzzy spheres of dimension $d=4$ and any $d\ge 3$ were first introduced respectively in \cite{GroKliPre96} and \cite{Ramgoolam}; other versions in $d=3,4$ or $d\ge 3$ 
 have been proposed in \cite{DolOCon03,DolOConPre03,Ste16,Ste17}.

The Hilbert space of a (zero-spin) quantum particle on configuration space $S^d$ and the space of continuous functions on $S^d$ carry (the same)  {\it reducible} representation of 
$O(D)$, with $D:=d\!+\!1$; this decomposes into  irreducible representations (irreps) as follows
\be
{\cal L}^2(S^d)\simeq\bigoplus\nolimits_{l=0}^\infty V_D^l\simeq C(S^d),
\label{directsum}
\ee
where the carrier space $V_D^l$ is an eigenspace of the quadratic Casimir $\bL^2$  with eigenvalue  
\be
E_l:=l(l\!+\!D\!-\!2)           \label{El}
\ee
($V_3^l\equiv V^l$). $C(S^d)$ can be seen as an algebra of bounded operators on ${\cal L}^2(S^d)$.
  On the contrary, the mentioned fuzzy hyperspheres (including the Madore-Hoppe FS)
are either based on sequences of irreps of $Spin(D)$ (so that $r^2$, which is 
proportional to $\bL^2$,  is identically 1) parametrized by $n$ \cite{Mad92,HopdeWNic,GroKliPre96,Ramgoolam,DolOCon03,DolOConPre03},
or on sequences of reducible representations that are the direct sums of small bunches of  such irreps \cite{Ste16,Ste17}. In either case, even excluding the $n$ for which the associated representation of $O(D)$ is only {\it projective}, the carrier space does not go to (\ref{directsum})
in the limit $n\to\infty$; we think this makes the interpretation of these fuzzy spheres
as fuzzy configuration spaces $S^d$ (and of the $x^i$ as spatial coordinates) questionable.
For the Madore-Hoppe FS such an interpretation is even more difficult, because relations  (\ref{FS}) are equivariant under $SO(3)$, but not under the whole $O(3)$, e.g. not under parity $x^i\mapsto -x^i$, while the ordinary sphere $S^2$ is; on the contrary, all the other mentioned fuzzy spheres  are $O(D)$-equivariant, because the commutators $[x^i,x^j]$ are Snyder-like \cite{Snyder}, i.e.
proportional to angular momentum components $L_{ij}$.

The purpose of this work is to complete the construction \cite{FioPis18,Pis20}  of new, fully $O(D)$-equivariant   fuzzy quantizations of spheres  $S^d$ of arbitrary dimension $d=D-1\in\NN$ (thought as configuration spaces) and of $T^*S^d$
(thought as phase spaces), in a sense that will be fully clarified at the end of section \ref{discuss}; in the commutative limit the involved  $O(D)$-representation 
goes to (\ref{directsum}). We also simplify and uniformize (with respect to $D$) the procedure of \cite{FioPis18,Pis20}.

We recall this procedure starting from the general underlying philosophy \cite{FioPis20PoS,FioPis18,FioPis18POS}. Consider a quantum theory $\T$ with Hilbert space  \ $\Hi$, \ algebra of observables on $\Hi$ (or with a domain dense in $\Hi$)
  \ $\A\equiv\mbox{Lin}(\Hi)$, \ Hamiltonian  \ $H\in\A$. For any subspace $\overline{\Hi}\subset\Hi$  preserved by the action of $H$, let $\overline{P}:\Hi\mapsto\overline{\Hi}$ \ be  the associated projector and
$$
\overline{\A}\equiv \mbox{Lin}\left(\overline{\Hi}\right)
=\{\overline{A}\equiv \overline{P}A\overline{P}\:\: |\: A\in\A\};
$$
the observable $\overline{A}\equiv\overline{P}A\overline{P}\in\overline{\A}$ will have  the same physical interpretation as $A$. By construction \ $\overline{H}=\overline{P}H=H\overline{P}$. \ 
The projected Hilbert space $\overline{\Hi}$, algebra of observables $\overline{\A}$ and Hamiltonian \ $\overline{H}$ \ provide a new quantum theory $\overline{\T}$. \ 
If $\overline{\Hi}$, $H$ are invariant under some  group $G$, then 
\ $\overline{P},\overline{\A},\overline{H},\overline{\T}$ \ will be as well.
In general, the relations among the generators of $\overline{\A}$ differ from those among the generators of $\A$. In particular,  if the theory $\T$ is based on commuting coordinates $x^i$
(commutative space) this will be in general no longer  true for $\overline{\T}$: 
\ $[\overline{x}^i,\overline{x}^j]\neq 0$, \ and we have generated a quantum theory on a NC space.

A physically relevant instance of the above projection mechanism occurs when
$\overline{\Hi}$ is the subspace of $\Hi$ characterized by energies  $E$ below
a certain cutoff, \ $E\le\overline{E}$; 
then $\overline{\T}$ is a {\it low-energy effective approximation} of $\T$. What it can be useful for? 
If \ $\overline{\A}$ \ contains all the observables corresponding to measurements that we can {\it really}
perform with the experimental apparati at our disposal, and the initial state of the system belongs  to $\overline{\Hi}$, then neither the dynamical evolution ruled by $H$, nor any measurement can map it out of $\overline{\Hi}$, and we can replace  $\T$ by
the effective theory \ $\overline{\T}$. Moreover, if at $E>\overline{E}$ we even expect new physics  not accountable by $\T$, then  $\overline{\T}$ 
may also help to figure out a new theory $\T'$ valid for all $E$.

\begin{figure}
\includegraphics[height=3.7cm]{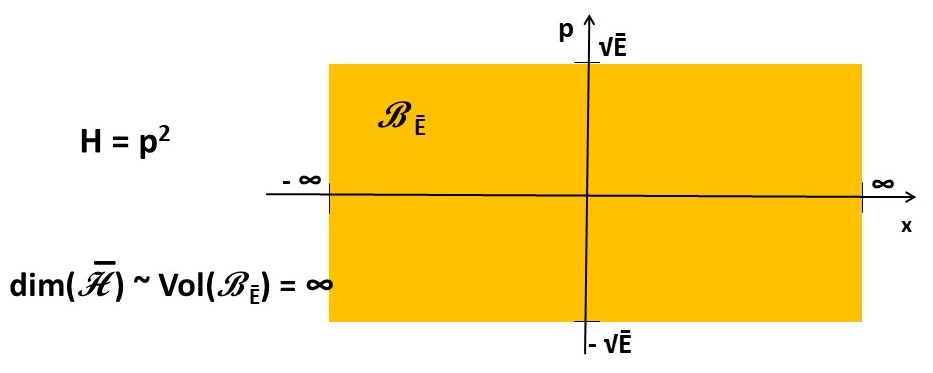}\hfill
\includegraphics[height=3.7cm]{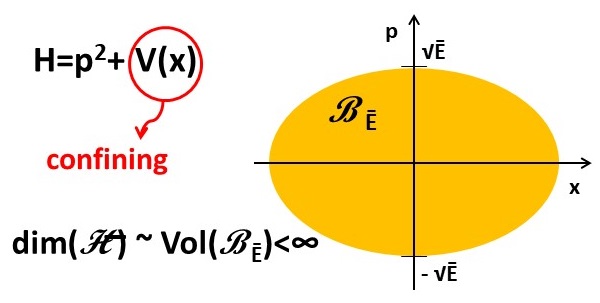}
\caption{Classical phase space regions with energy $E\le\overline{E}$}
\label{ClassicalRegions}
\end{figure}

\medskip
For an ordinary (for simplicity, zero-spin) quantum particle in the Euclidean (configuration) space $\RR^D$  it is  $\Hi={\cal L}^2(\RR^d)$. Fixed a Hamiltonian $H(x,p)$, 
by standard wisdom the dimension of $\overline{\Hi}$ is
\be
\mbox{dim}(\overline{\Hi})\approx\mbox{Vol}(\B_{\scriptscriptstyle\overline{E}})/h^D,
\label{dimH}
\ee
where $h$ is the Planck constant and $\B_{\scriptscriptstyle\overline{E}}\!\equiv\!\big\{\!(x,\!p)\!\in\!\RR^{2D}\:|\: H(x,\!p)\!\le\! \overline{E}\big\}\!$ is the classical phase space region with energy below $\overline{E}$. If $H$ consists only of the kinetic energy $T$, then this is infinite 
(fig. \ref{ClassicalRegions} left). If $H=T\!+\!V$, with a confining potential $V$, then this is finite 
(fig. \ref{ClassicalRegions} right) at least for sufficiently small  $\overline{E}$,
and also the classical region $v_{\overline{E}}\subset \RR^D$ in configuration space
 determined by the condition \ $V\le \overline{E}$ is bounded.
In the sequel we rescale $x,p,H,V$ so that they are dimensionless and, denoting by
$\Delta$ the Laplacian in  $\RR^D$,  we can write
\be
H=-\Delta + V.                                      \label{Ham}
\ee
The `dimensional reduction' $\RR^D\leadsto S^d$ of the configuration space is obtained:
\begin{enumerate}
\item Assuming in addition that $V=V(r)$ depends only on  the distance $r$ from the center of the sphere $S^d\subset\RR^D$ and has a very sharp minimum, parametrized by a very large $k\equiv V''(1)/4$, on the sphere  $S^d$ of equation $r=1$,  see fig. \ref{fig1}.

\item  Choosing $\overline{E}$ so low that all
radial quantum excitations are `frozen', i.e. excluded from $\overline{\Hi}\subset\Hi$; 
this makes $H$ coincide with the Laplacian $\bL^2$ on $S^d$, up to terms $O(1/\sqrt{k})$.

\item Making
both $k,\overline{E}$ depend on, 
grow and diverge with a natural number $\Lambda$.
Thereby we rename
$\overline{\Hi},\overline{P},\overline{\A}$ as
 $\Hi_{\Lambda},P_{\Lambda},\A_{\Lambda}$. 

 \end{enumerate}
As $H$ is $O(D)$-invariant,
so are $P_{\Lambda}$,  $\overline{H}=P_{\Lambda} H$, and the projected
theory is $O(D)$-equivariant. 

Technical details are given in sections \ref{genset}, \ref{xt-polynomials}. Section \ref{genset}
fixes the notation and contains preliminaries partly developed in \cite{Pis20}. The representation-theoretical results of section \ref{xt-polynomials}, which deserve attention also on their own, allow to explictly characterize the space $V_D^l$ as the space of
harmonic homogeneous polynomials of degree $l$ in the Cartesian coordinates $x^i$ of
$\RR^D$ restricted to the sphere $S^d$; we determine such polynomials 
constructing the trace-free completely symmetric projector of $\left(\RR^D\right)^{\otimes^l}$
and applying it to the homogeneous polynomials of degree $l$ in   $x^i$. The actions of the $L_{hk}$
and of the multiplication operators $x^i\cdot$ on such polynomials can be expressed
by general formulae valid for all $D,l$; this allows to avoid the rather complicated  actions of the $L_{hk}$ on spherical harmonics (which also span $V_D^l$) used in \cite{Pis20}.
It turns out that both $\Hi_{\Lambda}, V_{D+1}^{\Lambda}$ decompose into irreps of  $O(D)$ as follows:
\be
\Hi_{\Lambda}\simeq V_{D+1}^{\Lambda}=\bigoplus\limits_{l=0}^{\Lambda} V_D^l,
\label{Hi_Lambda-deco}
\ee
The second equality shows that, in contrast with the mentioned fuzzy hyperspheres, we recover (\ref{directsum}) in the limit $\Lambda\to\infty$. The first equality suggests that
 the unitary irrep of the $*$-algebra $\A_{\Lambda}$ on $\Hi_{\Lambda}$ is 
isomorphic to the irrep $\pi_\Lambda$ of $Uso(D+1)$ on $V_{D+1}^{\Lambda}$, what we in fact prove in section  \ref{isomorphism} (this had been proved for $D=2,3$ and conjectured for $D>3$ in \cite{FioPis18,Pis20}).
The relations fulfilled by   $\overline x^i,\overline L_{hk}$ are determined in section 
\ref{xLRel}: the commutators $[\overline x^i,\overline x^j]$ are also Snyder-like \cite{Snyder}, i.e.
are proportional to $\overline L_{hk}/k$, with a proportionality factor that is the same
constant on all of $\Hi_{\Lambda}$, except on the $l=\Lambda$ component of the latter,
where it is a slightly different constant. $\overline x^i$  generate  the whole 
$\A_{\Lambda}$. The square distance $\overline\bx^2\equiv \overline x^i\overline x_i$ is a function
of $\bL^2$ only, such that almost all its spectrum is very close to 1 and goes to 1 
in the limit $\Lambda\to\infty$.
In section \ref{Dconverge} we show in which sense  $\Hi_{\Lambda},\A_{\Lambda}$
go to $\Hi,\A$ as $\Lambda\to\infty$, in particular how one can recover $f\cdot\in C(S^d)\subset\A$, the multiplication operator of wavefunctions in ${\cal L}^2(S^d)$ by a continuous function $f$, as the strong limit of a suitable sequence $f_\Lambda\in\A_\Lambda$
 (again, this had been only conjectured in \cite{Pis20}).
In section \ref{discuss} we discuss our results and possible developments in comparison with the literature; in particular, we point out 
that our pair  $(\Hi_{\Lambda},\A_{\Lambda})$ can be seen as a fuzzy quantization
of a coadjoint orbit of $O(D)$ that can be identified with the cotangent space $T^*S^d$,
the classical phase space over the $d$-dimensional sphere. Finally, we have concentrated 
most proofs in the appendix  \ref{Appe}.

\section{General setting}
\label{genset}

We choose  a set of real Cartesian coordinates $x:=(x^1,...x^D)$ of $\RR^D$ and
abbreviate $\partial_i\equiv \partial/\partial x^i$. We normalize $x^i,\partial_i$ and $H$ itself so as to be dimensionless.  Then we can express $r^2:=\bx^2\equiv x_ix^i$, $\Delta:= \partial_i\partial^i$ (sum over repeated indices understood), where actually $x_i=x^i$ and $\partial^i=\partial_i$ because the coordinates  are real and Cartesian. 
  The self-adjoint operators $x^i,-i\partial_i$  
on ${\cal L}^2(\RR^D)$ fulfill the  canonical commutation relations
\bea
[x^i,x^j]=0,\qquad [-i\partial_i,-i\partial_j]=0,\qquad [x^i,-i\partial_j]= i\delta^i_j , \label{ccr}
\eea
which are equivariant under all orthogonal transformations $Q$ (including parity $Q=-I$)
\bea
x^i\mapsto x'{}^i=Q^i_j x^j, \qquad Q^{-1}=Q^T.          \label{ortho}
\eea
All scalars $S$, in particular $S=\Delta, r^2,V,H$, are invariant.
This implies $[S,L_{ij}]=0$, where
\be
L_{ij}:=i(x^j\partial_i-x^i\partial_j)
\ee
are the angular momentum components associated to $x$. These generate rotations 
of $\RR^D$, i.e.
\bea
[iL_{ij},v^h]= v^i\delta^h_j-v^j\delta^h_i     \label{Lvcr}
\eea 
hold for the components $v^h$ of all vector operators, in particular $v^h=x^h,\partial_h$, and close the commutation relations of $so(D)$,
\bea
[iL_{ij},iL_{hk}]=i\left(L_{ik}\delta_{hj}-L_{jk}\delta_{hi}-L_{ih}\delta_{kj}+L_{jh}\delta_{ki}\right) .         \label{LLcr}
\eea
The $D$ derivatives $\partial_i$ make up a globally defined basis for the linear space of smooth vector fields on $\RR^D$. 
As the  $L_{ij}$ are vector fields tangent to all spheres $r\!= $const, the set
$B=\{\partial_r,L_{ij}\: |\: i<j\}$ ($\partial_r:=\partial/\partial r$) is an alternative complete set that is singular  for $r=0$, but globally defined 
elsewhere; for $D=2$ it is a basis, while for $D>2$ it is redundant, because of the relations
\be
\varepsilon^{i_1i_2i_3....i_D}x^{i_1}L_{i_2i_3}=0.           \label{Lijrel}
\ee
This redundancy (unavoidable if $S^d$ is not parallelizable) will be no problem for our purposes.

\begin{figure}[htbp]
        \begin{minipage}[c]{.48\textwidth}
          \includegraphics[scale=0.23]{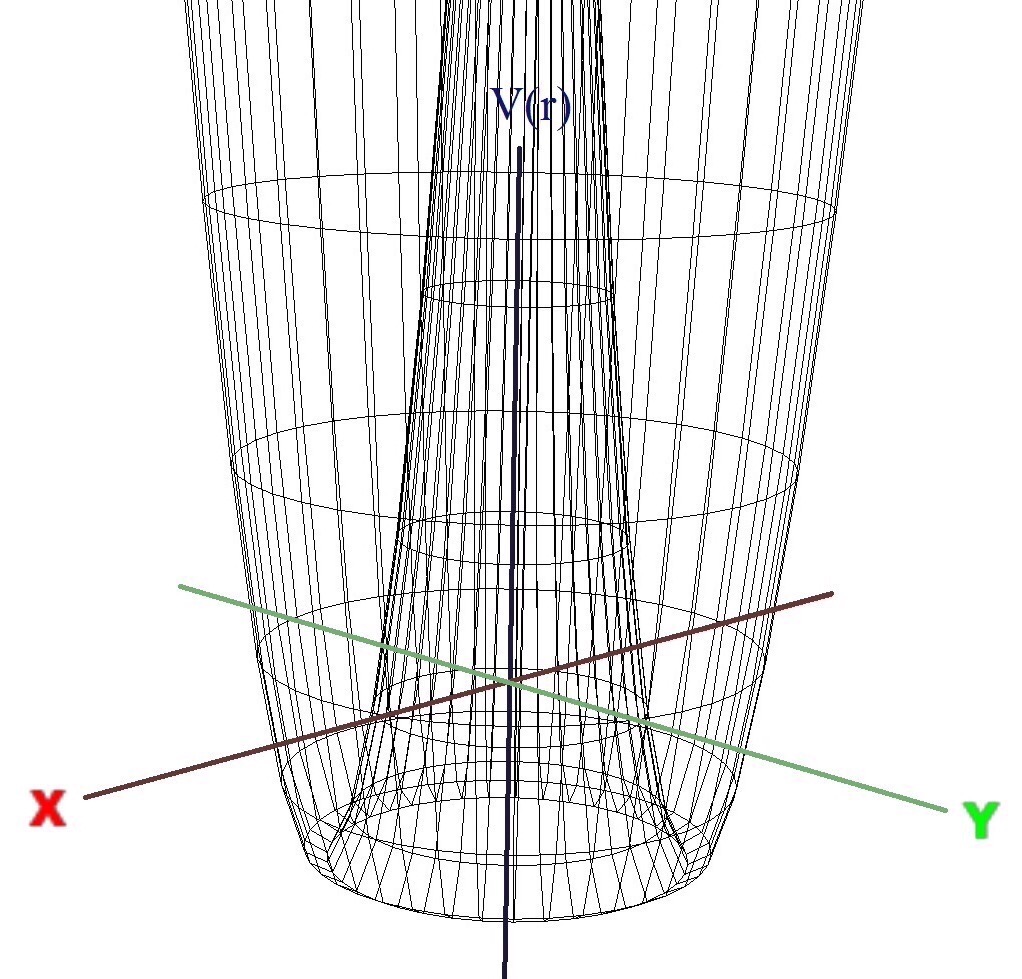}
          \caption{Three-dimensional plot of $V(r)$ \\ in dimension $D=2$.}
       \label{fig1}
 \end{minipage}%
        \hspace{5mm}%
        \begin{minipage}[c]{.48\textwidth}
\begin{center}          \includegraphics[scale=0.5]{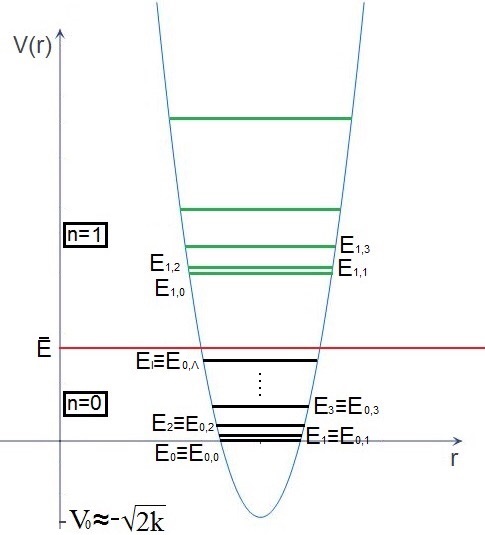}
          \caption{Two-dimensional plot of $V(r)$, \\ including the energy-cutoff.}
\end{center}        \end{minipage}
      \end{figure}

We shall assume that $V(r)$   has a very sharp minimum at \ $r=1$ with very large $k= V''(1)/4>0$, \ and fix 
$V_0:= V(1)$ so that the ground state $\bpsi_0$ has zero energy, i.e. $E_0=0$ (see fig. \ref{fig1}). We  choose an energy cutoff $\overline{E}$
fulfilling first of all the condition
\be
V(r)\simeq V_0+2k (r-1)^2\qquad \mbox{if $r$ fulfills}\quad V(r)\le  \overline{E},
\label{cond1}
\ee
so that we can neglect terms of order higher than two in the Taylor expansion of $V(r)$ around $1$
and approximate the potential as a harmonic one  in the classical region $v_{\overline{E}}\subset \RR^D$ 
 determined by the condition \ $V(r)\le \overline{E}$. \ 
By  (\ref{cond1}),  $v_{\overline{E}}$ is approximately the spherical shell
$|r\!-\!1|\le \sqrt{\frac{\overline{E}\!-\!V_0}{2k}}$; when both $\overline{E}\!-\!V_0$ and $k$ diverge, while their ratio goes to zero, then $v_{\overline{E}}$ reduces to the unit sphere $S^d$. 
We expect that
in this limit the dimension of  $\Hi_{\overline{E}}$ diverges, and we recover standard 
quantum mechanics on  $S^d$. As we shall see, this is the case.

Of course, the eigenfunctions of $H$ can be more easily determined in terms of polar coordinates $r,\theta_1,...,\theta_d$, recalling that the Laplacian in $D$ dimensions decomposes as follows
\be
\Delta=\partial_r^2+(D-1)\frac 1r\partial_r-\frac 1{r^2}\bL^2,  \label{LaplacianD}
\ee
(see section \ref{ComplSymProj}) where $\bL^2:=L_{ij}L_{ij}/2$ is the square angular momentum 
(in normalized units), i.e. the quadratic Casimir of $Uso(D)$ and the Laplacian on the sphere $S^d$; $\bL^2$ can be expressed in terms of angles $\theta_a$ and derivatives $\partial/\partial \theta^a$ only. The eigenvalues of $\bL^2$ are $l\left(l+D-2\right)$, see section \ref{ComplSymProj}; we denote by
 $V_D^l$ the $\bL^2=E_l$ eigenspace within ${\cal L}^2(S^d)$.
Replacing the Ansatz \ $\bpsi=f(r)Y_l(\btheta)$, \ with $f(r)= r^{-d/2}\, g(r)$, $Y_l\in V_D^l$, $\btheta\equiv(\theta_1,...,\theta_d)$, \ transforms the Schr\"odinger PDE \ $H\bpsi=E\bpsi$ \
into the Fuchsian ODE in the unknown  $g(r)$\footnote{The present treatment of the equations in the case $D=2$ is slightly different from the one adopted in Ref. \cite{FioPis18}, where  the independent radial variable had been changed  according to 
$r\mapsto \rho\equiv\ln r$; however this gives the same wavefunctions at lowest order in 
$\rho\simeq r\!-\!1$).}
\be
-g''(r)+\left[\frac{\left[D^2-4D+3+4l(l+D-2) \right]}{4r^2}+V(r)\right]g(r)=Eg(r).
\label{eqpolarD2}
\ee
The requirement $\bpsi\in {\cal L}^2(\RR^D)$ implies  that $g$
belongs to ${\cal L}^2(\RR^+)$, in particular goes to zero as $r\rightarrow\infty$. The self-adjointness of $H$ implies  that it must be $f(0)=0$; this is compatible \cite{Pis20} with Fuchs theorem
provided $r^2V(r)\overset{r\rightarrow 0^+}\longrightarrow T\in\mathbb{R}^+$ 
 [what is in turn compatible with (\ref{cond1})].
Since $V(r)$ is very large outside  the thin spherical shell $v_{\overline{E}}$   (a neighbourhood of $S^d$), $g,f,\bpsi$ become negligibly small there,  and, by condition (\ref{cond1}),  the lowest eigenvalues $E$ are at leading order those of the $1$-dimensional harmonic oscillator approximation \cite{Pis20} of (\ref{eqpolarD2})
\be
-g''(r)+g(r)k_{l}\left(r-\widetilde{r}_{l}\right)^2=\widetilde{E}_{l}g(r),
\label{eqpolarD3}
\ee
which is obtained neglecting terms $O\big((r\!-\!1)^3\big)$ in the Taylor expansions of
$1/r^2,V(r)$ about $r\!=\!1$.
Here 
\be \label{definizioni1}
\ba{l}
\widetilde{r}_{l}:=1+\frac{b(l,D)}{3b(l,D)+2k },\quad \widetilde{E}_{l}:=E-V_0\frac{2b(l,D)\left[k +b(l,D)\right]}{3b(l,D)+2k }, \\[6pt]
k_{l}:=2k+3b(l,D) ,\quad 
b(l,D):=\frac{D^2-4D+3+4l(l+D-2)}{4}.
\ea  
\ee 
The (Hermite functions) square-integrable solutions of (\ref{eqpolarD3}) 
$$
g_{n,l}(r)=M_{n,l}\hspace{0.15cm}e^{-\frac{\sqrt{k_{l}}}{2}\left(r-\widetilde{r}_{l}\right)^2}\cdot H_n\left((r-\widetilde{r}_{l})\sqrt[4]{k_{l}}\right)\quad\mbox{ with }n\in\mathbb{N}_0
$$
(here $M_{n,l}$ are   normalization constants and $H_n$ are the Hermite polynomials) lead to
\be\label{valuef}
f_{n,l}(r)=\frac{M_{n,l}}{r^{\frac{d}{2}}}\hspace{0.15cm}e^{-\frac{\sqrt{k_{l}}}{2}\left(r-\widetilde{r}_{l}\right)^2}\cdot H_n\left((r-\widetilde{r}_{l})\sqrt[4]{k_{l}}\right)\quad\mbox{ with }n\in\mathbb{N}_0.
\ee
The corresponding `eigenvalues' in (\ref{eqpolarD3})   
$\widetilde{E}_{n,l}=(2n+1)\sqrt{k_{l}}$   lead to energies
$$
E_{n,l}=(2n+1)\sqrt{k_{l}}+V_0+\frac{2b(l,D)[k +b(l,D)]}{3b(l,D)+2k }.
$$
As said, we fix $V_0$ requiring that the lowest one $E_{0,0}$ be zero; this implies
$$
V_0=-\sqrt{k_{0}}-\frac{2b(0,D)\left[k +b(0,D)\right]}{3b(0,D)+2k }
=-\sqrt{2k }-b(0,D)-\frac{3b(0,D)}{2\sqrt{2k }}+O\left(k ^{-\frac{1}{2}}\right),
$$
and the expansions of  $E_{n,l}$ and $\widetilde{r}_{l}$ at leading order in $k $ become
\be 
E_{n,l}=
l(l+D-2)+ 2n\sqrt{2k}+O\left(k ^{-\frac{1}{2}}\right),\qquad
\widetilde{r}_{l}=1+\frac{b(l,D)}{2k }+O\left(k ^{-2}\right).
\ee 
$E_{0,l}$ coincide at lowest order with  the desired eigenvalues $E_l$ of the Laplacian  $\bL^2$ on $S^d$, while  if $n>0$
$E_{n,l}$ diverge as $k \to\infty$;  to exclude all states with $n>0$ (i.e.,  to `freeze' radial oscillations; then all corresponding classical trajectories are circles) we  impose the cutoff
\bea
E_{n,l}\le \overline{E}(\Lambda):=\Lambda(\Lambda\!+\!D\!-\!2) <2\sqrt{2k},\qquad\Lambda\in\NN. \label{consistencyD}
\eea
The right inequality is satisfied prescribing a suitable dependence $k \left(\Lambda\right)$, e.g. $k \left(\Lambda\right)=\left[\Lambda(\Lambda\!+\!D\!-\!2) \right]^2$; the left one  is satisfied  setting $n=0$ and $l\le \Lambda$. Abbreviating   $ f_l\equiv f_{0,l}$,
we end up with eigenfunctions  and associated energies (at leading order in $1/\Lambda$)
\be
\bpsi_l(r,\btheta)= f_l(r)\,Y_l(\btheta), \qquad H\bpsi_l=E_{l}\,\bpsi_l,
\qquad\quad
l=0,1,...,\Lambda.\label{statopsi}
\ee
Thus $\Hi_{\Lambda}$ decomposes into irreps of $O(D)$ (and eigenspaces of $\bL^2,H$)
as follows
\be
\Hi_{\Lambda}=\bigoplus_{l=0}^{\Lambda}\Hi_{\Lambda}^l, \qquad  \Hi_{\Lambda}^l:=f_l(r)\, V_D^l.
\ee
We can express the projectors $P^l_{\Lambda}:\Hi_{\Lambda}\to\Hi_{\Lambda}^l$  as the following 
polynomials in $\overline{\bL}^2$:
\bea
P^l_{\Lambda}=\prod_{n=0, n\neq l}^{\Lambda} \frac{\overline{\bL}^2-E_n}
{E_l-E_n}.    \label{Proj-l}
\eea
In the commutative limit  $\Lambda \to\infty$ the spectrum $\{E_{0,l}\}_{l=0}^{\Lambda}$ of $\overline{H}$ goes to the whole spectrum $\{E_l\}_{l\in\mathbb{N}_0}$ of   $\bm{L}^2$.
If  $\bphi,\bphi'\in\Hi\equiv {\cal L}^2(\mathbb{R}^D)$ 
can be factorized  into radial parts $f(r),f'(r)$ and  angular parts $T,T'\in\mathcal{L}^2(S^d)$, i.e. $\bphi=f\,T$, $\bphi'=f'\,T'$, then so can be their scalar product:
\bea
\left(\bphi,\bphi'\right):=\int_{\mathbb{R}^D}d^Dx\,\bphi^*(x)\bphi'(x)
=\la T,T'\ra\: \int^\infty_0\!\!\! dr\,r^d\,f^*(r) \, f'(r).
\label{scalprodRD}
\eea
Here we have denoted by $\la \cdot,\cdot\ra$ the scalar product of $\mathcal{L}^2(S^d)$,
\bea
\la T,T'\ra:=\int_{S^d}\!\!\! d\alpha\: T^*\,T',
\label{scalprodSd}
\eea
where $d\alpha$ is the $O(D)$-invariant measure on $S^d$\footnote{In terms of the angles $\btheta$, \ 
$d\alpha=\left[\sin^{d-1}{(\theta_{d})} \sin^{d-2}{(\theta_{d-1})} \cdots \sin{(\theta_2)}\right] d\theta_1 d\theta_2\cdots d\theta_{d}$.
}.
Assume that  \ $\B:=\{Y_l^{\bm{m}}\}_{(l,\bm{m})\in {\sf I}}$, \ is an orthonormal basis of  $\mathcal{L}^2(S^d)$ \ consisting of eigenvectors of $\bL^2$ (e.g. spherical harmonics),
\bea\label{ArmSfe}
\bm{L}^2\,Y_l^{\bm{m}}\, =\,E_l\: Y_l^{\bm{m}},
\qquad \la Y_{l'}^{\bm{m'}},Y_l^{\bm{m}}\ra=\delta^{\bm{m}\bm{m'}}\, \delta_{ll'};
\eea
here $\bm{m}\!\in\! {\sf I}_l$ is a (multi-)index\footnote{If $D=2,3$, then ${\sf I}_l\subset \ZZ$; more precisely
${\sf I}_0=\{0\}$, while for $l>0$ it is ${\sf I}_l=\{+,-\}$ if $D=2$, ${\sf I}_l=\{-l,1\!-\!l,...,l\}$ if $D=3$. If $D>3$, then ${\sf I}_l\subset \ZZ^{D-2}$.} labelling
the elements of an orthonormal basis \ $\B_l\equiv\{Y_l^{\bm m}\}_{{\bm m}\in {\sf I}_l}$ \  of $V_D^l$, and $ {\sf I}:=\left\{(l,\bm{m})\:|\:  l\in\NN_0,\,\bm{m}\in {\sf I}_l\right\}$. \
Then, by appropriate choices of the normalization constants\footnote{Choosing them positive, one easily finds  \ $M_l=\sqrt[8]{k_l}/\sqrt[4]{\pi}$. In fact, by (\ref{ArmSfe}), (\ref{scalprodRD})  normalizing $\bpsi_{0,l}^{\bm{m}}$ amounts to $1=\int^\infty_0\!dr\,r^d\,f_l^2(r)=\int^\infty_0\!dr\,g_l^2(r)\simeq M_l^2\int^\infty_{-\infty}\!dr\,e^{- \sqrt{k_{l}} \left(r-\widetilde{r}_{l}\right)^2}$; here
$\simeq$ (due to the shift $0\mapsto-\infty$ of the left integration extreme) means equality up to terms of the order $1/\sqrt{2k}e^{-\sqrt{2k}}$, which has zero asymptotic expansion in $1/\sqrt{2k}$, see \cite{FioPis18}.}  of (\ref{valuef}), one obtains as
orthonormal bases  respectively of $\Hi$ and $\Hi_{\Lambda}^l$
$$
\B':=\left\{\bpsi_{n,l}^{\bm{m}}:=f_{n,l}(r)Y_l^{\bm{m}},\:\: | \:\:
n\in\mathbb{N}_0,\:\bm{l}\in \I\right\}, \qquad \B_l':=\{\bpsi_{0,l}^{\bm{m}}=f_l(r)\,Y_l^{\bm m}\}_{{\bm m}\in {\sf I}_l}.
$$
The  projector $\widetilde{P}_\Lambda^{l}:\Hi\to\Hi_\Lambda^{l}$ acts by
$$
\left(\widetilde{P}_\Lambda^{l}\bphi\right)(x)=\sum_{\bm{m}\in {\sf I}_l}
\bpsi_{0,l}^{\bm{m}}(x) \:\left(\bpsi_{0,l}^{\bm{m}},\bm{\phi} \right)
=\sum_{\bm{m}\in {\sf I}_l}
\bpsi_{0,l}^{\bm{m}}(x)\int_{\RR^D}\!\!d^Dx'\bpsi
_{0,l}^{\bm{m}*}(x')
\bm{\phi}(x').
$$
If $\bphi$ has the form  \ $\bm{\phi}(r,\btheta)= \Theta_{j}(\btheta)\phi(r)$,  with $\Theta_{j}\in V_D^{j}$,  \ then by (\ref{scalprodRD}), (\ref{ArmSfe})  this simplifies to 
\be
\left(\widetilde{P}_\Lambda^{l}\bm{\phi}\right)(r,\btheta)=\delta_{lj}\, 
\Theta_{j}(\btheta)\,f_l(r)\int^\infty_0\!\!\!\!r^d dr\, f_l^*(r)\phi(r),
\label{Proj_l}
\ee
which is zero if $l\neq j$, has the same angular dependence $\Theta_{j}(\btheta)$ as 
$\bm{\phi}$  if $l=j$.

In next section we provide an explicit characterization of elements  $\Theta_{l}\in V_D^l$ as polynomials
in the coordinates $t^i$ of points of the unit sphere $S^d$, which fulfill the relation
\be
t^it_i=1 \qquad\mbox{(sum over $i$)},   \label{t-relation}
\ee 
rather than as
 combinations  of spherical harmonics $Y_l^{\bm{m}}(\btheta)$, $\bm{m}\in {\sf I}_l$.

\section{Representations of $O(D)$  via polynomials in $x^i$, $t^i$}
\label{xt-polynomials}

The differential operator $\bL^2$ can be expressed as
\be
\bL^2=\eta(D\!-\!2\!+\!\eta)-r^2\Delta, \qquad \eta:=x^i\partial_i . \label{Casimir'}
\ee
The `dilatation operator' \ $\eta$ \  and the Laplacian $\Delta$ fulfill
\bea
&\eta x^i=x^i(\eta+1), \qquad &\eta \partial_i = \partial_i(\eta-1),\\[6pt]
&\Delta x^i=x^i \Delta+ 2 \partial^i,\qquad &\Delta r^2=r^2 \Delta+4\eta+ 2 D.
\label{Deltaxr}
\eea
In particular, the action of $\eta$ on monomials in the $x^i$ amounts to multiplication by their total degree.
In terms of polar coordinates it is \ $\eta=r\partial_r$, \ which replaced in (\ref{Casimir'}) 
gives  (\ref{LaplacianD}). 

\bigskip
Let  $\CC[x^1,...,x^D]$   be the space of complex polynomial functions on $\RR^D$ and, for all $l\!\in\!\NN_0$, let
$W_D^l $ be the subspace of homogeneous ones of degree $l$.
The monomials of degree $l$ $x^{i_1}x^{i_2}...x^{i_l}\in W_D^l$ 
 can be reordered in the form $(x^1)^{l_1}...(x^D)^{l_D}$ and make up a basis  of $W_D^l$:
$$
\B_{W_D^l}:=\left\{(x^1)^{l_1}...(x^D)^{l_D}\:| \: (l_1,...,l_D)\in\NN_0, \: \sum_{i=1}^Dl_i=l\right\}, \qquad\mbox{dim}(W_D^l)={D\!+\!l\!-\!1\choose l};
$$
 the dimension of $W_D^l$  is the number of elements of $\B_{W_D^l}$. Clearly $W_D^l$ carries a representation of  $O(D)$ as well as $Uso(D)$, but this is reducible if $l\ge 2$; in fact, the subspace $r^2W_D^{l-2}\subset W_D^l$ manifestly carries a smaller representation.  
We denote by $\check V_D^{l}$  the ``trace-free" component of  $W_D^l$, namely the subspace such that \ 
$W_D^l=\check V_D^l\oplus r^2W_D^{l-2}$. \ As a consequence,
\be
\mbox{dim}(\check V_D^l)=\mbox{dim}(W_D^l)-\mbox{dim}(W_D^{l-2})
=\frac{(l\!+\!D\!-\!3)...(l\!+\!1)}{(D\!-\!2)!}(D\!+\!2l\!-\!2).  \label{dimVDl} 
\ee
$\check V_D^{l}$ carries the irreducible representation (irrep) $\bpi_D^{l}$ of $Uso(D)$ and $O(D)$  characterized by the highest eigenvalue of $\bL^2$ within $W_D^l$, namely $E_l$  (the eigenvalues of all other $[D/2\!-\!1]$ Casimirs are determined by $l$). 
Abbreviating \ $X_{l,\pm}^{hk}:=(x^h\!\pm\!i x^k)^l$, \ 
this can be easily shown  observing that for all $h,k\in\{1,...,D\}$
 $X_{l,\pm}^{hk}\in W_D^l$ are annihilated by $\Delta$ and are eigenvectors of $\bL^2$ with that eigenvalue; moreover, they are eigenvectors of $L_{hk}$ with eigenvalue $\pm l$. Hence \
$X_{l,+}^{hk}$, $X_{l,-}^{hk}$ \ can be used as the highest and lowest weight vectors of $\check V_D^{l}$\footnote{In fact, in terms of Cartesian coordinates, using (\ref{Casimir'}),  (\ref{Deltaxr})   we immediately find the following commutation relations among  operators of multiplication $(x^h\!\pm\!i x^k)\cdot$ and differential operators 
\bea
&&(\partial_h\!+\!i\partial_k)(x^h\!+\!ix^k)=(x^h\!+\!ix^k)(\partial_h\!+\!i\partial_k),\quad
(\partial_h\!-\!i\partial_k)(x^h\!-\!ix^k)=(x^h\!-\!ix^k)(\partial_h\!-\!i\partial_k)  \nn[8pt]
&&\Delta(x^h\!\pm\!ix^k)=(x^h\!\pm\!ix^k)\Delta+2 (\partial_h\!\pm\!i\partial_k), 
\qquad 
\Delta\, X_{l,\pm}^{hk}=X_{l,\pm}^{hk}\,\Delta+2lX_{l-1,\pm}^{hk} (\partial_h\!+\!i\partial_k),\nn[8pt]
&& L_{hk}(x^h\!\pm\!i x^k)=(x^h\!\pm\!i x^k) \big(L_{hk}\pm 1\big), 
\qquad L_{hk}\, (L_{hj}\!\pm\!iL_{kj})= (L_{hj}\!\pm\!iL_{kj})\,\big[L_{hk}\pm1\big]. \label{Lxhkcr}
\eea
Consequently, we obtain the following functions at the rhs as results of the operator actions on functions at the lhs:  \ $\Delta X_{l,\pm}^{hk}=0$ \  and,  for all functions $g(r)$, 
\bea
&& \bL^2\, g(r)\, X_{l,\pm}^{hk} =g(r)\left[\eta(D\!-\!2\!+\!\eta)-r^2\Delta\right]X_{l,\pm}^{hk}=E_l\, g(r)\,X_{l,\pm}^{hk},
\label{HighestWeights}\\[8pt]
&& L_{hk}\,X_{l,\pm}^{hk}\, g(r)=\pm l\, X_{l,\pm}^{hk}\, g(r) ,  \qquad
L_{hk} \, X_{l,+}^{hk}  X_{m,-}^{hk}\, g(r) =(l\!-\!m)\, X_{l,+}^{hk}  X_{m,-}^{hk}\, g(r). \label{HighestWeights'}
\eea
Denoting by $\tau=\left[\frac D2\right]$ the rank of $so(D)$,
as a basis of a Cartan subalgebra of $so(D)$ one can take any set \
$\{H_1\equiv L_{i_1i_2},H_2\equiv L_{i_3i_4},...,H_\tau\equiv L_{i_{2\tau-1}i_{2\tau}}\}$,  \ with  $i_1,i_2,...,i_{2\tau}\in\{1,...,D\}$ all different from each other. If $(i_1,i_2)=(h,k)$,
by (\ref{HighestWeights}), (\ref{HighestWeights'})  $X_{l,+}^{hk}$,
$X_{l,-}^{hk}$ are the corresponding highest, lowest weight vectors, in the sense 
\bea
H_1 \,X_{l,\pm}^{hk} g(r) =\pm l \,X_{l,\pm}^{hk} g(r) ,\qquad
H_a \,X_{l,\pm}^{hk} g(r) =0\:\:\mbox{if } a>1.  \label{HighestWeights''}
\eea}.
Since all the $L_{ij}$ commute with $\Delta$, $\check V_D^{l}$  can be characterized also as the subspace of $W_D^l$ which is annihilated by $\Delta$. 
A complete set
in $\check V_D^{l}$ consists of trace-free homogeneous polynomials
$X_l^{i_1i_2...i_l}$, which we will obtain below
applying the completely symmetric trace-free projector $\Ps^l$ to the $x^{i_1}x^{i_2}...x^{i_l}$'s.

We slightly enlarge $\CC[x^1,...x^D]$ introducing as new generators $r,r^{-1}$ subject
to the relations \ $r^2=x^ix_i$ (sum over $i$), $rr^{-1}=1$.   Inside this enlarged algebra the elements 
\be
t^i:=\frac{x^i}r
\ee
fulfill the relation  (\ref{t-relation})
characterizing the coordinates of points of the unit sphere $S^d$. 
Choosing $g(r)=r^{-l}$ in (\ref{Lxhkcr}-\ref{HighestWeights''}) we obtain the same relations with \ $x^i,\,X_{l,\pm}^{hk}$ replaced by $t^i,\,  T_{l,\pm}^{hk}:=(t^h\!\pm\!i t^k)^l$. \
We shall denote by $Pol_D$  the algebra of complex polynomials in such $t^i$, by $Pol_D^{\Lambda}$ the subspace
of polynomials up to degree $\Lambda$, by $P^{\Lambda}:Pol_D\to Pol_D^{\Lambda}$ the corresponding projector.  $Pol_D$ endowed with the scalar product \ 
$\la T,T'\ra:=\int_{S^d}d\alpha\, T^*T'$ \ is a pre-Hilbert space; its completion is 
${\cal L}^2(S^d)$. We  extend $P^{\Lambda}$ to all of ${\cal L}^2(S^d)$ by continuity
in the norm of the latter. Also $Pol_D^\Lambda, V_D^l$  are Hilbert subspaces of ${\cal L}^2(S^d)$.
$Pol_D^{\Lambda}=W_D^{\Lambda}r^{-\Lambda}\oplus W_D^{\Lambda-1}r^{1-\Lambda}$ carries a unitary reducible representation of $O(D)$ [and $Uso(D)$]  which splits 
via \ $Pol_D^{\Lambda}=\bigoplus_{l=0}^\Lambda  V_D^l$ \ into irreps
carried by $V_D^l:=\check V_D^{l}/r^{l}$.  
Its dimension is thus
\bea
\mbox{dim}\left(Pol_D^{\Lambda}\right)&=& \sum_{l=0}^\Lambda\mbox{dim}(\check V_D^l)= \sum_{l=0}^\Lambda\mbox{dim}(W_D^l)-\sum_{l=2}^\Lambda\mbox{dim}(W_D^{l-2})\nn
&=& \mbox{dim}(W_D^\Lambda)+\mbox{dim}\left(W_D^{\Lambda-1}\right)\nn[6pt]
&=&\frac{(D\!+\!\Lambda\!-\!1)...(\Lambda\!+\!1)}{(D\!-\!1)!}+\frac{(D\!+\!\Lambda\!-\!2)...\Lambda}{(D\!-\!1)!}\nn
&=& \frac{(D\!+\!\Lambda\!-\!2)...(\Lambda\!+\!1)}{(D\!-\!1)!}\left(D\!+\!2\Lambda\!-\!1\right)=:N    \label{defN}\\
&\stackrel{(\ref{dimVDl} )}{=}&\mbox{dim}\left(V_{D+1}^\Lambda\right) .
\eea
This suggests that $\Hi_\Lambda\simeq Pol_D^{\Lambda}\simeq  V_{D+1}^\Lambda$
 as $Uso(D)$ (reducible) representations. We  have proved the first isomorphism
in section  \ref{genset} and will prove the second in section \ref{Embeddings}.

\subsection{$O(D)$-irreps via trace-free  completely symmetric projectors}
\label{ComplSymProj}

Let $(\pi,\E)$ be the $D$-dimensional irreducible unitary representation of $Uso(D)$ and
$O(D)$; the carrier space $\E$ is isomorphic to $V_D^1$. As a vector space $\E\simeq \RR^D$;
the set of coordinates $x:=(x^1,...x^D)\in\RR^D$ can be seen as the set of components
of an element of $\E$ with respect to (w.r.t.) an orthonormal basis. The permutator on \ 
$\E^{\otimes^2}\equiv \E \otimes \E$ \ is defined via $\Pe(u\otimes v)=v\otimes u$
and linearly extended. In all bases it is represented by the $D^2\times D^2$
matrix \  $\Pe^{hi}_{jk}=\delta^{h}_{k}\delta^{i}_{j}$. 
The symmetric and antisymmetric  projectors $\Ps^+,\Ps^-$ on $\E^{\otimes^2}$ are obtained as
\bea
\Ps^\pm =\frac 12\left(\1_{D^2}\pm \Pe\right);                \label{sym-antisym_projectors}
\eea
here and below we denote by $\1_{D^l}
$ the identity operator on $\E^{\otimes^l}$, which in all bases  is represented by the $D^l\times D^l$ matrix
$\1_{D^l}{}^{h_1...h_l}_{i_1...i_l}:=\delta^{h_1}_{i_1}...\delta^{h_l}_{i_l}$.
The antisymmetrized tensor product $\Ps^-\E^{\otimes^2}$ is an irrep under $O(D)$, while the symmetrized one $\Ps^+\E^{\otimes^2}$  contains two irreps:
the 1-dim trace one and the trace-free symmetric one. The matrix representation of the 
1-dim  projector $\Pt$ on the former is
\be 
\Pt{}_{kl}^{ij} = \frac 1{D} g^{ij}g_{kl}           \label{Pt} 
\ee 
where the $D \times D$ metric matrix $g_{ij}$  (in the chosen basis)
is a $so(D)$-isotropic symmetric tensor,  and $g^{ij}g_{jh}=\delta^i_h$, whence $g^{ij}g_{ij}=D$. Here we shall use an orthonormal basis of $\E$, whence $g_{ij}=g^{ij}=\delta_{ij}$, and indices of vector components can be raised or lowered freely, e.g. $x_i=x^i$. The
$\frac12(D\!-\!1)(D\!+\!2)$-dim trace-free symmetric projector $\Ps^s$  is given by
\be
\Ps^s:=\Ps^+-\Pt=\frac 12\left(\1_{D^2}+ \Pe
\right)-\Pt \:.                  \label{projectorpm2}
\ee
These projectors satisfy the equations 
\be 
\Ps^{\alpha}\Ps^{\beta} = \Ps^\alpha \delta^{\alpha\beta}, \qquad 
\sum_\alpha\Ps^{\alpha}= \1_{D^2},            \label{projector1} 
\ee 
where $\alpha,\beta = -,s,t$.  In the sequel we shall abbreviate $\Ps\equiv\Ps^s$. This implies in particular \ $\Ps\Ps^{'}=0$, \ where 
we have introduced the new projector \ $\Ps^{'}:=\Ps^{-}+\Pt$. \
$\Pe,\Pt$ are symmetric matrices, i.e. invariant under transposition
${}^T$, and therefore also the other projectors are:
\be
\Pe^T=\Pe,\qquad\qquad \Ps^{\alpha}{}^T={\cal P}^{\alpha}.   \label{sym}  
\ee
Given a (linear) operator $M$  on  $\E^{\otimes^n}$,
for all integers $l,h$ with $l>n$,  and $1\le h\le l\!+\!1\!-\!n$ we denote by 
$M_{h(h\!+\!1)...(h\!+\!n\!-\!1)}$ the operator  on  $\E^{\otimes^l}$ acting
as the identity on the first $h\!-\!1$ and the last $l\!+\!1\!-\!n\!-\!h$ tensor factors,
and as $M$ in the remaining central ones. For instance, if $M=\Pe$ and $l=3$ 
we have $\Pe_{12} = \Pe \otimes \1_D$,  
$\Pe_{23} =  \1_D\otimes \Pe $.
It is straightforward to check 
\begin{prop}
All the projectors 
$A=\Ps^+,\Ps^-,\Ps,\Pt, \Ps^{'}$ fulfill the ``braid" relation  
\be 
A_{12}\,\Pe_{23}\,\Pe_{12}  
=\Pe_{23}\,\Pe_{12}\,A_{23}.   \label{braid1} 
\ee 
Moreover, 
\bea
D\: \Pt_{23}\Pt_{12}=\Pe_{12}\Pe_{23}\Pt_{12},\quad\qquad
D\Pe_{12}\Pt_{23}\Pt_{12}=\Pe_{23}\Pt_{12}, \label{useful1}  \\[8pt] 
D\: \Pt_{12}\Pt_{23}=\Pe_{23}\Pe_{12}\Pt_{23} ,\quad\qquad
D\: \Pe_{23}\Pt_{12}\Pt_{23}=\Pe_{12}\Pt_{23,} \label{useful1'}  \\[8pt] 
D\:\Pt_{23} \Pt_{12}=\Pt_{23}\Pe_{12}\Pe_{23} ,\quad\qquad
D\: \Pt_{23}\Pt_{12}\Pe_{23}=\Pt_{23}\Pe_{12}; \label{useful1''}  
\eea
Eq. (\ref{braid1}-\ref{useful1''}) hold also for   $l>3$, e.g. for all $2\le h\le l-1$
\be 
A_{(h-1)h}\,\Pe_{h(h+1)}\,\Pe_{(h-1)h}  
=\Pe_{h(h+1)}\,\Pe_{(h-1)h}\,A_{h(h+1)}. \label{braid2} 
\ee 
\end{prop}
\bp{}
Since $A=\1_{D^2},\Pe,\Pt$ fulfill (\ref{braid1}), then also $A=\Ps^+,\Ps^-,\Ps,\Pt, \Ps^{'}$ do.  One can immediately check the first equality  in (\ref{useful1})  via direct calculation;
left multiplying the first by $\Pe_{12}$  one obtains the second. Eq. (\ref{useful1'}) are obtained 
from (\ref{useful1}) exchanging $1\leftrightarrow 3$ and using the symmetry of $\Pe,\Pt$ under the flip. Eq.  (\ref{useful1''}) are obtained 
from (\ref{useful1'})  by transposition. \ep

Next, we define and determine the completely symmetric  trace-free projector 
$\Ps^{l}$ on $\E^{\otimes^l}$ generalizing $\Ps^2\equiv\Ps$ to $l> 2$.
It  projects the tensor product of  $l$ copies of $\E$ to the carrier space 
of the $l$-fold completely symmetric irrep  of $Uso(D)$, isomorphic to $V^{l}_D$,
therein contained. It is uniquely characterized by the following properties:
\bea 
&&\ba{l}\Ps^{l}\Ps^-_{n(n\!+\!1)}=0,\quad \Ps^{l}\Pt_{n(n\!+\!1)}=0,  \\[8pt]  
\Ps^-_{n(n\!+\!1)}\Ps^{l}=0,\quad
\Pt_{n(n\!+\!1)}\Ps^{l}=0,\ea\qquad\qquad n=1,...,l\!-\!1\label{Plproj1}  \\[8pt]
&&\left(\Ps^{l}\right)^2=\Ps^{l},   \label{Plproj2}       
\eea
Consequently, it is also \ $\mbox{tr}_{1\ldots l}\!\left({\cal
P}^{l}\right) =\mbox{dim}(V^{l}_D)$, which 
guarantees that $\Ps^{l}$ acts as the identity (and not
as a proper projector) on $V^{l}_D$. \ The right relations in (\ref{Plproj1}) amount to
\be
\Ps^{l}{}^{i_1...i_l}_{j_1...j_l}\delta^{j_nj_{n+1}}=0,\quad
\delta_{i_ni_{n+1}}\Ps^{l}{}^{i_1...i_l}_{j_1...j_l}=0,\qquad\qquad n=1,...,l\!-\!1.
 \label{Plprojg} 
\ee
Clearly the whole of (\ref{Plproj1}) can be summarized
as $\Ps^{l}\Ps^{'}_{n(n\!+\!1)}=0=\Ps^{'}_{n(n\!+\!1)}\Ps^{l}$.
It is straightforward to prove that the above properties imply also the ones
\be
 \Ps^{l}\Pe_{n(n\!+\!1)}=\Ps^{l}, \quad n=1,...,l\!-\!1;\qquad
\Ps^{l}\Ps^{h}_{(i+1)...(i+h)}=\Ps^{l},\quad  h<l, \:\: 0\le i\le l\!-\!h.
\label{furtherprop}
\ee

\begin{prop} The projector $\Ps^{l\!+\!1}$ can be expressed
as a polynomial in the permutators $\Pe_{12},...,\Pe_{(l\!-\!1)l}$  and trace projectors
$\Pt_{12},...,\Pt_{(l\!-\!1)l}$ through either recursive relation
\bea
\Ps^{l\!+\!1}&=&
\Ps^{l}_{12...l}M_{l(l\!+\!1)}\Ps^{l}_{12...l},
\label{ansatz1} \\[8pt]
&=& \Ps^{l}_{2...(l\!+\!1)}
M_{12}\Ps^{l}_{2...(l\!+\!1)}, \label{ansatz2} 
\eea
\be
\mbox{where } \quad M\equiv M(l\!+\!1) =\frac 1{l\!+\!1} \!\left[\1_{D^2}+ l\,\Pe -
\frac{2Dl}{D\!+\!2l\!-\!2} \Pt\right]                \label{Ml}
\ee
As a consequence, the $\Ps^{l}$ are symmetric, $(\Ps^{l})^T=\Ps^{l}$.
\label{symmetrizers}
\end{prop}
This the analog of Proposition 1  in \cite{Fio04JPA} for the quantum group $U_qso(D)$ covariant symmetric projectors; the proof is in the appendix. 
By a straightforward computation one checks that
\be
\Ps^{l+1}{}^{hi_1...i_l}_{hj_1...j_l}= \frac 1{l\!+\!1} \!\left[D+ l -
\frac{2l}{D\!+\!2l\!-\!2}\right] \Ps^l{}^{i_1...i_l}_{j_1...j_l} 
\label{contractPs}
\ee
($h$ is summed over).
Using (\ref{Deltaxr}),   (\ref{Plprojg}) one easily shows that the homogeneous polynomials
\be
X_l^{i_1...i_l}:=\Ps^l{}^{i_1...i_l}_{j_1...j_l} x^{j_1} ... x^{j_l}\in \check V^l_D 
\label{defXD}
\ee
are harmonic, i.e. satisfy $\Delta X_l^{i_1...i_l}=0$;  using (\ref{Casimir'}), we find that they are eigenvectors of $\bL^2$,
\be
\bL^2\,X_l^{i_1...i_l}=\,E_l\,X_l^{i_1...i_l},
\label{LeigenvectorsX}
\ee 
with eigenvalues (\ref{El}).
They make up a complete set in $\check V_D^l$, which  can be thus also characterized  as the subspace  of  $W_D^l$ that is annihilated by $\Delta$, whereas $\Delta \phi\neq 0$ for all $\phi\in r^2W_D^{l-2}$. The $X_l^{i_1...i_l}$ are not all independent, because they are invariant under 
permutations of $(i_1...i_l)$ and by (\ref{Plprojg}) fulfill the linear dependence relations
\be
\delta_{i_ni_{n+1}}X_l^{i_1...i_l}=0,\qquad\qquad n=1,...,l\!-\!1.              
\label{gX=0} 
\ee

\begin{prop} In a compact notation,
\bea
\ba{l}
\left(\Ps^{l}_{1...l}-\Ps^{l+1}\right)x_1...x_{l+1}=\zeta_{l+1}\,\Ps^{l}_{1...l}\Pt_{l(l\!+\!1)} x_1...x_{l+1},\\[8pt]
\left(\Ps^{l}_{2...(l\!+\!1)}-\Ps^{l+1}\right)x_1...x_{l+1} =\zeta_{l+1}\,
\Ps^{l}_{2...(l\!+\!1)}\Pt_{12} x_1...x_{l+1},
\ea
\qquad \zeta_{l+1}=\frac {D\, l}{D\!+\!2l\!- \!2}.                   \label{difference-l}
\eea
\label{Difference-l}
\end{prop}
The proof is in Appendix \ref{ProofDifference-l}. More explicitly, (\ref{difference-l}) becomes 
\bea
x^hX_l^{i_1...i_l}-X_{l+1}^{hi_1...i_l}=\frac{\zeta_{l+1}}Dr^2\,\Ps^l{}^{i_1i_2...i_l}_{hj_2...j_l}x^{j_2} ... x^{j_l} =\frac{\zeta_{l+1}}Dr^2\,\Ps^l{}^{i_1i_2...i_l}_{hj_2...j_l}X_{l-1}^{j_2...j_l}         \label{difference-l'}
\eea
Contracting the previous relation with $\delta_{hi_1}$ and using (\ref{Plprojg}) we obtain
\bea
x^hX_l^{hi_2...i_l}=\frac{\zeta_{l+1}}Dr^2\,\Ps^l{}^{hi_2...i_l}_{hj_2...j_l}X_{l-1}^{j_2...j_l}\stackrel{(\ref{contractPs})}{=} \frac {r^2}{D\!+\!2l\!-\!2}
\left[D\!+\! l \!-\!1\!-\!
\frac{2l\!-\!2}{D\!+\!2l\!-\!4}\right]
X_{l-1}^{i_2...i_l}.        \label{xXcontraction}
\eea
In the Appendix we also prove

\begin{prop} The maps $L_{hk}:  \check V_D^l\to    \check V_D^l$ explicitly act as follows:
\bea
\ba{lll}
iL_{hk}X_l^{i_1...i_l} &= &\displaystyle l\,\frac{\zeta_{l+2}}{\zeta_{l+1}}
\left(\Ps^{l+1}{}^{hi_1...i_l}_{kj_1...j_l}
-\Ps^{l+1}{}^{ki_1...i_l}_{hj_1...j_l}\right)X_l^{j_1...j_l}, \\[8pt]
&= & \displaystyle l\,\Ps^l{}^{i_1...i_l}_{j_1...j_l}
\left(\delta^{kj_1}X_{l}^{hj_2...j_l}-\delta^{hj_1}X_{l}^{kj_2...j_l} \right).
\ea     \label{LonX}
\eea
 \label{LOnX}
\end{prop}

Dividing (\ref{LeigenvectorsX}), (\ref{gX=0}),  (\ref{difference-l'}), (\ref{xXcontraction}), (\ref{LonX}) by the appropriate powers of $r$ we find

\begin{prop} The \ $T_l^{i_1i_2...i_l}:=X_l^{i_1i_2...i_l}/r^l=\Ps^l{}^{i_1...i_l}_{j_1...j_l} t^{j_1} ... t^{j_l}$ \ belong to $V_D^l$, because
\bea
\bL^2\,T_l^{i_1...i_l}  &=& E_l\,T_l^{i_1...i_l};
\label{LeigenvectorsT}
\eea
they make up a complete set $\T_l$ in it, but not a basis, because they are invariant under 
permutations of $(i_1...i_l)$ and by (\ref{gX=0}) fulfill the linear dependence relations
\be
\delta_{i_ni_{n+1}}T_l^{i_1...i_l}=0,\qquad\qquad n=1,...,l\!-\!1.              
\label{gT=0} 
\ee
The actions of the operators \ $t^{h}\cdot$,  $iL_{hk}$ on the $T_l^{i_1...i_l}$ explicitly read
\bea
&& t^{h}\,T_{l}^{i_1...i_l} = T_{l+1}^{hi_1...i_l}+\frac {l}{D\!+\!2l\!- \!2}\,
\Ps^{l}{}^{i_1i_2...i_l}_{hj_2...j_l} T_{l-1}^{j_2...j_l}\: \in\:   V_D^{l+1}\oplus   V_D^{l-1},   \label{tTdeco}\\[6pt]
&& t^iT_l^{ii_2...i_l}=  \frac {1}{D\!+\!2l\!-\!2}
\left[D\!+\! l \!-\!1\!-\!
\frac{2l\!-\!2}{D\!+\!2l\!-\!4}\right]
T_{l-1}^{i_2...i_l}        \: \in\:      V_D^{l-1},     \label{tTcontraction}\\[10pt]
&&\ba{lll}
iL_{hk}T_l^{i_1...i_l} &= &\displaystyle l\,\frac{\zeta_{l+2}}{\zeta_{l+1}}\left(\Ps^{l+1}{}^{hi_1...i_l}_{kj_1...j_l}
-\Ps^{l+1}{}^{ki_1...i_l}_{hj_1...j_l}\right)T_l^{j_1...j_l}, \\[8pt]
&= & \displaystyle l\,\Ps^l{}^{i_1...i_l}_{j_1...j_l}
\left(\delta^{kj_1}T_{l}^{hj_2...j_l}-\delta^{hj_1}T_{l}^{kj_2...j_l} \right).
\ea     \label{LonT}
\eea
\label{propT}
\end{prop}

For all $\bphi\in\Hi_s\equiv {\cal L}^2(S^d)$ let 
\be
\bphi=\sum_{l=0}^{\infty}\sum_{\bm{m}\in {\sf I}_l}\phi^l_{\bm{m}} Y_l^{\bm{m}}
=\sum_{l=0}^{\infty}\sum_{i_1,....,i_l}\phi^l_{i_1...i_l} T_l^{i_1...i_l}    \label{decophi}
\ee
be its decompositions  in the basis of spherical harmonics and in the complete set
$\T:=\bigcup_{l=0}^{\infty}\T_l$; here the two sets of coefficients are related by
$\phi^l_{i_1...i_l}=\sum_{\bm{m}\in {\sf I}_l}\phi^l_{\bm{m}}A^{\bm{m}}_{i_1...i_l}$, where $A^{\bm{m}}_{i_1...i_l}$ are 
such that $Y_l^{\bm{m}}=\sum A^{\bm{m}}_{i_1...i_l}T_l^{i_1...i_l}$.
The $\phi^l_{i_1...i_l}$ are uniquely determined if, as we shall assume, we choose them trace-free and completely symmetric, i.e. fulfilling
\be
\phi^l_{i_1...i_l}= \phi^l_{j_1...j_l} \,\Ps^l{}_{i_1...i_l}^{j_1...j_l}, 
\label{phi-coeff-prop}
\ee
whence $\phi^l_{i_1...i_l}\delta^{i_ni_{n+1}}=0$ for $n=1,...,l\!-\!1$. Then (\ref{decophi}) can
be also written in the form
\be
\bphi=\sum_{l=0}^{\infty}\sum_{i_1,....,i_l}\phi^l_{i_1...i_l}t^{i_1}...\,t^{i_l}.    \label{decophi'}
\ee
The projector $P_ \Lambda$ acts  by truncation, \ 
$P_ \Lambda\bphi\equiv\bphi_ \Lambda:=\sum_{l=0}^{\Lambda}\sum_{\bm{m}\in {\sf I}_l}
\phi^l_{\bm{m}} Y_l^{\bm{m}}=\sum_{l=0}^{\Lambda}
\phi^l_{i_1...i_l} T_l^{i_1...i_l}$ (sum over $i_1,....,i_l$).
Clearly 
$\phi_ \Lambda\stackrel{\Lambda\to\infty}{\longrightarrow}\phi$
in the $\Hi_s$-norm $\Vert\,\Vert$.

\medskip
All completely symmetric, $O(D)$-isotropic tensors
of even rank ${\overline{N}}$ are proportional to 
\bea
G_{\overline{N}}^{i_1...i_{\overline{N}}}:=\delta^{i_1i_2}\delta^{i_3i_4}...\delta^{i_{{\overline{N}}-1}i_{\overline{N}}}\,+
\mbox{permutations of }\: (i_1,...,i_{\overline{N}}).     \label{DefGN}
\eea

\begin{prop}
The `trace' of $G_{\overline{N}}$, \
$\mbox{tr}\left(G_{\overline{N}}\right):=\delta_{i_1i_2}\delta_{i_3i_4}...\delta_{i_{{\overline{N}}-1}i_{\overline{N}}}G_{\overline{N}}^{i_1...i_{\overline{N}}}$, \
is equal to
\be
\mbox{tr}\left(G_{\overline{N}}\right)=\overline{N}!!\,D(D\!+\!2)...(D\!+\!{\overline{N}}\!-\!2).  \label{TraceG}
\ee
The $O(D)$-invariant integral over $S^d$ of 
the tensor $H_{\overline{N}}^{i_1...i_{\overline{N}}}:=t^{i_1}...t^{i_{\overline{N}}}$
is equal to
\be
\int_{S^d} d\alpha\: t^{i_1}...t^{i_{\overline{N}}}=C_{\overline{N}}\, G_{\overline{N}}^{i_1...i_{\overline{N}}},
\qquad C_{\overline{N}}=\frac{\mbox{mis}(S^d)}{\overline{N}!!\,D(D\!+\!2)...(D\!+\!{\overline{N}}\!-\!2)}.
\label{inv-t-integral}
\ee
In terms of the decompostions (\ref{decophi}b)-(\ref{phi-coeff-prop}) the scalar product of  generic $\bphi,\bpsi \subset\Hi_s$  is equal to
\bea
\la\bphi,\bpsi\ra = \sum_{l=0}^{\infty} Q_l 
\big(\phi^l_{i_1...i_l}\big)^* \psi^l_{i_1...i_l},
\qquad Q_l:=C_{2l} \,2^l(l!)^2 =\frac{\mbox{mis}(S^d)\,l!\,(D\!-\!2)!!}{(D\!+\!2l\!-\!2)!!}.
\label{productphipsi}
\eea
\label{traceG}
\end{prop}

In particular, (\ref{productphipsi}) implies \ $\left\la T_l^{i_1...i_l},T_n^{j_1...j_n}\right\ra = 
\delta_{ln}Q_l \, \Ps^l{}_{i_1...i_l}^{j_1...j_l}$ \ and more generally
\bea
 \left\la T_l^{i_1...i_l},\bphi\right\ra = 
Q_l \, \Ps^l{}_{i_1...i_l}^{j_1...j_l}\phi^l_{j_1...j_l}\stackrel{(\ref{phi-coeff-prop})}{=}Q_l \,\phi^l_{i_1...i_l},
\label{productTl-phi}
\eea
the second equality holds only
if (\ref{phi-coeff-prop}) holds. Then, we have also
\bea
&& \Vert\bphi\Vert^2 \equiv \la\bphi ,\bphi\ra=  \sum_{l=0}^{\infty}
Q_l \sum_{i_1,...,i_l}\!\!\big|\phi^l_{i_1...i_l}\big|^2.
\label{square-norm-phi}
\eea

We now determine the decomposition of $T_l^{i_1...i_l}T_m^{j_1...j_m}$.

\begin{theorem} The product $T_l^{i_1...i_l}T_m^{j_1...j_m}$ decomposes as follows
into  $V_D^n$ components:
\bea
T_l^{i_1...i_l}T_m^{j_1...j_m}= \sum\limits_{n\in L^{lm}}\, 
S^{i_1...i_l,j_1...j_m}_{k_1...k_n}T_n^{k_1...k_n}, 
\label{TTdeco}
\eea
where  $L^{lm}:=\{|l\!-\!m|,|l\!-\!m|\!+\!2,...,l\!+\!m\}$ and,  defining \ 
$\displaystyle r\!:=\!\frac{l\!+\!m\!-\!n}2\in\{0,1,...,m\}$, 
\bea
\ba{l}
\displaystyle
S^{i_1...i_l,j_1...j_m}_{k_1...k_n}= N^{lm}_n \,V^{i_1...i_l,j_1...j_m}_{k_1...k_n}, \qquad
N^{lm}_n = \frac{(D\!+\!2n\!-\!2)!!\,l!\,  m!}{(D\!+\!2n\!+\!2r\!-\!2)!!\, (l\!-\!r)!\,(m\!-\!r)!}\\[10pt]
V^{i_1...i_l,j_1...j_m}_{k_1...k_n} =  
 \Ps^l{}_{a_1...a_rc_{1}...c_{l-r}}^{i_1...i_l}\Ps^m{}_{a_1...a_r c_{l-r+1}...c_n}^{j_1...j_rj_{r+1}...j_m} \Ps^n{}_{c_1...c_n} ^{k_1...k_n}. 
\ea\label{Scondi''}
\eea
\label{propTTdeco}
\end{theorem}

The coefficients $S^{i_1...i_l,j_1...j_m}_{k_1...k_n}$ are the analogs 
of the Clebsch-Gordon coefficients, which appear in the decomposition of a product of two spherical harmonics into a combination of spherical harmonics for $D=3$. 
The first term of the sum  (\ref{TTdeco}) is \ $T_{l+m}^{i_1...i_lj_1...j_m}$. This is consistent with the first term in the  iterated application of  (\ref{tTdeco}). 
If $r\!=\!m\!=\!1$, since $\Ps^1{}_{a_1}^{j_1}=\delta^{j_1}_{a_1}$, $n=l\!-\!1$, the result is consistent  with  the second term in  (\ref{tTdeco}):
\bea
S^{i_1...i_l,j_1}_{k_1...k_{l\!-\!1}}
 =\frac{l}{D\!+\!2l\!-\!2}\, \Ps^l{}_{j_1k_1...k_{l-1}}^{i_1...i_l}.\nonumber
\eea

\subsection{Embedding in $\RR^{D+1}$,  isomorphism
$\mbox{End}\!\left(Pol_D^\Lambda\right)\simeq {\bf \pi}_{D+1}^{\Lambda}\big[Uso(D\!+\!1)\big]$}
\label{Embeddings}

We naturally embed $\CC[\RR^D]\hookrightarrow\CC[\RR^\bD]$, 
where $\bD\equiv D+1$.
Henceforth we use real Cartesian coordinates $(x^i)$ for $\RR^D$ and $(x^I)$ for $\RR^\bD$; $h,i,j,k\in\{1,...,D\}$, $H,I,J,K\in\{1,...,\bD\}$. We naturally embed $O(D)\hookrightarrow SO(\bD)$ identifying
$O(D)$ as the subgroup  of $SO(\bD)$ which is the little group of the  $\bD$-th axis; its Lie algebra, isomorphic to $so(D)$,  is 
generated by the $L_{hk}$. 
We shall add $\bD$ as a subscript to distinguish objects in this enlarged dimension from their counterparts in dimension $D$, e.g. the distance $r_\bD$   from the origin in $\RR^\bD$, 
from its counterpart $r\equiv r_D$ in $\RR^D$, $\Ps^l_\bD$ from  $\Ps^l\equiv \Ps^l_D$, 
and so on.

We look for the decomposition of each $\check V_\bD^\Lambda$ into irreps of such a $Uso(D)$.
Clearly, $\check V_\bD^0\simeq\CC\simeq V_D^0$, while $\check V_\bD^1= \check V_{D,1}^0\oplus \check V_{D,1}^1$, where $\check V_{D,1}^0\simeq  V_D^0$ is spanned by $x^\bD$,  and  $\check V_{D,1}^1\simeq V_D^1$ is spanned by the $x^i$.
The $X_{\bD,2}^{IJ}=x^Ix^J-r_\bD^2\delta^{IJ}/\bD$ span $\check V_\bD^2$; the set of elements
\bea
X_{\bD,2}^{\bD\bD}=x^\bD x^\bD-\frac{r_\bD^2}{\bD}
, \qquad X_{\bD,2}^{i\bD}=x^i x^\bD, \qquad X_{2}^{ij}=X_{\bD,2}^{ij}+\frac{\delta^{ij}}{D}X_{\bD,2}^{\bD\bD}
\eea
span carrier spaces  $\check V_{D,2}^0,\check V_{D,2}^1,\check V_{D,2}^2$ respectively isomorphic to 
$V_D^0,V_D^1,V_D^2$, and \ $\check V_\bD^2=\check V_{D,2}^2\oplus \check V_{D,2}^1\oplus \check V_{D,2}^0$. \  
More generally, $\check V_\bD^\Lambda$ is spanned by the 
the $X_{\bD,\Lambda}^{I_1...I_\Lambda}$, which 
are homogeneous polynomials of degree $\Lambda$ in the $x^I$ obtained by (\ref{defXD}) in dimension $\bD$, i.e.
\be
X_{\bD,l}^{I_1...I_l}:=\Ps_\bD^l{}^{I_1...I_l}_{J_1...J_l} x^{J_1} ... x^{J_l}\in \check V^l_\bD 
\label{defXbD}
\ee 
where the projectors $\Ps_\bD^l$ are constructed as in Proposition \ref{symmetrizers}, but with $D$ replaced by $\bD$. 
Any pair of indices $I_a,I_b$ appears either through the product $x^{I_a}x^{I_b}$ or through $r_\bD^2\delta^{I_aI_b}$.
If we introduce $r_\bD$ as a further generator constrained by the relation $r_\bD^2=x^Ix^I$ (sum over $I$), then the
$X_{\bD,\Lambda}^{I_1...I_\Lambda} $ can be seen also as homogeneous polynomials of degree $\Lambda$ in  $x^I,r_\bD$. \ Since $\delta^{j_a\bD}=0$,  in $X_{\bD,\Lambda}^{j_1...j_l\bD...\bD}$:  any pair of indices $\bD\bD$ appears    either through the product $x^{\bD}x^{\bD}$ or through 
$r_\bD^2\delta^{\bD\bD}=r_\bD^2$;  any pair of indices $j_a,\bD$ appears through the product $x^{j_a}x^{I_b}$; any pair of indices $j_a,j_b$ appears either through the product $x^{j_a}x^{j_b}$ or through $r_\bD^2\delta^{j_aj_b}$. By property (\ref{Plprojg}), the latter terms
completely disappear in any combination
$\Ps^l{}^{i_1...i_l}_{j_1...j_l}X_{\bD,\Lambda}^{j_1...j_l\bD...\bD}$, $l\in\{2,3,...,\Lambda\}$. \
 Therefore such a combination can be factorized as follows
\be
\check F^{i_1...i_l}_{\bD,\Lambda}\: := \: \Ps^l{}^{i_1...i_l}_{j_1...j_l}X_{\bD,\Lambda}^{j_1...j_l\bD...\bD} \: = \:
 \check p_{\Lambda,l}\,X_{l}^{i_1...i_l}, 
\label{cFDeco}
\ee
where $\check p_{\Lambda,l}$ is a homogeneous polynomial of degree $\Lambda-l$ in  $x^\bD,r_\bD$ of the form
\be
\check p_{\Lambda,l}=\left(x^\bD\right)^{\Lambda-l}+\left(x^\bD\right)^{\Lambda-l-2}r_\bD^2\, b_{\Lambda,l+2}
+\left(x^\bD\right)^{\Lambda-l-4} r_\bD^4\, b_{\Lambda,l+4}+...\:\:  ;  \label{cp_h}
\ee
the coefficients $b_{\Lambda,h}$ can be determined from $ \Delta_\bD\, \check F^{i_1...i_l}_{\bD,\Lambda}\!=\!0$, which follows from
$\Delta_\bD\, X_{\bD,\Lambda}^{I_1...I_\Lambda} \!=\!0$.

\begin{prop}
$\check V_\bD^\Lambda$ decomposes into the following irreducible components
of  $Uso(D)$: 
\be
\check V_\bD^\Lambda=\bigoplus_{l=0}^\Lambda \check V_{D,\Lambda}^l,
\ee
where $ \check V_{D,\Lambda}^l\simeq V_D^l$ is spanned by
the $ \check F^{i_1...i_l}_{\bD,\Lambda}$, since the latter are eigenvectors of $\bL^2$:
\bea
\bL^2 \check F^{i_1...i_l}_{\bD,\Lambda}=E_l \check F^{i_1...i_l}_{\bD,\Lambda}.
\label{L^2_DcF}
\eea
Denoting by $[a]$ the integral part of $a\in\RR^+$, the coefficients of (\ref{cp_h}) are given by
\bea
b_{\Lambda,l+2k}=(-)^k\frac{(\Lambda\!-\!l)!\,(2\Lambda\!-\!4\!-\!2k\!+\! \bD)!!}{(\Lambda\!-\!l\!-\!2k)!\,(2k)!!\,(2\Lambda\!-\!4\!+\! \bD)!!}, \qquad k=1,2,....\left[\frac{\Lambda\!-\!l}{2}\right] . \label{bcoeff}
\eea
The $\check F^{i_1...i_l}_{\bD,\Lambda}$ transform under $L_{hk}$ as the $T^{i_1...i_l}_l$, and under $L_{h\bD}$ as follows:
\bea
 iL_{h\bD}\check F^{i_1...i_l}_{\bD,\Lambda}=
 ( \Lambda\!-\!l)\,\check F_{\bD,\Lambda}^{hi_1...i_l}-\frac {l(\Lambda\!+\!l\!+\!D\!-\!2)}{D\!+\!2l\!-\!2}\: \Ps^l{}^{i_1i_2...i_l}_{hj_2...j_l}\,\check F_{\bD,\Lambda}^{j_2...j_l}\, .
\label{L-su-cF}
\eea
\label{cVbDDeco}
\end{prop}
The proof is in Appendix \ref{Proof-cVbDDeco}.
We now determine the decomposition of $V_\bD^\Lambda:=\check V_\bD^\Lambda/(r_\bD)^\Lambda$ into  irreps of $Uso(D)$.
$V_\bD^0\simeq V_D^0\simeq\CC$ is spanned by 1. \ $V_\bD^1= V_{D,1}^0\oplus V_{D,1}^1$, where $V_{D,1}^0\simeq  V_D^0$ is spanned by $t^\bD$,  and  $V_{D,1}^1\simeq V_D^1$ is spanned by the $t^i$ (here $t^I:=x^I/r_\bD$). $V_\bD^2$ is spanned by the $T_{\bD,2}^{IJ}=t^It^J\!-\! \frac{\delta^{IJ}}{\bD}$; more explicitly,
\bea
T_{\bD,2}^{\bD\bD}=t^\bD t^\bD-\frac{1}{\bD}=\frac{D}{\bD}-t^ht_h
, \qquad T_{\bD,2}^{i\bD}=t^i t^\bD, \qquad T_{\bD,2}^{ij}=t^it^j-\delta^{ij}\frac{1}{\bD}; \nonumber
\eea
$F_{\bD,2}:=T_{\bD,2}^{\bD\bD}$ spans an irrep $V_{D,2}^0$
 isomorphic to $V_D^0$, the $F_{\bD,2}^i:=T_{\bD,2}^{i\bD}$ span an irrep
$V_{D,2}^1$  isomorphic to $V_D^1$, while
the $F_{D,2}^{ij}:=\Ps^{ij}_{hk}T_{\bD,2}^{hk}$ (which fulfill $F_{D,2}^{ij}\delta_{ij}=0$) span an irrep $V_{D,2}^2$ isomorphic  $V_D^2$. The  $F_{D,2}^{ij}$ can be expressed as combinations of the $T_{\bD,2}^{IJ}$:
\bea
F_{D,2}^{ij} 
=T_{\bD,2}^{ij}+\frac{\delta^{ij}}{D}T_{\bD,2}^{\bD\bD}
=T_{\bD,2}^{ij}-\frac{\delta^{ij}}{D}T_{\bD,2}^{hh}\: (\mbox{sum over } h)
=t^it^j- \frac{\delta^{ij}}{D}t^ht_h
\nonumber
\eea
This shows
the decomposition \ $V_\bD^2= V_{D,2}^2\oplus V_{D,2}^1\oplus V_{D,2}^0\simeq Pol_D^2$ \  explicitly. 
More generally, let
\bea
F^{i_1...i_l}_{\bD,\Lambda}:=\check F^{i_1...i_l}_{\bD,\Lambda}/(r_\bD)^{\Lambda}=\Ps^l{}^{i_1...i_l}_{j_1...j_l}T_{\bD,\Lambda}^{j_1...j_l\bD...\bD}=
p_{\Lambda,l}\Ps^l{}^{i_1...i_l}_{j_1...j_l}t^{j_1}...t^{j_l}=:
p_{\Lambda,l}T_{l}^{i_1...i_l},          \label{TiDdeco}
\eea
where  we have introduced a polynomial $p_{\Lambda,l}$  of degree $h=\Lambda-l$ in $t^\bD$  
(containing only powers of the same parity as $h$) by $p_{\Lambda,l}:=\check p_{\Lambda,l}(x^\bD,r_\bD)\, r_\bD^{l-\Lambda}$; more explicitly the latter reads
\be
 p_{\Lambda,l}=\left(t^\bD\right)^{\Lambda-l}+\left(t^\bD\right)^{\Lambda-l-2}  b_{\Lambda,l+2}
+\left(t^\bD\right)^{\Lambda-l-4}  b_{\Lambda,l+4}+...\:\:  .  \label{p_h}
\ee
As a direct consequence of Propositions \ref{cVbDDeco}, 
dividing all relations by $(r_\bD)^\Lambda$, we find
\begin{corollary}
$V_\bD^\Lambda$ decomposes into the following irreducible components
of  $Uso(D)$: 
\be
V_\bD^\Lambda=\bigoplus_{l=0}^\Lambda V_{D,\Lambda}^l,
\ee
where $ V_{D,\Lambda}^l\simeq V_D^l$ is spanned by
the $ F^{i_1...i_l}_{\bD,\Lambda}$. The latter are eigenvectors of $\bL^2$,
\bea
\bL^2 F^{i_1...i_l}_{\bD,\Lambda}=E_l F^{i_1...i_l}_{\bD,\Lambda},
\label{L^2_DF}
\eea
 transform under $L_{hk}$ as the $T^{i_1...i_l}_l$, and under $L_{h\bD}$ as follows:
\bea
 iL_{h\bD} F^{i_1...i_l}_{\bD,\Lambda}=
 ( \Lambda\!-\!l)\, F_{\bD,\Lambda}^{hi_1...i_l}-\frac {l(\Lambda\!+\!l\!+\!D\!-\!2)}{D\!+\!2l\!-\!2}\: \Ps^l{}^{i_1i_2...i_l}_{hj_2...j_l}\, F_{\bD,\Lambda}^{j_2...j_l}\, .
\label{L-su-F}
\eea
\label{VbDDeco}
\end{corollary}

For convenience, we slightly enlarge $Uso(D)$ by introducing the new generator 
 \be
\lambda=\frac 12 \left[\sqrt{(D\!-\!2)^2+4\bL^2}-D+2\right],
\ee
which fulfills \ $\lambda(\lambda\!+\!D\!-\!2)=\bL^2$, \ so that $V_D^l$ is a $\lambda=l$ eigenspace, and \ $\lambda \, F^{i_1...i_l}_{\bD,\Lambda}
=l \,F^{i_1...i_l}_{\bD,\Lambda}$. 

\begin{prop}
There exist a $Uso(D)$-module isomorphism 
$\hat\varkappa_\Lambda:Pol_D^\Lambda\rightarrow V_\bD^\Lambda$ and a $Uso(D)$-equivariant
algebra  isomorphism  $\hat\kappa_\Lambda:\mbox{End}\!\left(Pol_D^\Lambda\right)\to \bpi_\bD^\Lambda\big[Uso(\bD)\big]$ such that
\be
\hat\varkappa_\Lambda(aT)=\hat\kappa_\Lambda(a)\hat\varkappa_\Lambda(T), \qquad\forall\:T\in Pol_D^\Lambda,
\quad  a\in\mbox{End}\!\left(Pol_D^\Lambda\right)\:.
\label{compatibilityCond}
\ee
On the $T_{l}^{i_1...i_l}$  (spanning  $Pol_D^\Lambda$) and on generators $L_{hi},P^{\Lambda} t^i\cdot$ of $\mbox{End}\!\left(Pol_D^\Lambda\right)$ they act by
\bea
\hat\varkappa_\Lambda\left(T_{l}^{i_1...i_l}\right):= \hat a_{\Lambda,l}F^{i_1...i_l}_{\bD,\Lambda}=\hat a_{\Lambda,l} p_{\Lambda,l}T_{l}^{i_1...i_l},\qquad l=0,1,...,\Lambda ,       \label{Tcorr}\\[8pt]
\hat\kappa_\Lambda\left(L_{hi}\right):= L_{h i},\qquad \hat\kappa_\Lambda\left(P^{\Lambda} t^i\cdot\right):= \hat m_\Lambda^*(\lambda)L_{\bD i} \hat m_\Lambda(\lambda),                 \label{Opcorr}
\eea
where
\bea\label{mO(D)}
\hat m_{\Lambda}(l)=
 \sqrt{\frac{\Gamma\!\left(\frac {\Lambda+l+D-1}2\right)\,\Gamma\!\left(\frac {\Lambda-l+1}2\right)}{2\,\Gamma\!\left(\frac {\Lambda+l+D}2\right)\, \Gamma\!\left(\frac {\Lambda-l}2+1\right)}},  \quad
\hat a_{\Lambda,l}=\hat a_{\Lambda,0}\,i^l \sqrt{\frac{\Lambda(\Lambda\!-\!1)...(\Lambda\!-\!l\!+\!1)}{(\Lambda\!+\!D\!-\!1)(\Lambda\!+\!D)...(\Lambda\!+\!l\!+\!D\!-\!2)}}.
\eea
\label{Algebra-isomorphism}
\end{prop}

\noindent
The proposition and its proof are obtained from Theorem \ref{Algebra-isomorphism'} and the associated proof by fixing $\Lambda$, taking $k$ independent of $\Lambda$ and letting the $k\to\infty$.

\section{Relations among the $\overline x^i,\overline L_{hk}$}
\label{xLRel}

Since for all fixed $l=0,1,...,\Lambda$ the $T_{l}^{i_1i_2...i_l}$, make  up a complete set 
$\T_l$ in $V_D^l$, then  the  functions
\be
\bpsi_l^{i_1i_2...i_l}:=T_{l}^{i_1i_2...i_l} f_{l}                \label{Defpsi} 
\ee
($i_h\in\{1,...,D\}$ for $h\in\{1,...,l\}$) make up a complete set $\Ss^l_{D,\Lambda}$ in the eigenspace 
$\Hi_\Lambda^l$ of $H$, $\bL^2$,  with eigenvalues $E_{0,l}$, $E_l$. They are not linearly independent, because they are completely symmetric under permutation of the indices and fulfill the relations
\be
\delta_{i_ni_{n+1}}\bpsi_l^{i_1...i_l}=0,\qquad\qquad n=1,...,l\!-\!1.              
\label{gpsi=0} 
\ee
$\Ss_{D,\Lambda}:=\cup_{l=0}^\Lambda\Ss^l_{D,\Lambda}$ is  a complete set in $\Hi_\Lambda$. By (\ref{Defpsi}), (\ref{LonT}) the $\overline L_{hk}$ act on the $\bpsi
_l^{i_1i_2...i_l}$ via
\bea
i\overline L_{hk}\bpsi_l^{i_1i_2...i_l}
&= & \displaystyle l\,\Ps^l{}^{i_1...i_l}_{j_1...j_l}
\left(\delta^{kj_1}\bpsi_{l}^{hj_2...j_l}-\delta^{hj_1}\bpsi_{l}^{kj_2...j_l} \right).
  \label{Lonpsi}
\eea
 By (\ref{Proj_l}),  (\ref{tTdeco}),
applying $P_{\Lambda}=\sum_{l=0}^\Lambda P^l_{\Lambda}$ 
to $x^i\bpsi^{i_1i_2...i_l}=t^{i}\,T_{l}^{i_1i_2...i_l} \, r f_{l}(r)$ we find
\bea
P_{\Lambda}\left(x^i\bpsi_l^{i_1i_2...i_l}\right) &=&c_{l+1}\,\bpsi_{l+1}^{ii_1...i_l}+
c_{l}\, \frac{\zeta_{l+1}}D\,\Ps^{l}{}^{i_1i_2...i_l}_{ij_2...j_l} \bpsi_{l-1}^{j_2...j_l}, \nn[4pt]
\mbox{where }\quad c_{l} &:=& \left\{\!\!\ba{ll}
\sqrt{1+\frac{(2D\!-\!5)(D\!-1\!)}{2k}+\frac{(l\!-\!1)(l\!+\!D\!-\!2)}{k}} \:&\mbox{if }
1 \le l\le\Lambda, \\[6pt]
0 \:&
\mbox{otherwise,}\ea\right. \quad
\label{defc}
\eea
up to order $O\left(k^{-3/2}\right)$, see appendix \ref{scpD}. Hence, at the same order,
\bea
\overline{x}^i\bpsi
_l^{i_1i_2...i_l}=c_{l+1}\,\bpsi_{l+1}^{ii_1...i_l}+
\frac{c_{l}\, l}{D\!+\!2l\!-\!2}\,\Ps^{l}{}^{i_1i_2...i_l}_{ij_2...j_l}\, \bpsi_{l-1}^{j_2...j_l}.
\label{xonpsi}
\eea
The $O\left(k^{-3/2}\right)$ corrections depend 
on the terms proportional to $(r\!-\!1)^k$, $k>2$, in the Taylor expansion of $V$. 
These could be set rigorously equal to zero by a suitable choice of $V$. Henceforth we  adopt (\ref{Lonpsi}-\ref{xonpsi}) as exact {\it definitions} of  $\overline{L}_{hk},\overline{x}^i$. In the appendix we prove

\begin{prop}\label{genDpropo}
The $\overline{x}^i,\overline{L}_{hk}$  are self-adjoint operators generating the 
$N^2$-dimensional $*$-algebra 
\ $\A_{\Lambda}:=End(\Hi_{\Lambda})\simeq M_{N}(\CC)$ \ of observables  on  $\Hi_{\Lambda}$; \   $N$ is given by (\ref{defN}). Abbreviating \ $\overline{\bx}^2:=\overline{x}^i\,\overline{x}^i$, $\overline{\bL}^2:=\overline{L}_{ij}\,\overline{L}_{ij}/2$, $B:=(2D\!-\!5)(D\!-1\!)/2$, \ they  fulfill, at leading order in $1/\sqrt{k}$,
\bea
&& [i\overline{L}_{ij},\overline{x}^h]= \overline{x}^i\delta^h_j\!-\!\overline{x}^j\delta^h_i ,\label{linea1}\\[8pt]
&& [i\overline{L}_{ij},i\overline{L}_{hk}]=i\left(\overline{L}_{ik}\delta^j_h\!-\!\overline{L}_{jk}\delta^i_h\!-\!\overline{L}_{ih}\delta^j_k\!+\!\overline{L}_{jh}\delta^i_k\right), 
\label{linea4}\\[8pt]
&& \varepsilon^{i_1i_2i_3....i_D}\overline{x}^{i_1}\overline{L}_{i_2i_3} =0,  \qquad D\ge 3,  \label{linea3}\\[8pt]
&& (\overline{x}^h\!\pm\!i\overline{x}^k)^{2\Lambda+1}=0, \quad
(\bar L^{hj}\!+\!i\bar L^{kj})^{2\Lambda+1}=0,   \qquad
\mbox{if }h\neq j\neq k\neq h, \label{linea5}\\[8pt]
&&  \left[\overline{x}^i,\overline{x}^j\right] = i\overline{L}_{ij} \left(\!-\frac{I}{k}\!+\!K\,P_{\Lambda}^{\Lambda}\!\!\right), \quad 
\ba{l} K:=
\frac{1}{k}\!+\!\frac{1}{D\!+\!2\Lambda\!- \!2}\left[1\!+\!\frac{B}{k}\!+\!
\frac{(\Lambda\!-\!1)(\Lambda\!+\!D\!-\!2)}{k}\right],\ea
\label{linea2}\\[8pt]
&& \overline{\bx}^2=1\!+\!\frac {\overline{\bL}^2}k \!+\!\frac{B}{k}
-\frac{\Lambda\!+\!D\!-\!2}{2\Lambda\!+\!D\!-\!2}
\left[1\!+\!\frac{B}{k}\!+\!\frac{\Lambda (\Lambda\!+\!D\!-\!1)}{k}\right]  P_{\Lambda}^{\Lambda}  
=:\chi(\bL^2). \label{bx^2}
\eea
\end{prop}
This is the analog of Proposition 4.1 in \cite{FioPis18}. 
We obtain a fuzzy sphere  choosing $k$ as a function $k(\Lambda)$ fulfilling (\ref{consistencyD}), e.g. \ $k=\Lambda^2(\Lambda\!+\!D\!-\!2)^2/4$; 	\ the commutative limit is \
$\Lambda\rightarrow\infty$. 

\medskip
{\bf Remarks}:
\begin{enumerate}[label=\thesection.\alph*]

\item Eq. (\ref{linea3}) is the analog of (\ref{Lijrel}).
By (\ref{linea2}), it  can be reformulated also in the form $\varepsilon^{i_1i_2i_3....i_D}\overline{x}^{i_1}\overline{x}^{i_2}\overline{x}^{i_3} =0$.

\item  $\overline{\bx}^2$
is not a constant, but by   (\ref{bx^2}), (\ref{Proj-l})$_{l=\Lambda}$ can be expressed as a polynomial $\chi$ in $\overline{\bL}^2$ only, with the same eigenspaces $\Hi_{\Lambda}^l$. All its eigenvalues $r^2_l$, except $r^2_\Lambda$,
are close to 1, slightly (but strictly) grow with $l$ and collapse to 1 as $\Lambda\to \infty$. Conversely,
$\overline{\bL}^2$ can be expressed as a polynomial $\upsilon$ in $\overline{\bx}^2$, via \
$\overline{\bL}^2=\sum_{l=0}^\Lambda E_l  P^l_{\Lambda}$ and 
$ P^l_{\Lambda}=\prod_{n=0, n\neq l}^{\Lambda} \frac{\overline{\bx}^2-r^2_n}
{r^2_l-r^2_n}$.

\item By (\ref{linea2}),  (\ref{Proj-l})$_{l=\Lambda}$ the commutators $[\overline{x}^i,\overline{x}^j]$  are Snyder-like, i.e. of the form $\alpha \overline{L}_{ij}$; also $\alpha$ depends only on the $\overline{L}_{hk}$,
more precisely can be expressed as a polynomial in $\overline{\bL}^2$.

\item  Using (\ref{linea1}), (\ref{linea4}), (\ref{linea2}), all polynomials in $\overline{x}^i,\overline{L}_{hk}$  can be expressed as combinations of  monomials in $\overline{x}^i,\overline{L}_{hk}$ in any prescribed order, e.g. in the natural one
\be
\left(\overline{x}^1\right)^{n_1}...\left(\overline{x}^D\right)^{n_D}
\left(\overline{L}_{12}\right)^{n_{12}}\left(\overline{L}_{13}\right)^{n_{13}}...\left(\overline{L}_{dD}\right)^{n_{dD}}, \qquad n_i,n_{ij}\in\NN_0; \label{monomials}
\ee
 the coefficients, which can be put at the right of these monomials, are complex combinations of 1 and $P^{\Lambda}_{\Lambda}$.   Also $P^{\Lambda}_{\Lambda}$ can be expressed as a polynomial in $\overline{\bL}^2$ via (\ref{Proj-l})$_{l=\Lambda}$.
Hence a suitable subset (depending on $\Lambda$) of such ordered monomials makes up a   basis of the $N^2$-dim  vector space \ $\A_\Lambda$.

\item Actually,  $\overline{x}^i$ {\it generate} the $*$-algebra $\A_\Lambda$, because 
also the $\overline{L}_{ij}$ can be expressed as {\it non-ordered} polynomials in the $\overline{x}^i$: 
by  (\ref{linea2}) $\overline{L}_{ij}=[\overline{x}^j,\overline{x}^i]/\alpha$, and also $1/\alpha$, which depends only on $  P^{\Lambda}_{\Lambda}$, can be expressed itself as a polynomial in $\overline{\bx}^2$, as shown above.

\item  Eq. (\ref{linea1}-\ref{bx^2})  are equivariant under the whole group $O(D)$, including
the inversion  $\overline{x}^i\!\mapsto\!-\overline{x}^i$
of one axis, or more  (e.g. parity), contrary to Madore's and Hoppe's FS.


\end{enumerate}

The  operator norm of $\overline{\bx}^2$ equals its highest eigenvalue, \
$\Vert\overline{\bx}^2\Vert_{op}=r^2_{\Lambda-1}=1\!+\![B\!+\!(\Lambda\!-\!1)(\Lambda\!+\!D\!-\!3)]/k$. \
For all $\bpsi\in\Hi_{\Lambda}$ and  $i=1,...,D$, we find 
$\Vert\overline{x}^i\bpsi
\Vert^2=\la\overline{x}^i\bpsi
,\overline{x}^i\bpsi
\ra\le\sum_{i=1}^D\la\overline{x}^i\bpsi
,\overline{x}^i\bpsi
\ra=\la\bpsi
, \overline{\bx}^2\bpsi
\ra\le\Vert\overline{\bx}^2\Vert_{op}\,
\Vert\bpsi
\Vert^2$, whence
\be
\Vert\overline{x}^i\Vert_{op}\le \sqrt{1\!+\!\frac{B\!+\!(\Lambda\!-\!1)(\Lambda\!+\!D\!-\!3)}{k}}\le 1\!+\!\epsilon,
\quad \epsilon:=\frac{B\!+\!(\Lambda\!-\!1)(\Lambda\!+\!D\!-\!3)}{2k}
\label{Norm-bx^2}
\ee

\section{Isomorphisms  of $\Hi_{\Lambda},\A_{\Lambda}$, 
and $*$-automorphisms of $\A_{\Lambda}$}
\label{isomorphism}

\begin{theorem}
There exist a $O(D)$-module isomorphism 
$\varkappa_\Lambda:\Hi_\Lambda\rightarrow V_\bD^\Lambda$ and a $O(D)$-equivariant
algebra  map  $\kappa_\Lambda:\A_\Lambda\equiv\mbox{End}\!\left(\Hi_\Lambda\right)\to \bpi_\bD^\Lambda\big[Uso(\bD)\big]$, $\bD\equiv D\!+\!1$, such that
\be
\varkappa_\Lambda(a\bpsi)=\kappa_\Lambda(a)\varkappa_\Lambda(\bpsi), \qquad\forall\:\bpsi
\in \Hi_\Lambda,
\quad  a\in\A_\Lambda\:.
\label{compatibilityCond'}
\ee
On the  $\bpsi_{l}^{i_1...i_l}$ (spanning $\Hi_\Lambda$) and on generators $L_{hi}, \overline{x}^i\cdot$ of $\A_\Lambda$ they respectively act as follows:
\bea
&& \varkappa_\Lambda\left(\bpsi_{l}^{i_1...i_l}\right):= a_{\Lambda,l}F^{i_1...i_l}_{\bD,\Lambda}=a_{\Lambda,l} \, p_{\Lambda,l}\, T_{l}^{i_1...i_l},\qquad l=0,1,...,\Lambda ,       \label{Tcorr'}\\[8pt]
&&  \kappa_\Lambda\left(\overline{L}_{hi}\right):= \bpi_\bD^\Lambda (L_{h i})\,,\qquad \kappa_\Lambda\left(\overline{x}^i\right):= \bpi_\bD^\Lambda\big[m_\Lambda^*(\lambda) \, X^i\, m_\Lambda(\lambda)\big]\,,                                  \label{Opcorr'}
\eea
where \ $X^i:=L_{\bD i}$, \ $p_{\Lambda,l}=p_{\Lambda,l}\!\left(t^\bD\right)$ \ are the polynomials (\ref{p_h}), and
\bea
&& a_{\Lambda,l}=a_{\Lambda,0}\,i^l \sqrt{\frac{\Lambda(\Lambda\!-\!1)...(\Lambda\!-\!l\!+\!1)}{(\Lambda\!+\!D\!-\!1)(\Lambda\!+\!D)...(\Lambda\!+\!l\!+\!D\!-\!2)}}, \label{mO(D)''}\\[10pt]
&&  m_{\Lambda}(s)=
\sqrt{\frac{\Gamma\!\left(\frac {\Lambda+s+d}2\right)\,\Gamma\!\left(\frac {\Lambda-s+1}2\right)\,\Gamma\!\left(\frac {s+1+d/2+iA}2\right)\,\Gamma\!\left(\frac {s+1+d/2-iA}2\right)}{\Gamma\!\left(\frac {\Lambda+s+D}2\right)\, \Gamma\!\left(\frac {\Lambda-s}2+1\right)\,\Gamma\!\left(\frac {s+d/2+iA}2\right)\,\Gamma\!\left(\frac {s+d/2-iA}2\right)\,\sqrt{k}}} ;\label{mO(D)'}
\eea
here  \ $A:=\sqrt{k+(D\!-\!1)(D\!-\!3)3/4}$, \ and $\Gamma$ is Euler gamma function.
\label{Algebra-isomorphism'}
\end{theorem}
The proof is in appendix \ref{ProofAlgebra-isomorphism'}. \ 
The theorem extends Propositions 3.2, 4.2 of \cite{FioPis18} to $d>2$.
 The claims for $d>2$ were partially formulated, but not proved,  in \cite{Pis20}.
 

\medskip
As already recalled, the group of  $*$-automorphisms of $\A_{\Lambda}\simeq M_N(\CC)$ is inner and isomorphic to $SU(N)$, i.e. of the type 
\be
a\mapsto g\, a \, g^{-1}, \qquad a\in  M_N(\CC) ,   \label{autom}
\ee
with $g$ an unitary $N\times N$ matrix with unit determinant. We can identify in $SU(N)$ a subgroup $\simeq SO(\bD)$, acting via the $N$-dimensional representation 
$(V_\bD^\Lambda,\bpi_\bD^\Lambda)$; namely, it consists of matrices of the form $g=\bpi_\bD^\Lambda\left[e^{i\alpha}\right]$,
where $\alpha\in so(\bD)$.
Choosing 
$\alpha\in so(D)\subset so(\bD)$ the automorphism amounts to 
  a $SO(D)\subset SO(\bD)$ transformation, i.e. a rotation in the $x\equiv(x^1,...,x^D)\in\RR^D$ space. 
$O(D)\subset SO(\bD)$ transformations
with determinant  $-1$ in this space
keep the same form in the $\overline  X\equiv (X^1,...,X^D)$ space (where $X^i\equiv L_{\bD i}$)
and by (\ref{Opcorr'}) also in the $\overline  x\equiv (\overline x^1,...,\overline x^D)$ space.
In particular, those inverting one or more axes of $\RR^D$ (i.e. changing the sign
of one or more $x^i$, and thus also of $X^i,\overline  x^i$), e.g. parity,
can be also realized as  $SO(\bD)$ transformations, i.e. rotations in $\RR^\bD$.
This shows that (\ref{Opcorr'}) is equivariant
under parity and the whole
$O(D)$, which plays the role of isometry group of this fuzzy sphere.

\section{Fuzzy spherical harmonics, and limit $\Lambda\to\infty$}
\label{Dconverge}

In this section we suppress Einstein's summation convention over repeated indices.
The  previous results allow to define $Uso(D)$-module Hilbert space isomorphisms 
\bea
\ba{lccccc}
\sigma_\Lambda\: : \: & Pol_D^\Lambda \:=\: \displaystyle\bigoplus\limits_{l=0}^\Lambda
&\:\: V_D^l &\to  &
\Hi_\Lambda  \:=\: \displaystyle\bigoplus\limits_{l=0}^\Lambda &\:\: \Hi_\Lambda^l, \\
& &\:\vin &   && \:\vin \\
&& T_l^{i_1...i_l},Y_l^{\bm{m}} &\mapsto && \bpsi_l^{i_1...i_l},\bpsi_l^{\bm{m}};
\ea
\eea
here $\bpsi_l^{\bm{m}}:=\sum A^{\bm{m}}_{i_1...i_l}\bpsi_l^{i_1...i_l}$. The objects
at the right of the arrows are our fuzzy analogs of the objects at the left.
 In the limit $\Lambda\to\infty$ the above decomposition
of $\Hi_\Lambda$ into irreducible components under $O(D)$ 
 becomes isomorphic to the decomposition of $Pol_D\simeq \Hi_s$.

We define the $O(D)$-equivariant embedding \   ${\cal I}:\Hi_\Lambda\hookrightarrow \Hi_s$ \   by  setting \  ${\cal I}\left(\bpsi_l^{i_1...i_l}\right):=T_l^{i_1...i_l}$ \
and applying linear extension. Below we drop $\I$  and identify $\bpsi_l^{i_1...i_l}=T_l^{i_1...i_l}$, or equivalently
$\bpsi_l^{\bm{m}}=Y_l^{\bm{m}}$,  as elements of the Hilbert space $\Hi_s$. 
For all $\bphi\equiv\sum_{l=0}^{\infty}\phi^l_{i_1...i_l}T_l^{i_1...i_l}\in {\cal L}^2(S^2)$ and $\Lambda\in\NN$ let 
$\bphi_ \Lambda:=P_\Lambda\bphi=\sum_{l=0}^{\Lambda}\phi^l_{i_1...i_l}T_l^{i_1...i_l}$ be its projection to $\Hi_\Lambda$ (or $\Lambda$-th truncation). Clearly $\bphi_ \Lambda\to\bphi$
in the $\Hi_s$-norm $\Vert\,\Vert$:
in this simplified notation, \ $\Hi_\Lambda$ `invades' $\Hi_s$ as $\Lambda\to\infty$;  

${\cal I}$ induces the $O(D)$-equivariant embedding of operator algebras \ 
${\cal J}\!:\!\A_\Lambda\!\hookrightarrow\! B\left(\Hi_s\right)$ \
by setting\ $\J(a)\,\I(\bpsi):=\I(a\bpsi)$;  \ here
$B\left(\Hi_s\right)$ stands for the $*$-algebra of bounded operators on $\Hi_s$.
By construction $\A_\Lambda$ annihilates $\Hi_\Lambda^\perp$. \ 
In particular,  $\J\!\left(\overline{L}_{hk}\right)=L_{hk}P^{\Lambda}$, and
 $\overline{L}_{hk}\bphi\stackrel{\Lambda\to\infty}{\longrightarrow} L_{hk}\bphi$ \  
 for all $\bphi\!\in\! D(L_{hk})\equiv$ the domain of $L_{hk}$.
More generally, 
$f(\overline{L}_{hk})\to f(L_{hk})$ strongly on $D[f(L_{hk})]\subset\Hi_s$, for all measurable functions $f(s)$.

Continuous functions $f$ on $S^d$, acting as multiplication operators 
$f\cdot:\bphi\in\Hi_s\mapsto f\bphi\in\Hi_s$, make up a subalgebra  $C(S^d)$
of $B\left(\Hi_s\right)$. \
Clearly, $f$ belongs also to  $\Hi_s$. Since $Pol_D$ is dense in both $\Hi_s$,   $C(S^d)$,  
$f_N$ converges to $f$  as $N\to \infty$  in both the $\Hi_s$  and  the $C(S^d)$  norm.

 We   define the  $\Lambda$-th fuzzy analogs  of the  $T_l^{i_1...i_l}$  (seen as an element of $C(S^d)$, i.e. acting by multiplication on $\bpsi\in\Hi_s$), $l\le\Lambda$, by replacing $t^i\cdot\mapsto  \overline{x}^i$ in the definition of $T_l^{i_1...i_l}$, i.e. by
\be
 \widehat{T}_l^{i_1...i_l} :=
\Ps^l{}^{i_1...i_l}_{j_1...j_l} \overline{x}^{j_1} ... \overline{x}^{j_l},                 \label{defhatT}
\ee
sum over repeated indices (cf. Proposition \ref{propT}); the $\widehat{T}_l^{i_1...i_l}$ fulfill again (\ref{gT=0}).
Having identified \ $\bpsi_{l}^{i_1...i_l} \equiv T_{l}^{i_1...i_l}$, 
we rewrite (\ref{tTdeco}), (\ref{xonpsi}) in the form
\bea
&& t^{h}\,T_{l}^{i_1...i_l} = T_{l+1}^{hi_1...i_l}+d_l\,
\Ps^{l}{}^{i_1i_2...i_l}_{hj_2...j_l} T_{l-1}^{j_2...j_l}, \qquad  
d_l:=\frac {l}{D\!+\!2l\!- \!2}\label{tTdeco'}\\[6pt]
&& \overline{x}^hT_l^{i_1i_2...i_l}=c_{l+1}\,T_{l+1}^{hi_1...i_l}+
c_{l}\: d_l\,\Ps^{l}{}^{i_1i_2...i_l}_{hj_2...j_l}\, T_{l-1}^{j_2...j_l}.
\label{xonT}
\eea
Using these formulae in appendix \ref{proof-of-prophTTdeco} we prove the fuzzy version of Proposition \ref{propTTdeco}:

\begin{theorem} The action of the $\widehat{T}_l^{i_1...i_l}=\Ps^l{}^{i_1...i_l}_{j_1...j_l} \overline{x}^{j_1} ... \overline{x}^{j_l} $ on $\Hi_\Lambda$ is given by
\bea
\widehat{T}_l^{i_1...i_l}T_m^{j_1...j_m}=\sum\limits_{n\in L}\,\widehat{N}^{lm}_n\,\Ps^l{}_{a_1...a_rc_{1}...c_{l-r}}^{i_1...i_l}\Ps^m{}_{a_1...a_r c_{l-r+1}...c_n}^{j_1...j_rj_{r+1}...j_m} \Ps^n{}_{c_1...c_n} ^{k_1...k_n}\, T_n^{k_1...k_n},
\label{hTTdeco}
\eea
with suitable coefficients $\widehat{N}^{lm}_n$ related to their classical limits $ N^{lm}_n>0$ of formula (\ref{TTdeco}) by
\be
\widehat{N}^{lm}_n=0\quad\mbox{{\rm if} }\:\:l\!-\!m>\Lambda,\qquad\quad
N^{lm}_n\le \widehat{N}^{lm}_n\le N^{lm}_n (c_\Lambda)^l
\quad\mbox{{\rm otherwise}}.      \label{hN}
\ee
\label{prophTTdeco}
\end{theorem}

As a fuzzy analog  of the vector space $C(S^d)$ we adopt
\be
{\cal C}_\Lambda:=\left\{\hat f_{2\Lambda}:=\sum_{l=0}^{2\Lambda}\sum_{i_1,....,i_l}f^l_{i_1...i_l}  \widehat{T}_l^{i_1...i_l}\:,\: f^l_{i_1...i_l}\in\CC\right\}
\subset\A_\Lambda\subset B\big(\Hi_s\big);
\label{def_CLambdaD}
\ee
here the highest $l$ is $2\Lambda$ because by (\ref{hN}) the $\widehat{T}_l^{i_1...i_l}$ annihilate $\Hi_\Lambda$ if $l>2\Lambda$.  By construction,
\be
{\cal C}_\Lambda=\bigoplus\limits_{l=0}^{2\Lambda}  \widehat{V}_D^l,\qquad\qquad  \widehat{V}_D^l:=
\left\{\sum_{i_1,....,i_l}f^l_{i_1...i_l}  \widehat{T}_l^{i_1...i_l}\:,\: f^l_{i_1...i_l}\in\CC\right\} \label{deco2}
\ee
is the decomposition
of ${\cal C}_\Lambda$ into irreducible components under $O(D)$. $\widehat{V}_D^l$ is trace-free
for all $l>0$. 
In the limit $\Lambda\to\infty$ (\ref{deco2}) becomes the decomposition of $C(S^d)$. 
As a fuzzy analog  of $f\in C(S^d)$
we adopt the sum \ $\hat f_{2\Lambda}$ \ appearing in (\ref{def_CLambdaD})
with the coefficients of the expansion (\ref{decophi}) of $f$ up to $l=2\Lambda$.
In appendix \ref{convergD} we prove

\begin{theorem} \label{convergD}
For all $f,g\in C(S^d)$ the following strong limits as $\Lambda\rightarrow \infty$ hold: $\hat{f}_{2\Lambda}\rightarrow f\cdot,\widehat{\left(fg\right)}_{2\Lambda}\rightarrow fg$ and $\hat{f}_{2\Lambda}\hat{g}_{2\Lambda}\rightarrow fg\cdot$. 
\end{theorem}
\noindent
The last statement says that the product in $\mathcal{A}_{\Lambda}$ of the approximations $\widehat{f}_{2\Lambda}$, $\widehat{g}_{2\Lambda}$ goes to the product in $B\left(\Hi_s\right)$ 
(the algebra of bounded operators on $\Hi_s\equiv{\cal L}^2(S^d)$) of $f\cdot,g\cdot$. 
We point out that  $\hat f_{2\Lambda}$  {\it does not} converge  to $f$ {\it in operator norm},
because the operator $\hat f_{2\Lambda}$ (a polynomial in the $\overline{x}^i$)   annihilates $\Hi_{\Lambda}^\perp$ (the orthogonal complement of $\Hi_{\Lambda}$),  since so do
the  $\overline{x}^i=P^\Lambda x^i\cdot P^\Lambda$.

Essentially the same claims of this theorem were proved for $d=1,2$ in \cite{FioPis18} 
requiring that $k(\Lambda)$ diverges much faster than
required by (\ref{consistencyD}), and  were formulated (without proof) for $d>2$ in Theorem 7.1 of \cite{Pis20} with the same strong assumptions on  
the divergence of $k(\Lambda)$.

\section{Outlook, discussion and conclusions}
\label{discuss}

In this paper we have completed our construction \cite{FioPis18,Pis20} 
of fuzzy spheres $S^d_{\Lambda}$  that be equivariant under the {\it full} orthogonal group $O(D)$, $D\equiv d\!+\!1$, for {\it all} $d\in\NN$.
The construction procedure consists (sections \ref{intro}, \ref{genset}) in  starting  with a quantum particle in $\RR^{D}$ configuration space  subject to a $O(D)$-invariant potential $V(r)$  with a very sharp minimum on the sphere of  radius \ $r=1$ and projecting the Hilbert space  $\Hi={\cal L}^2(\RR^D)$ to the subspace 
$\Hi_{\overline{E}}$ with energy below a suitable  cutoff $\overline{E}$; \ $\overline{E}$ is sufficiently low to exclude all excited radial modes of $\Hi$
(this can be considered as a {\it quantum} version of the   {\it constraint} $r=1$),
so that on $\Hi_{\overline{E}}$ the Hamiltonian essentially reduces to the square angular momentum $\bL^2$
(the Laplacian, i.e. the free Hamiltonian, over the sphere $S^d$). By making both 
the confining parameter $k\equiv V''(1)/4$ and $\overline{E}$ depend on $\Lambda\in\NN$, and diverge with it,  we have
obtained a sequence $\left\{(\Hi_\Lambda,\A_\Lambda)\right\}_{\Lambda\in\NN}$ of $O(D)$-equivariant approximations of a quantum particle on  $S^d$. $\Hi_\Lambda$ is the $\Lambda$-th projected Hilbert space of states
and $\A_\Lambda\!\equiv$End$(\Hi_\Lambda)$ is the associated  $*$-algebra of observables.
The projected Cartesian coordinates $\overline x^i$ no longer commute (section \ref{xLRel});
their commutators $[\overline x^i,\overline x^j]$  are of Snyder type, i.e. proportional to the angular momentum components $\overline L_{ij}$.
$\A_\Lambda$ is spanned by ordered monomials (\ref{monomials}) in $\overline{x}^i,\overline{L}_{ij}$  (of appropriately bounded degrees), in the same way as 
 the algebra $\A_s$ of observables on $\Hi_s$  is spanned by ordered monomials in $t^i,L_{ij}$. 
However, while $\overline x^i$ generate the whole $\A_\Lambda$ because $[\overline x^i,\overline x^j]\propto\overline L_{ij}$, this has no analog $\A_s$.
The square distance $\overline{\bx}^2$ from the origin is not identically 1, but
a function of $\bL^2$ such that  its spectrum is very close to 1 and collapses to 1 as $\Lambda\to\infty$.
We have also constructed (section \ref{Dconverge}) the subspace $\C_{\Lambda}\subset\A_{\Lambda}$ 
of completely symmetrized trace-free  polynomials in the $\overline x^i$; this is also spanned by the
fuzzy analogs of spherical harmonics (thought as multiplication operators on $\Hi_s$).
$\Hi_\Lambda,\A_\Lambda,\C_{\Lambda}$ carry reducible representations of $O(D)$; 
as $\Lambda\to\infty$ their decompositions into irreps respectively go to the decompositions of 
$\Hi_s\equiv {\cal L}^2(S^d)$,  of $\A_s$ and of $C(S^d)\subset\A_s$ (the abelian subalgebra of continuous functions on $S^d$ acting as operators on $\Hi_s$) [see (\ref{Hi_Lambda-deco}),  (\ref{deco2})].
There are natural embeddings \ $\Hi_\Lambda\hookrightarrow \Hi_s$, 
$\C_\Lambda\hookrightarrow C(S^d)$ and $\A_\Lambda\hookrightarrow \A_s$ such that
$\Hi_\Lambda\to \Hi_s$ in the norm of $\Hi_s$, while
$\C_\Lambda\to C(S^d)$,  $\A_\Lambda\to \A_s$ strongly  as $\Lambda\to\infty$ (section \ref{Dconverge}). 

A basis of $\A_\Lambda$ consists of a suitable ($\Lambda$-dependent) subset $S_\Lambda$ of ordered monomials (\ref{monomials}).
 Since  $\overline L_{ij}\bpsi_0=0$ for all $i,j\le D$, the subset $S_\Lambda'$ of $S_\Lambda$ with all $n_{ij}=0$ is a basis of ${\cal C}_\Lambda$, and $\Hi_\Lambda={\cal C}_\Lambda\bpsi_0$; 
$\Hi_\Lambda,{\cal C}_\Lambda,Pol_D^\Lambda$ carry the same reducible representation of $O(D)$.
As $\Lambda\to\infty$: i) $S_\Lambda$ becomes
a basis $S$ of $\A_s$ consisting of ordered monomials in $t^h,L_{ij}$; ii) 
$S_\Lambda'$ becomes a basis $S'$ of $C(S^d)$ consisting of ordered monomials in $t^h$; iii) $S_\Lambda'\bpsi_0$ becomes a basis of ${\cal L}^2(S^d)=C(S^d)\bpsi_0$.

The structure of the  pairs $(\Hi_\Lambda,\A_\Lambda)$ is made transparent by the discovery
(section \ref{isomorphism}) that these are isomorphic to $\left(V^\Lambda_{\bD},\bpi_\Lambda[Uso(\bD)]\right)$, $\bD\!\equiv\!D\!+\!1$, also as $O(D)$-modules;
\ $\bpi_\Lambda$ is the irrep of $Uso(\bD)$ on the space 
$V^\Lambda_{\bD}$ of harmonic polynomials of degree $\Lambda$ on $\RR^{\bD}$,
restricted to $S^D$. 

If we reintroduce $\hbar$ and the physical angular momentum components
$l_{ij}:=\hbar L_{ij}$, and we define  as usual
the quantum Poisson bracket as \ $\{f,g\}=[f,g]/i\hbar$, \ then in the $\hbar\to 0$ limit
$\A_s$ goes to the (commutative) algebra $\F$ of (polynomial) functions on the classical phase space $T^*S^d$, which is generated by $t^i,l_{ij}$. We can directly obtain $\F$ from $\A_\Lambda$  adopting a suitable $\Lambda$-dependent $\hbar$ going to zero as 
$\Lambda\to\infty$\footnote{More precisely, to obtain the classical Poisson brackets
from (\ref{linea1}-\ref{linea2}) it suffices that $\hbar(\Lambda)k(\Lambda)$ keeps diverging; if e.g. $k=\Lambda^2(\Lambda\!+\!D\!-\!2)^2/4$, then $\hbar(\Lambda)=O(\Lambda^{-\alpha})$ with $0<\alpha<4$ is enough. Setting $\overline{l}_{ij}:=\hbar \overline{L}_{ij}$, in this limit $\overline{x}^i,\overline{l}_{ij}\to t^i,l_{ij}$ respectively.}.  
Using the isomorphism $\A_\Lambda\simeq\bpi_\bD^\Lambda\big[Uso(\bD)\big]$,
we now show that, more formally, 
we can see $\{\A_\Lambda\}_{\Lambda\in\NN}$ as a fuzzy
quantization of a coadjoint orbit of $O(\bD)$ that goes to the classical phase space $T^*S^d$.
We recall that given a Lie group $G$, a coadjoint orbit 
$\mathcal {O}_{\blambda }$, for $\blambda$ in the dual space $\mathfrak {g}^{*}$ of 
the Lie algebra $\mathfrak {g}$ of $G$, may be defined either extrinsically, as the actual orbit $\mathrm {Ad} _{G}^{*}\blambda $ of the coadjoint action $\mathrm {Ad} _{G}^{*} $ inside $\mathfrak {g}^{*}$ passing through $\blambda$, 
or intrinsically as the homogeneous space $G/G_{\blambda }$, where $G_{\blambda }$ is the stabilizer of $\blambda$ with respect to the coadjoint action (this distinction is worth making since the embedding of the orbit may be complicated). Coadjoint orbits are naturally endowed with a symplectic structure arising from the group action.
If $G$ is compact semisimple, identifying  $\mathfrak{g}^*$  
with   $\mathfrak{g}$ via the (nondegenerate) Killing form, we can resp. rewrite 
these definitions  in the form 
\bea
\mathcal {O}_{\blambda }:=\left\{g \blambda g^{-1}\:|\: g\in G\right\}\subset\mathfrak{g}^* ,\qquad\mathcal {O}_{\blambda }:= G/G_{\blambda } \quad \mbox{where }\: G_{\blambda }:=\left\{ g\in G \:|\: g \blambda g^{-1}= \blambda\right\}.
\eea
Clearly,  $G_{\Lambda\blambda }=G_{\blambda }$ for all $\Lambda\in \CC\setminus \{0\}$. Denoting as $\Hi_{\blambda}$ the (necessarily finite-dimensional) 
carrier space of the irrep with  highest weight $\blambda$, one can regard 
(see e.g. \cite{Haw99}) the sequence of
$\{\A_\Lambda\}_{\Lambda\in\NN}$, with
$\A_\Lambda:=\mbox{End}\left(\Hi_{\Lambda\blambda}\right)$,  
as a fuzzy quantization of the symplectic space $\mathcal {O}_{\blambda }\simeq G/G_{\blambda }$.
We recall that the Killing form $B$ of $so(\bD)$ gives \ 
$B(L_{HI},L_{JK})=2(\bD\!-\!2)\left(\delta^H_J\delta^I_K-\delta^H_K\delta^I_J\right)$ \
for all $H,I,J,K\in\{1,2,...,\bD\equiv D\!+\!1\}$.
Let  $\sigma:=\left[\frac{\bD}2\right]=$ rank of $so(\bD)$. As the basis of
the Cartan subalgebra $\mathfrak{h}$ of $so(\bD)$ we choose $\{H_a\}_{a=1}^\sigma$,
where 
\bea
H_\sigma:=L_{D\bD},\quad H_{\sigma-1}:=L_{(d-1)d},\quad ...,\quad H_1=\left\{\!\!\ba{ll} 
L_{12}\:\: &\mbox{if }\bD=2\sigma \\[6pt]
L_{23}\:\:  &\mbox{if }\bD=2\sigma\!+\!1\ea\right.
\eea
We choose the irrep of $Uso(\bD)$ on $ V^\Lambda_\bD\simeq\Hi_\Lambda$; as the highest weight vector 
$\Omega^\Lambda_\bD\in V^\Lambda_\bD$ we
choose $\Omega^\Lambda_\bD:=(t^D\!+\!it^\bD)^\Lambda$ (for brevity we do not 
write down the associated partition of roots of $so(\bD)$ into positive and negative).
The associated
weight in the chosen basis, i.e. the joint spectrum of $H:=(H_1,...,H_\sigma)$, is 
$\bLambda=(0,...,0,\Lambda)$.
Identifying weights $\blambda\in \mathfrak{h}^*$  
with elements $H_\blambda\in \mathfrak{h}$ via the Killing form, we
find that $H_\bLambda \propto H_\sigma=L_{D\bD}$.
The stabilizer of the latter in $SO(\bD)$ is
$SO(2)\times SO(d)$, where  $SO(2)$, $SO(d)$ have Lie algebra
respectively spanned by $L_{D\bD}$ and by
the $L_{ij}$ with $i,j<D$.
Therefore the corresponding coadjoint orbit $SO(\bD)/\big(SO(2)\times SO(d)\big)$
has dimension
$$
\frac{D(D+1)}2-1-\frac{(D-2)(D-1)}2=2(D-1)=2d,
$$
which is also the dimension of $T^*S^{d}$, the cotangent space of the 
$d$-dimensional sphere $S^d$ (or phase space over $S^d$). This is consistent with the interpretation of 
$\A_\Lambda$ as the algebra of observables (quantized phase space)  on the fuzzy sphere.
It would have not been the case if we had chosen some other generic irrep of
$Uso(\bD)$: the coadjoint orbit would have been some other equivariant bundle over 
$S^d$ \cite{Haw99}.

For instance, the  4-dimensional fuzzy spheres introduced in \cite{GroKliPre96}, as well as the ones of dimension
$d\ge 3$  considered in \cite{Ramgoolam,DolOCon03,DolOConPre03}, are  based on $End(V^\Lambda)$, 
where the $V^\Lambda$ carry {\it irreducible representations} of both $Spin(D)$ and $Spin(D+1)$, and therefore of both $Uso(D)$ and  $Uso(\bD)$. Then: \ i) i\ for some $\Lambda$ these may be only {\it projective} representations of
$O(D)$;  \ ii) \ in general (\ref{linea3}) will not be satisfied; \ iii) \
as $\Lambda\to \infty$ $V^\Lambda$ does not go to ${\cal L}^2(S^d)$ as a representation of $Uso(D)$, in contrast with our $\Hi_\Lambda\simeq V^\Lambda_{\bD}$. The $X^i:=L_{i\bD}$ play the role of fuzzy coordinates.
As $\bx^2\equiv X^iX^i$ is central, it can be set $\bx^2=1$ identically. The commutation relations are also $O(D)$-covariant and Snyder-like, except for the case of the Madore-Hoppe fuzzy sphere  \cite{Mad92,HopdeWNic}. 
The corresponding coadjoint orbit for $d=4$ is the 6-dimensional $\CC P^3$ \cite{Ste16,Ste17}, which can be seen as a $so(5)$-equivariant $S^2$ bundle over $S^4$ (while \cite{Ramgoolam} does not identify coadjoint orbits for generic $d$).

In \cite{Ste16,Ste17} the authors consider also constructing a fuzzy 4-sphere $S^4_N$
through a {\it reducible} representation of $Uso(5)$ on a Hilbert space $V$ obtained decomposing
 an irrep $\pi$ of $Uso(6)$ characterized by a  highest weight triple
 $(N,0,n)$ with respect to $(H'_{\lambda_1},H'_{\lambda_1}H'_{\lambda_1})$, where
$$
H'_{\lambda_1}:=\frac i2 (L_{34}+L_{12}+L_{56}),\quad H'_{\lambda_2}:=iL_{56},\quad H'_{\lambda_3}:=\frac i2 (L_{34}-L_{12}+L_{56}).
$$ 
The $X^i=L_{i6}$ ($i=1,...,5$), which make up a basis of the vector space \ $so(6)\setminus so(5)$, \ still play the role of noncommuting Cartesian coordinates.  
If $n=0$ then $\bx^2\equiv 1$ ($V$ carries an irrep of $O(5)$), and one recovers the 
($so(5)$-equivariant $S^2$ bundle over the)  fuzzy 4-sphere of   \cite{GroKliPre96}.
If $n>0$, then  the $O(5)$-scalar $\bx^2=X^iX^i$ is no longer central, but its spectrum is still very close to 1  provided $N\gg n$, because then $V$ decomposes only in few irreducible $SO(5)$-components, all with eigenvalues of $\bx^2$ very close to 1.  The associated coadjoint orbit is 10-dimensional  and can be seen as a $so(5)$-equivariant $\CC P^2$ bundle over $\CC P^3$, or a $so(5)$-equivariant twisted bundle over either $S^4_N$ or $S^4_n$. On the contrary, with respect to $(H'_{\lambda_1},H'_{\lambda_1}H'_{\lambda_1})$ the highest weight triple  of the irrep $V^\Lambda_{6}$   considered here is  $(\Lambda,\Lambda,\Lambda)$; as said, $\bx^2\equiv \overline x^i\overline x^i\simeq 1$  is   guaranteed  by adopting as noncommutative Cartesian coordinates the  $\overline x^i=m_\Lambda(\bL^2)\,X^i\,m_\Lambda(\bL^2)$, with a suitable function $m_\Lambda$, rather than the $X^i$, and the associated coadjoint orbit has
dimension 8, which  is also the dimension of $T^*S^{4}$, as wished.

\smallskip
We now clarify in which sense we have provided a
$O(D)$-equivariant fuzzy quantization of  $T^*S^d$ and $S^d$ -  the phase space and
the configuration space of our  particle.

Although  $\A_s$ is generated by all the $t^h,L_{ij}$ with $h\!\le\! D$, $i\!<\! j\!\le\!  D$ (subject to the relations   (\ref{Lvcr}),  (\ref{LLcr}), (\ref{t-relation}), 
$\varepsilon^{i_1i_2i_3....i_D}t^{i_1}L_{i_2i_3}=0$ due to (\ref{Lijrel}),
$t^it^h=t^ht^i$), and $C(S^d)$ is generated  by the  $t^h$ alone,
the  $\overline{x}^i$ (or the simpler generators $X^i$) alone generate\footnote{That the $\overline{x}^i$ do has been explained in the Remarks after Proposition \ref{genDpropo}; that the $X^i=L_{\bD i}$ do follows from (\ref{LLcr}),
which implies $L_{ij}=i[X^j,X^i]$, and Proposition \ref{Algebra-isomorphism'}.} 
the whole $\A_\Lambda\simeq\bpi_\bD^\Lambda\big[Uso(\bD)\big]$, which contains 
${\cal C}_\Lambda$ as a {\it proper} subspace, but not as a subalgebra.
Thus the Hilbert-Poincaré series of the algebra  generated by the 
$\overline{x}^i$ (or $X^i$), $\A_\Lambda$, is larger than that of   $Pol_D^\Lambda$
and  ${\cal C}_\Lambda$.
If by a ``quantized space"  we understand
a noncommutative deformation of the {\it algebra} of functions on that space
{\it preserving the Hilbert-Poincaré series}, then $\{\A_\Lambda\}_{\Lambda\in\NN}$  is a 
($O(D)$-equivariant, fuzzy) quantization of $T^*S^d$, the  phase space on $S^d$, while
$\{{\cal C}_\Lambda\}_{\Lambda\in\NN}$  is not a quantization of $S^d$, nor are the other fuzzy spheres,
except the Madore-Hoppe fuzzy 2-dimensional sphere: all the others, as ours, have the same Hilbert-Poincaré series of a suitable equivariant bundle on $S^d$, i.e. a manifold with a dimension $n>d$ (in our case, $n=2d$). (Incidentally, in our opinion also for the Madore-Hoppe fuzzy sphere the most natural interpretation is of a quantized phase space, because
the $\hbar\to 0$ limit of the quantum Poisson bracket endows its algebra with
a nontrivial Poisson structure.)

Therefore we understand  $\Hi_\Lambda,{\cal C}_\Lambda$ as fuzzy  ``quantized" $S^d$ in the following weaker sense. $\Hi_\Lambda$  is the quantization of the space  ${\cal L}^2(S^d)$ of square integrable functions, and  the space ${\cal C}_\Lambda$ of fuzzy spherical harmonics is the quantization of the space $C(S^d)$ of continuous functions, seen as operators acting on the former, because the whole
$\Hi_\Lambda$ is obtained applying to the ground state $\bpsi_0$ (or any other
$\bpsi\in\Hi_\Lambda$)  the polynomials in the $\overline{x}^i$ alone, or
equivalently (by Proposition 5.1) the polynomials in the $X^i=L_{\bD i}$ alone, or the space $\C_\Lambda$, in the same way as the Hilbert ${\cal L}^2(S^d)$ is obtained (modulo completion) by applying $C(S^d)$ or $Pol_D$ , i.e. the polynomials in the $t^i=x^i/r$, to the ground state, i.e. the constant function on $S^d$. These quantizations are $O(D)$-equivariant because $\Hi_\Lambda,{\cal C}_\Lambda$ not only
 have the same dimension, but carry also the same reducible representation of $O(D)$, that of the space (and commutative algebra) $Pol_D^\Lambda$ of polynomials of degree $\Lambda$ in the $t^i=x^i/r$.  Identifying $\Hi_\Lambda,{\cal C}_\Lambda$ with $Pol_D^\Lambda$ as $O(D)$-modules, in the $\Lambda\to\infty$ the latter
becomes dense in both $C(S^d)$, ${\cal L}^2(S^d)$, and its decomposition into irreps
of  $O(D)$ becomes that (\ref{directsum}) of $C(S^d)$, ${\cal L}^2(S^d)$.
This is not the case for the other fuzzy spheres.


\medskip
Many  aspects of these new fuzzy spheres deserve  investigations: e.g.   
space uncertainties, optimally localized states and coherent states also for $d>2$, as done in \cite{FioPis20JPA,FioPis20LMP,FioPis20PoS} for $d=1,2$; a distance between optimally  localized states
(as done e.g. in \cite{DanLizMar14} for the FS); extending the construction to particles with spin\footnote{This should be possible adopting as the starting Hilbert space
$\Hi\simeq{\cal L}^2(\RR^D)\otimes \CC^n$,  $n:=2\big[\frac D2\big]$.}; QFT based on our fuzzy spheres; application of our fuzzy spheres to problems in quantum gravity, or condensed matter physics; etc.
It would be also interesting to investigate whether our procedure can be applied (or generalized) to other symmetric compact 
submanifolds \footnote{By (\ref{dimH}), if $S$ is not compact then the corresponding Hilbert spaces $\Hi_\Lambda$ will have infinite dimension, and therefore will not lead to fuzzy spaces in the sense given in the Introduction.}  $S\subset\RR^D$ that are level sets of smooth or polynomial function(s) $\rho(x)$.

\medskip
Finally, we point out that a  different approach to the construction of 
noncommutative submanifolds of noncommutative $\RR^D$, equivariant with respect to a `quantum group' (twisted Hopf algebra) has been proposed in  \cite{FioreWeber,FioFraWebquadrics}; 
it is based on a systematic use of Drinfel'd twists.

\subsubsection*{Acknowledgments}

We thank F. Pisacane, H. Steinacker for useful discussions in the early stages of the work.
Work done also within the activities of GNFM.

\section{Appendix}
\label{Appe}

\subsection{Proof of Proposition \ref{symmetrizers}}

Our Ansatz is (\ref{ansatz1})
with $M\equiv M(l\!+\!1)$ a $O(D)$-invariant matrix to be determined.
The most general one is
\be
M(l\!+\!1)=\alpha_{l\!+\!1}\left(\1_{D^2}+\beta_{l\!+\!1}\Pe
+\gamma_{l\!+\!1}\Pt\right)          \label{m2}
\ee
We first determine the coefficients
$\beta_{l\!+\!1},\gamma_{l\!+\!1}$ by imposing the conditions
(\ref{Plproj1}). By the recursive assumption, only the condition with $m=l$ is
not fulfilled automatically and must be imposed by hand. Actually, it suffices
to impose just (\ref{Plproj1}a), due
to the symmetry of the Ansatz (\ref{ansatz1}) and of the matrices 
$\Ps^{l}$ under transposition. 
Abbreviating $M'\equiv M(l)$ this amounts to
\bea
0&\stackrel{!}{=}&\Ps^{l\!+\!1}\Ps^{'}_{l(l\!+\!1)}
\stackrel{(\ref{ansatz1})}{=} \Ps^{l}_{1\ldots
l}M_{l(l\!+\!1)}\Ps^{l}_{1\ldots l} \Ps^{'}_{l(l\!+\!1)}\nn[4pt]
&\stackrel{(\ref{ansatz1})_{l\mapsto l\!-\!1}}{=} &\Ps^{l}_{1\ldots
l}M_{l(l\!+\!1)}\Ps^{l\!-\!1}_{1...(l\!-\!1)}
M_{(l\!-\!1)l}' \Ps^{l\!-\!1}_{1...(l\!-\!1)} \Ps^{'}_{l(l\!+\!1)}\nn[4pt]
 &=& \Ps^{l}_{1\ldots l}\Ps^{l\!-\!1}_{1...(l\!-\!1)}
M_{l(l\!+\!1)}M_{(l\!-\!1)l}'  \Ps^{l\!-\!1}_{1\ldots
(l\!-\!1)} \Ps^{'}_{l(l\!+\!1)} = 
\Ps^{l}_{1\ldots l}M_{l(l\!+\!1)}M_{(l\!-\!1)l}'  \Ps^{l\!-\!1}_{1\ldots
(l\!-\!1)} \Ps^{'}_{l(l\!+\!1)} \nn[4pt]
&\propto & \Ps^{l}_{1\ldots l}
\left(\1_{D^{l\!+\!1}}+\beta_{l\!+\!1}\Pe_{l(l\!+\!1)}
+\gamma_{l\!+\!1}\Pt_{l(l\!+\!1)}\right)
\left(\1_{D^{l\!+\!1}}+\beta_{l}\Pe_{(l\!-\!1)l}
+\gamma_{l}\Pt_{(l\!-\!1)l}\right)\Ps^{l\!-\!1}_{1\ldots
(l\!-\!1)} \Ps^{'}_{l(l\!+\!1)} \nn[4pt]
 &=&\Ps^{l}_{1\ldots l}\left[\1_{D^{l\!+\!1}}\!+\!\beta_{l\!+\!1}
\underset{(\ref{useful2})}{\Pe_{l(l\!+\!1)}}\!+\!\gamma_{l\!+\!1}\Pt_{l(l\!+\!1)}\!+\! \beta_l\underset{(\ref{Plproj1} )}{\Pe_{(l\!-\!1)l}}\!+\!
\gamma_l \underset{(\ref{Plproj1})}{\cancel{\Pt_{(l\!-\!1)l}}}
\!+\! \beta_l\beta_{l\!+\!1}\underset{(\ref{braid2}),\, (\ref{Plproj1})}{\cancel{\Pe_{l(l\!+\!1)}
\Pe_{(l\!-\!1)l}}}\right.
 \nonumber 
\eea
\bea %
&&\left. \!+\!\gamma_l\gamma_{l\!+\!1}
\Pt_{l(l\!+\!1)}\Pt_{(l\!-\!1)l}\!+\!\beta_l\gamma_{l\!+\!1} \underset{(\ref{useful3})}{\Pt_{l(l\!+\!1)}\Pe_{(l\!-\!1)l}}\!+\!\gamma_l\beta_{l\!+\!1}
\underset{(\ref{useful4})}{\Pe_{l(l\!+\!1)}\Pt_{(l\!-\!1)l}} \right]
\Ps^{l\!-\!1}_{1\ldots (l\!-\!1)} \Ps^{'}_{l(l\!+\!1)}\nn[4pt]
&=& \Ps^{l}_{1\ldots l}\left\{\1_{D^{l\!+\!1}}\!+\!\beta_{l\!+\!1}
\left[2\Pt_{l(l\!+\!1)}\!-\! \1_{D^{l\!+\!1}}\right]
\!+\!\gamma_{l\!+\!1}\Pt_{l(l\!+\!1)}\!+\!
\beta_l\1_{D^{l\!+\!1}}\!+\!\gamma_l\gamma_{l\!+\!1}
\Pt_{l(l\!+\!1)}\Pt_{(l\!-\!1)l}\right. \nn[2pt]
&& \left. +\beta_l\gamma_{l\!+\!1} D  \Pt_{l(l\!+\!1)}\Pt_{(l\!-\!1)l}\left[2 
\underset{(\ref{useful5})}{\Pt_{l(l\!+\!1)}}\!-\! \1_{D^{l\!+\!1}}\right]
\!+\! \gamma_l\beta_{l\!+\!1}D \Pt_{l(l\!+\!1)}\Pt_{(l\!-\!1)l} \right\}
\Ps^{l\!-\!1}_{1\ldots (l\!-\!1)}\Ps^{'}_{l(l\!+\!1)}\nn
&=& \Ps^{l}_{1\ldots l}\left\{\1_{D^{l\!+\!1}}\left[1\!-\! \beta_{l\!+\!1}\!+\!
\beta_l\right]\!+\!\Pt_{l(l\!+\!1)} \left[2
\beta_{l\!+\!1}\!+\!
\gamma_{l\!+\!1}\!+\!\beta_l\gamma_{l\!+\!1}
\frac{2}{D } \right]\right. \nn[4pt]
&&\left. + \Pt_{l(l\!+\!1)}\Pt_{(l\!-\!1)l}
\left[\gamma_l\gamma_{l\!+\!1}
\!-\!\beta_l\gamma_{l\!+\!1} D   
\!+\! \gamma_l\beta_{l\!+\!1} D \right]
\right\} \Ps^{l\!-\!1}_{1\ldots (l\!-\!1)} \Ps^{'}_{l(l\!+\!1)}.
\nonumber
\eea
where we have used also the relations
\bea
&& \Pe \Ps^{'}=\Pe \left[\Pt\!+\!\Ps^-\right]\Ps^{'}=\left[\Pt\!-\!\Ps^-\right]\Ps^{'}=\left[2\Pt\!-\!  \1_{D^2}\right]\Ps^{'}     \label{useful2} \\[8pt]
&& \Pt_{l(l\!+\!1)}\Pe_{(l\!-\!1)l}\Ps^{'}_{l(l\!+\!1)}\stackrel{(\ref{useful1''})}{=}
D  \Pt_{l(l\!+\!1)}\Pt_{(l\!-\!1)l}\Pe_{l(l\!+\!1)}\Ps^{'}_{l(l\!+\!1)} \nn
&& \qquad\qquad \qquad\qquad \stackrel{(\ref{useful2})}{=} D \Pt_{l(l\!+\!1)}\Pt_{(l\!-\!1)l}\!\left[2\Pt_{l(l\!+\!1)}\!-\! \1_{D^{l\!+\!1}}\right]\!\Ps^{'}_{l(l\!+\!1)} \label{useful3} \\[8pt]
&& \Ps^{l}_{1\ldots l}\Pe_{l(l\!+\!1)}\Pt_{(l\!-\!1)l} \stackrel{(\ref{useful1})}{=} D  \Ps^{l}_{1\ldots l}\Pe_{(l\!-\!1)l}\Pt_{l(l\!+\!1)}\Pt_{(l\!-\!1)l}  \stackrel{(\ref{Plproj1})}{=}  D \Ps^{l}_{1\ldots l}\Pt_{l(l\!+\!1)}\!\Pt_{(l\!-\!1)l}\label{useful4} \\[8pt] 
&& \Pt_{l(l\!+\!1)}\Pt_{(l\!-\!1)l}\Pt_{l(l\!+\!1)}
=\frac 1{D ^2}\Pt_{l(l\!+\!1)}.\label{useful5} 
\eea
The conditions that the three square brackets vanish
\bea
&& 1\!+\!\beta_l\!-\!\beta_{l\!+\!1}=0,\nn
&& 2\beta_{l\!+\!1}\!+\!
\gamma_{l\!+\!1}\!+\!\beta_l\gamma_{l\!+\!1}
\frac{2}{D }=0, \nn
&& \gamma_l\gamma_{l\!+\!1}\!-\!\beta_l\gamma_{l\!+\!1} D   
\!+\!\gamma_l\beta_{l\!+\!1} D =0,\nonumber
\eea
are recursively solved,  starting from $l=1$ with initial input
$\beta_1=0=\gamma_1$ (since $\Ps^{1}=\1_D$),  by
$$
\beta_{l\!+\!1}=l, \qquad
\gamma_{l\!+\!1}=-\frac{2Dl} {D\!+\!2l\!- \!2}.
$$
We determine the coefficient $\alpha_{l\!+\!1}$ by imposing the
condition (\ref{Plproj2}). This gives
\bea
0&\stackrel{!}{=}&\Ps^{l\!+\!1}\left(\Ps^{l\!+\!1}-\1_{D^{l\!+\!1}}\right) \stackrel{(\ref{ansatz1})}{=}\Ps^{l\!+\!1}\left( \Ps^{l}_{1\ldots l}M_{l(l\!+\!1)}
\Ps^{l}_{1\ldots l}-\1_{D^{l\!+\!1}}\right)\nn
&\stackrel{(\ref{furtherprop}),(\ref{Ml})}{=} & \Ps^{l\!+\!1}\left[
\alpha_{l\!+\!1}\left(1+ \beta_{l\!+\!1}\right)\Ps^{l}_{1\ldots l}-1\right]
\stackrel{(\ref{furtherprop})}{=}\Ps^{l\!+\!1}\left[
\alpha_{l\!+\!1}\left(1+ \beta_{l\!+\!1}\right)-1\right].
\nonumber  
\eea
The condition that the square bracket  vanishes is recursively solved,  
starting from $l=0$ with initial input $\alpha_0=1$,  by $\alpha_{l\!+\!1}=1/(l\!+\!1)$.
This makes (\ref{m2}) into (\ref{Ml}) [yielding back (\ref{projectorpm2}) if $l=2$].
We have thus proved that the Ansatz (\ref{ansatz1}) fulfills (\ref{Plproj1}-\ref{Plproj2}). Similarly one proves  that also the Ansatz (\ref{ansatz2}) does the same job. 

\subsection{Proof of Proposition (\ref{Difference-l})}
\label{ProofDifference-l}

For $l=1$ it is $\zeta_2=1$, $\Ps^1=\1_D$, $\Ps^2=\Ps$, and the claim (\ref{difference-l}) is true by (\ref{projectorpm2}),
 (\ref{ccr}). We now show that  (\ref{difference-l})$_{l\mapsto l-1}$ implies  (\ref{difference-l}):
\bea
\left(\Ps^{l}_{1...l}-\Ps^{l+1}\right)x_1...x_{l+1}\stackrel{(\ref{ansatz1})}{=}
\Ps^{l}_{1...l}\!\left[\1^{\otimes^{l+1}}\!\!-\frac1{l\!+\!1}\!\left(\1^{\otimes^{l+1}}\!\!+l\,\Pe_{l({l\!+\!1)}}\!+\!\gamma_{l+1} \Pt_{l(l\!+\!1)}\right)\!\right]\!\Ps^{l}_{1...l}x_1...x_{l+1}\nonumber
\eea
\bea
=\frac {1}{l\!+\!1}\Ps^{l}_{1...l}\!\left[l\left(\1^{\otimes^{l+1}}\!\!-\Pe_{l(l\!+\!1)}\right)\!-\! \gamma_{l+1}\Pt_{l(l\!+\!1)} \right]\!\left(\Ps^{l}_{1...l}-\Ps^{l\!-\!1}_{1...(l\!-\!1)}+\Ps^{l\!-\!1}_{1...(l\!-\!1)}\right)x_1...x_{l+1}\nn
\stackrel{(\ref{difference-l})_{l\mapsto l-1}}{=}\frac {1}{l\!+\!1}\Ps^{l}_{1...l}\!\left[l\left(\1^{\otimes^{l+1}}\!\!-\Pe_{l(l\!+\!1)}\right)\!-\! \gamma_{l+1}\Pt_{l(l\!+\!1)} \right]\!
\Ps^{l\!-\!1}_{1...(l\!-\!1)}\left(\1^{\otimes^{l+1}}\! - \zeta_l\,\Pt_{(l\!-\!1)l}\right)x_1...x_{l+1}\nn
=\frac {1}{l\!+\!1}\Ps^{l}_{1...l}\!\left[l\left(\1^{\otimes^{l+1}}\!\!-\Pe_{l(l\!+\!1)}\right)\!-\! \gamma_{l+1}\Pt_{l(l\!+\!1)} \right]\!
\left(\1^{\otimes^{l+1}}\! - \zeta_l\,\Pt_{(l\!-\!1)l}\right)x_1...x_{l+1}\nn
=\frac {1}{l\!+\!1}\Ps^{l}_{1...l}\!\left[l\,\underset{
x_1x_2-\Pe x_1x_2=0}{\cancel{\left(\!\1^{\otimes^{l+1}}\!\!-\Pe_{l(l\!+\!1)}\!\right)}}\!-\! \gamma_{l+1}\Pt_{l(l\!+\!1)}\!- \!\zeta_l l\!\left(\!\1^{\otimes^{l+1}}\!\!-\Pe_{l(l\!+\!1)}\!\right)\!\Pt_{(l\!-\!1)l}\!+ \!\zeta_l\,\gamma_{l+1}\Pt_{l(l\!+\!1)}\Pt_{(l\!-\!1)l} \right]\!
x_1...x_{l+1}\nn
=\frac {1}{l\!+\!1}\Ps^{l}_{1...l}\!\left[ -\gamma_{l+1}\Pt_{l(l\!+\!1)}\!- \!\zeta_l l  \big(\underset{(\ref{Plproj1})}{\cancel{\Pt_{(l\!-\!1)l}}}\!\!-\Pe_{l(l\!+\!1)}\Pt_{(l\!-\!1)l}\big)\!\!+ \!\zeta_l\,\gamma_{l+1}\Pt_{l(l\!+\!1)}\Pt_{(l\!-\!1)l}\Pe_{l(l\!+\!1)} \right]\!
x_1...x_{l+1}\nn
\stackrel{(\ref{useful1''})}{=}\frac {1}{l\!+\!1}\Ps^{l}_{1...l}\!\left[-\gamma_{l+1}\Pt_{l(l\!+\!1)}\!+ \!\zeta_l l\,\Pe_{(l\!-\!1)l}\Pe_{l(l\!+\!1)}\Pt_{(l\!-\!1)l}\!+ \!\frac{\zeta_l}{D}\,\gamma_{l+1}\Pt_{l(l\!+\!1)}\Pe_{(l\!-\!1)l} \right]\!x_1...x_{l+1}\nn
\stackrel{(\ref{braid1})}{=}\frac {1}{l\!+\!1}\Ps^{l}_{1...l}\!\left[-\gamma_{l+1}\Pt_{l(l\!+\!1)}\!+ \!\zeta_l l\,\Pt_{l(l\!+\!1)}\Pe_{(l\!-\!1)l}\Pe_{l(l\!+\!1)}\!+ \!\frac{\zeta_l}{D}\,\gamma_{l+1}\Pt_{l(l\!+\!1)}\Pe_{(l\!-\!1)l} \right]\!x_1...x_{l+1}\nn
\stackrel{Px_1x_2=x_1x_2}{=}\frac {1}{l\!+\!1}\Ps^{l}_{1...l}\!\left[-\gamma_{l+1}\Pt_{l(l\!+\!1)}\!+ \!\zeta_l l\,\Pt_{l(l\!+\!1)}\!+ \!\frac{\zeta_l}{D}\,\gamma_{l+1}\Pt_{l(l\!+\!1)}\right]\!x_1...x_{l+1}\nn
=\frac {1}{l\!+\!1}\!\left[- \gamma_{l+1}\!+ \!\zeta_l l\!+ \!\frac{\zeta_l}{D}\,\gamma_{l+1}\right]\!\Ps^{l}_{1...l}\Pt_{l(l\!+\!1)}x_1...x_{l+1}=:\zeta_{l+1}\,\Ps^{l}_{1...l}\Pt_{l(l\!+\!1)}x_1...x_{l+1}\nonumber
\eea
namely the left equality in (\ref{difference-l}) is fulfilled if the $\zeta_l$ satisfy the recursion relation
$$
\zeta_{l+1}=\frac {1}{(l\!+\!1)}\!\left[- \gamma_{l+1}\!+ \!\zeta_l l\!+ \!\frac{\zeta_l}{D}\,\gamma_{l+1}\right]=
\frac {l}{(l\!+\!1)(D\!+\!2l\!- \!2)}\!\left[2D\!+ \!\zeta_l (D\!+\!2l\!- \!4)\right],
$$
which setting $\zeta_2=1$ is actually solved by $\zeta_l=\frac {D(l\!-\!1)}{(D\!+\!2l\!- \!4)}=-\frac 12\gamma_l$, as claimed.

\subsection{Proof of Proposition \ref{LOnX}}

Using Proposition \ref{Difference-l} we easily find 
\bea
iL_{hk}X_l^{i_1...i_l}=(x^h\partial^k\!-\!x^k\partial^h)\Ps^l{}^{i_1...i_l}_{j_1...j_l} x^{j_1} ... x^{j_l} = l (x^hg^{kj_1}\!-\!x^kg^{hj_1}) \Ps^l{}^{i_1...i_l}_{j_1...j_l} x^{j_2} ... x^{j_l} \qquad  \label{interm1}\\
\stackrel{(\ref{difference-l'})}{=}\frac{lD}{\zeta_{l+1}\, r^2}\left[x^h\!\left(\!x^kX_l^{i_1...i_l}
\!-\! X_{l+1}^{ki_1...i_l}\!\right)\!-\!x^k\!\left(\!x^hX_l^{i_1...i_l}\!-\! X_{l+1}^{hi_1...i_l}\!\right)\!\right]
=\frac{lD}{\zeta_{l+1}}\frac{x^kX_{l+1}^{hi_1...i_l}-
x^hX_{l+1}^{ki_1...i_l}}{r^2}\nn
\stackrel{(\ref{difference-l'})}{=}\frac{lD}{\zeta_{l+1}\, r^2}\left[\cancel{X_{l+2}^{khi_1...i_l}}- \cancel{X_{l+2}^{hki_1...i_l}}
+\frac{\zeta_{l+2}}Dr^2\left(g^{kj}\Ps^{l+1}{}^{hi_1...i_l}_{jj_1...j_l}
-g^{hj}\Ps^{l+1}{}^{ki_1...i_l}_{jj_1...j_l}\right)X_l^{j_1...j_l} \right]\nn
=\frac{l\zeta_{l+2}}{\zeta_{l+1}}\left(g^{kj}\Ps^{l+1}{}^{hi_1...i_l}_{jj_1...j_l}
-g^{hj}\Ps^{l+1}{}^{ki_1...i_l}_{jj_1...j_l}\right)X_l^{j_1...j_l}.
\qquad\qquad    \nonumber
\eea
Alternatively, from (\ref{interm1}) we obtain also  (\ref{LonX}), because
\bea
iL_{hk}X_l^{i_1...i_l}\stackrel{(\ref{furtherprop}b)}{=} l (x^hg^{kj_1}\!-\!x^kg^{hj_1}) \Ps^l{}^{i_1...i_l}_{j_1...j_l} X_{l-1}^{j_2...j_l} \nn
\stackrel{(\ref{difference-l'})}{=}  \Ps^l{}^{i_1...i_l}_{j_1...j_l} \,l 
g^{kj_1} \!\left(\!X_{l}^{hj_2...j_l}\!+\!
\frac{\zeta_l}Dr^2 g^{hk_2}\Ps^{l-1}{}^{j_2...j_l}_{k_2...k_l}X_{l-2}^{k_3...k_l}
\!\right)- (h\leftrightarrow k) \nonumber
\eea
\bea 
=\Ps^l{}^{i_1...i_l}_{j_1...j_l} \,l 
g^{kj_1}X_{l}^{hj_2...j_l}+
l\frac{\zeta_l}Dr^2\, g^{kj_1} g^{hk_2}\Ps^l{}^{i_1i_2...i_l}_{j_1k_2...k_l}X_{l-2}^{k_3...k_l}- (h\leftrightarrow k)\nn
=l\,\Ps^l{}^{i_1...i_l}_{j_1...j_l} 
\left(g^{kj_1}X_{l}^{hj_2...j_l}-g^{hj_1}X_{l}^{kj_2...j_l} \right). \nonumber
\qquad\qquad    
\eea
Multiplying the previous relations by $1/r^l$, which commutes with $L_{hk}$, (\ref{LonX}) give (\ref{LonT}). 

\subsection{Proof of Proposition \ref{traceG}  and Theorem \ref{propTTdeco}}
\label{prop-traceG}

The right-hand side (rhs) of (\ref{DefGN}) is the sum of ${\overline{N}}!$ terms; in particular,
 $G_2^{ij}=\delta^{ij}+\delta^{ji}=2\delta^{ij}$.  $G_{\overline{N}},G_{{\overline{N}}-2}$ are related by the recursive relation
\bea
G_{\overline{N}}^{i_1...i_{\overline{N}}}=(\delta^{i_1i_2}\!+\!\delta^{i_2i_1}) G_{{\overline{N}}-2}^{i_3...i_{\overline{N}}}+
(\delta^{i_1i_3}\!+\!\delta^{i_3i_1}) G_{{\overline{N}}-2}^{i_2i_4...i_{\overline{N}}}+...+
(\delta^{i_{{\overline{N}}-1}i_{\overline{N}}}\!+\!\delta^{i_{\overline{N}}i_{{\overline{N}}-1}}) G_{{\overline{N}}-2}^{i_1...i_{{\overline{N}}-2}}.
\label{recursiveG}
\eea
The rhs is the sum of ${\overline{N}}({\overline{N}}\!-\!1)$ products $\delta^{\cdots}G_{{\overline{N}}-2}^{\cdots}$.
The `trace' of $G_{\overline{N}}$
equals  $\mbox{tr}\left(G_2\right)=2D$ for ${\overline{N}}=2$ and by (\ref{recursiveG}) fulfills  the recursive relation $\mbox{tr}(G_{\overline{N}})=\overline{N}(D\!+\!{\overline{N}}\!-\!2)\, \mbox{tr}(G_{{\overline{N}}-2})$.
In fact, each of the ${\overline{N}}$ products $\delta^{i_1i_2}G_{{\overline{N}}-2}^{i_3...i_{\overline{N}}}$, $\delta^{i_2i_1} G_{{\overline{N}}-2}^{i_3...i_{\overline{N}}}$, $\delta^{i_3i_4}G_{{\overline{N}}-2}^{i_1i_2i_5...i_{\overline{N}}}$, $\delta^{i_4i_3}G_{{\overline{N}}-2}^{i_1i_2i_5...i_{\overline{N}}}$,..., $\delta^{i_{\overline{N}}i_{{\overline{N}}-1}} G_{{\overline{N}}-2}^{i_1...i_{{\overline{N}}-2}}$ in (\ref{recursiveG}) contributes by $D\,\mbox{tr}\left(G_{{\overline{N}}-2}\right)$, while each of the ${\overline{N}}(\!{\overline{N}}\!-\!2)$ remaining ones contributes by $\mbox{tr}\left(G_{{\overline{N}}-2}\right)$. The recursion relation is solved by (\ref{TraceG}).  
The  integral over $S^d$ of 
 $H_{\overline{N}}^{i_1...i_{\overline{N}}}:=t^{i_1}...t^{i_{\overline{N}}}$
is $O(D)$-invariant and therefore must be proportional to $G_{\overline{N}}^{i_1...i_{\overline{N}}}$; the proportionality coefficient
 $C_{\overline{N}}$ is found by consistency
contracting both sides with $\delta_{i_1i_2}\delta_{i_3i_4}...\delta_{i_{{\overline{N}}-1}i_{\overline{N}}}$,
and using (\ref{t-relation}), (\ref{TraceG}).
The scalar product of $\bphi^l,\bpsi^l\in V_D^l\subset\Hi_s$ is given by
\bea
\la\bphi^l,\bpsi^l\ra &=& \int_{S^d} d\alpha\: \bphi^l{}^*\bpsi^l
=
\big(\phi^l_{i_1...i_l}\big)^* \psi^l_{j_1,...,j_l} \int_{S^d} d\alpha\: t^{i_1}...t^{i_l}    t^{j_1}...t^{j_l} \nn
 &\stackrel{(\ref{inv-t-integral})}{=}& C_{2l} 
\big(\phi^l_{i_1...i_l}\big)^* \psi^l_{j_1...j_l}  G_{2l}^{i_1...i_lj_1...j_l}; \nonumber
\eea
the sum has $(2l)!$ terms.  In fact, all terms where both indices of 
at least one Kronecker $\delta$ contained in $G_{2l}$ are contracted with
the two indices of the coefficients $\phi^l_{i_1...i_l}$, or of 
the  $\psi^l_{i_1...i_l}$, vanish,
by (\ref{phi-coeff-prop}). The remaining $2^l(l!)^2$ terms 
arise from the $(l!)^2$ products contained in $G_{2l}$ of the type
$\delta^{i_{\pi(1)}j_{\pi'(1)}}...\delta^{i_{\pi(l)}j_{\pi'(l)}}$,
where $\pi,\pi'$ are permutations of $(1,...,l)$, and the other ones which are obtained
exchanging the order of the indices $i_{\pi(h)}j_{\pi'(h)}\mapsto  j_{\pi'(h)}i_{\pi(h)}$
in one or more of these  Kronecker $\delta$'s; they 
are all equal, again by (\ref{phi-coeff-prop}). Hence,
\bea
\la\bphi^l,\bpsi^l\ra = Q_l 
\big(\phi^l_{i_1...i_l}\big)^* \psi^l_{i_1...i_l}, \qquad
\label{productphil-psil}
\eea
By the orthogonality $V_D^l\perp V_D^{l'}$ for $l\neq l'$ we find that the scalar product
of  generic $\bphi,\bpsi \subset\Hi_s$ is given by (\ref{productphipsi}).
This concludes the proof of Proposition \ref{traceG}.


Applying $m$ times  (\ref{tTdeco}) and absorbing into a suitable combination of $T_n$'s ($n\in L$) the $m$-degree monomials in  $t^i$ whose combination gives
$T_m^{j_1...j_m}$, we find that (\ref{TTdeco}) must hold with suitable coefficients $S^{i_1...i_l,j_1...j_m}_{k_1...k_n}\in\RR$. We determine these coefficents   faster
using (\ref{TTdeco})  as an Ansatz,  making its scalar product   with $T_h^{k_1...k_h}$ and using
(\ref{productTl-phi}).  One finds
\bea
Q_n \,S^{i_1...i_l,j_1...j_m}_{k_1...k_n} &=& \left\la T_n^{k_1...k_n}, T_l^{i_1...i_l}T_m^{j_1...j_m}\right\ra \nn&=&
 \Ps^l{}_{a_1...a_l}^{i_1...i_l}\Ps^m{}_{b_{1}...b_m}^{j_1...j_m}
\Ps^h{}_{c_1...c_n} ^{k_1...k_n} 
\int_{S^d}\!\!\! d\alpha\: t^{a_1}...t^{a_l}    t^{b_1}...t^{b_m} t^{c_1}...t^{c_n}\nn
&=&  \Ps^l{}_{a_1...a_l}^{i_1...i_l}\Ps^m{}_{b_{1}...b_m}^{j_1...j_m}
\Ps^n{}_{c_1...c_n} ^{k_1...k_n}  \, C_{\overline{N}}\, G_{\overline{N}}^{a_1...a_lb_1... b_mc_1...c_n},   \label{Scondi}
\eea
where $\overline{N}=l\!+\!m\!+\!n=n\!+\!r=:2s$ is even.
Due to the form of $G_{\overline{N}}
$,
the sum has $\overline{N}!$ terms, each containing a product of $s$ Kronecker $\delta$'s.  All terms where both indices of 
some Kronecker $\delta$ contained in $G_{2s}$ are contracted with
two indices of  $\Ps^l$, or $\Ps^m$, or $\Ps^n$,  vanish,
by (\ref{Plprojg}). 

As a warm-up, consider first the case $n=l\!+\!m$. Renaming for convenience $b_1,...,b_m$ as $a_{l+1},...a_n$, the remaining $2^n (n!)^2$ terms  arise from the $(n!)^2$ products  of the type
$$
\delta^{c_{\pi(1)}a_{\pi'(1)}}\: ...\: \delta^{c_{\pi(n)}a_{\pi'(n)}}
$$
contained in  $G_{\overline{N}}$,  where $\pi,\pi'$ are   permutations of $(1,...,h)$,  and the other ones which are obtained
exchanging the order of the indices $c_{\pi(h)}a_{\pi'(h)}\mapsto  a_{\pi'(h)}c_{\pi(h)}$,
in one or more of these $n$ Kronecker $\delta$'s; by the complete symmetry of $\Ps^n$, they 
are all equal to the term where $\pi,\pi'$ are the trivial permutations; correspondingly, 
the product is \ $\delta^{c_1a_1}\: ...\: \delta^{c_na_n}$. \ Hence,
\bea
Q_n \,S^{i_1...i_l,j_1...j_m}_{k_1...k_n} = C_{2n}\, 2^n(n!)^2\,
 \Ps^l{}_{a_1...a_l}^{i_1...i_l}\Ps^m{}_{a_{l+1}...a_n}^{j_1...j_m}
\Ps^n{}^{a_1...a_n}_{k_1...k_n}= C_{2n}\, 2^n(n!)^2\,
\Ps^n{}^{i_1...i_lj_1...j_m}_{k_1...k_n} 
\nonumber
\eea
implying \ $S^{i_1...i_l,j_1...j_m}_{k_1...k_n} = 
\Ps^n{}^{i_1...i_lj_1...j_m}_{k_1...k_n}$, \ i.e. the term of highest rank
at the rhs(\ref{TTdeco}) is \ $T_n^{i_1...i_lj_1...j_m}$. This is consistent with the first term in
the (even iterated) application of  (\ref{tTdeco}).

For generic $n\in\ L$ (\ref{Scondi}) becomes
\bea
\ba{l}
Q_{n} \,S^{i_1...i_l,j_1...j_m}_{k_1...k_n} =C_{2s}\,  F^{l,m}_n
\Ps^l{}_{a_1...a_l}^{i_1...i_l}\Ps^m{}_{b_{1}...b_m}^{j_1...j_m}
\Ps^n{}_{c_1...c_n} ^{k_1...k_n}  \: \times  \\[6pt]
\delta^{a_1b_1}\, \delta^{a_2b_2}\,...\, \delta^{a_{r} b_r} \:
\delta^{a_{r+1}c_{1}}\, ...\, \delta^{a_{l}c_{l-r}}\:\delta^{b_{r+1}c_{l-r+1}}
\, ...\, \delta^{b_mc_n},
\ea  
 \label{Scondi'}
\eea
where  \ $r\!:=\!\frac{l\!+\!m\!-\!n}2$, $s=\frac{l\!+\!m\!+\!n}2$, $F^{l,m}_n=\frac{ 2^s\,s!\,l!\, m!\,(n!)^2}{(l\!-\!r)!\,(m\!-\!r)!\,(s\!-\!r)!}=\frac{ 2^s\,s!\,l!\, m!\,n!}{(l\!-\!r)!\,(m\!-\!r)!}$. \ 
In fact,  one of the
products of Kronecker $\delta$'s contained in  $G_{2s}$ and yielding a nonzero contribution is displayed in the second line. The number $F^{l,m}_n$ of such products can be determined as follows, starting
from the same product with all indices removed: there are $l$ ways to pick out the first index
from the set $A_l:=\{a_1,...,a_l\}$, $m$ ways to pick out the second index
from the set $B_m:=\{b_{1},...,b_m\}$, hence $lm$ ways to pick out the first index
from  $A_l$ and the second 
from $B_m$; similarly, there are $lm$ ways to pick out the second index
from  $A_l$ and the first from $B_m$;
altogether, there are $2lm$ ways  to pick out one of the first two indices  
from  $A_l$ and the other one from $B_m$. After anyone of these choices
the corresponding sets of indices $A_{l-1}$, $B_{m-1}$  will have $l\!-\!1,m\!-\!1$ indices respectively; therefore there are
$2(l\!-\!1)(m\!-\!1)$ ways  to pick out one among the third, fourth indices from
$A_{l-1}$  and the other from $B_{m-1}$. And so on. Therefore
there are $\frac{2^r\,l!\, m!}{(l\!-\!r)!\,(m\!-\!r)!}$ ways to
pick out $r$ indices appearing in the first $r$ $\delta$'s (one for each  $\delta$) from $A_l$ 
 and the $r$ remaining ones from $B_m$.
After anyone of these choices
the corresponding sets of indices $A_{l-r}$ and 
 $B_{m-r}$  will have $l\!-\!r,m\!-\!r$ indices respectively, and 
$D_n:=A_{l-r}\cup B_{m-r}$ will have $l\!-\!r\!+\!m\!-\!r=n$
indices.
Reasoning as in the case $n=l\!+\!m$, we find that there are $2^n (n!)^2$ ways to
pick out $n$ indices appearing in the remaining $n$ $\delta$'s  (one for each  $\delta$) from $D_n$  and the remaining $h$ ones from  $C_n:=\{c_{1},...,c_n\}$. Consequently, so far there are
$\frac{2^{n+r}\,l!\, m!}{(l\!-\!r)!\,(m\!-\!r)!} (n!)^2$ ways to do these operations.
The remaining ways are obtained allowing that the $r$ pair of indices picked one out of $A_l$ and  the other out of $B_m$ appear not necessarily in the first $r$ $\delta$'s, but in any subgroup of $r$ 
$\delta$'s out of the totality of $s$; hence we have to multiply the previous number by 
the number $\frac{s!}{(s-r)!}$ of such subgroups, and we finally obtain 
$F^{l,m}_n=\frac{ s!\,2^{s}\,l!\, m!\,(n!)^2}{(s\!-\!r)!\,(l\!-\!r)!\,(m\!-\!r)!}$ (because $n\!+\!r=s$), as claimed. Thus  (\ref{Scondi'}) implies the following relation, whiche gives (\ref{Scondi''}):
\bea
&&S^{i_1...i_l,j_1...j_m}_{k_1...k_n} =  \frac{(D\!+\!2n\!-\!2)!!}{(D\!+\!2s\!-\!2)!!\,n!\,(2s)!!}
\frac{ 2^s\,s!\,l!\, m!\,n!}{(l\!-\!r)!\,(m\!-\!r)!}\, \Ps^l{}_{a_1...a_l}^{i_1...i_l}\Ps^m{}_{b_{1}...b_m}^{j_1...j_m} \Ps^n{}_{c_1...c_n} ^{k_1...k_n}  \nn
&&\qquad\qquad\times\:\:\delta^{a_1b_1}\, \delta^{a_2b_2}\,...\, \delta^{a_{r} b_r} \:
\delta^{a_{r+1}c_{1}}\, ...\, \delta^{a_{l}c_{l-r}}\:\delta^{b_{r+1}c_{l-r+1}}
\, ...\, \delta^{b_mc_n} \nonumber
\eea

\subsection{Proof of Proposition \ref{cVbDDeco}}
\label{Proof-cVbDDeco}

Since $L_{hk}$ commute with scalars, and $[L_{hk},x^\bD]=0$,
$\check p_{\Lambda,l}$ commutes with all the $L_{hk}$ and therefore
also with $\bL^2$. Hence, using (\ref{LeigenvectorsX}), we find $\bL^2 \check p_{\Lambda,l}X_{l}^{i_1...i_l}=\check p_{\Lambda,l} \bL^2 X_{l}^{i_1...i_l}=E_l\check p_{\Lambda,l}X_{l}^{i_1...i_l}$, i.e (\ref{L^2_DcF}).
To compute the coefficients $b_{\Lambda,l+2k}$ we preliminarly note that
\bea
&& \Delta (x^i)^h= (x^i)^h \Delta +2h (x^i)^{h-1}\partial^i+h(h\!-\!1) (x^i)^{h-2},\nn[8pt]
&& \Delta r^h=r^h \Delta +h r^{h-2}(2\eta\!+\! D \!+\!h\!-\!2) \nn[8pt]
&& \Delta_\bD (x^I)^h= (x^I)^h \Delta +2h (x^I)^{h-1}\partial^I+h(h\!-\!1) (x^I)^{h-2},\nn[8pt]
&& \Delta_\bD r_\bD^h=r_\bD^h \Delta_\bD +h r_\bD^{h-2}(2\bar\eta\!+\! \bD \!+\!h\!-\!2) \nn[8pt]
&& \partial^j\check p_{\Lambda,l}=x^j\left[2\left(x^\bD\right)^{\Lambda-l-2}\, b_{\Lambda,l+2}
+4\left(x^\bD\right)^{\Lambda-l-4}r_\bD^2\, b_{\Lambda,l+4}+...\right]\nonumber
\eea
\bea
&& \Delta_\bD\,\check p_{\Lambda,l}|=(\Lambda\!-\!l)(\Lambda\!-\!l\!-\!1)\left(x^\bD\right)^{\Lambda-l-2} +\left[r_\bD^2\Delta_\bD+2(2\bar\eta\!+\! \bD )\right]\left(x^\bD\right)^{\Lambda-l-2}\, b_{\Lambda,l+2}\nn
&& +\left[r_\bD^4\Delta_\bD+4r_\bD^2(2\bar\eta\!+\! \bD\!+\!2 )\right]\left(x^\bD\right)^{\Lambda-l-4}\, b_{\Lambda,l+4}+...\nn
&& =(\Lambda\!-\!l)(\Lambda\!-\!l\!-\!1)\left(x^\bD\right)^{\Lambda-l-2}\nn
&& +\left\{r_\bD^2(\Lambda\!-\!l\!-\!2)(\Lambda\!-\!l\!-\!3) +2\big[2(\Lambda\!-\!l\!-\!2)\!+\! \bD \big]\left(x^\bD\right)^2\right\}\left(x^\bD\right)^{\Lambda-l-4} b_{\Lambda,l+2}\nn
&& +\left\{r_\bD^4(\Lambda\!-\!l\!-\!4)(\Lambda\!-\!l\!-\!5)+4r_\bD^2\left(x^\bD\right)^2\big[2(\Lambda\!-\!l\!-\!4)\!+\! \bD\!+\!2\big]\right\} \left(x^\bD\right)^{\Lambda-l-6}\,b_{\Lambda,l+4}+...\nn
&& =\left(x^\bD\right)^{\Lambda-l-2}\left\{(\Lambda\!-\!l)(\Lambda\!-\!l\!-\!1)+2b_{\Lambda,l+2}\big[2(\Lambda\!-\!l\!-\!2)\!+\! \bD \big]\right\}\nn
&& + r_\bD^2\left(x^\bD\right)^{\Lambda-l-4}\left\{(\Lambda\!-\!l\!-\!2)(\Lambda\!-\!l\!-\!3) 
 b_{\Lambda,l+2}+4\,b_{\Lambda,l+4}\big[2(\Lambda\!-\!l\!-\!4)\!+\! \bD\!+\!2\big]\right\}\nn
&& +r_\bD^4  \left(x^\bD\right)^{\Lambda-l-6}\left\{(\Lambda\!-\!l\!-\!4)(\Lambda\!-\!l\!-\!5)+6b_{\Lambda,l+6}\big[2(\Lambda\!-\!l\!-\!6)\!+\! \bD\!+\!4\big]\right\}+...\nonumber
\eea
We now impose that the $\check F^{i_1...i_l}_{\bD,\Lambda}$ are harmonic in dimension $\bD$. By a direct calculation 
\bea
&& 0= \Delta_\bD\, \check F^{i_1...i_l}_{\bD,\Lambda}=
\Ps^l{}^{i_1...i_l}_{j_1...j_l}\left\{ x^{j_1}...x^{j_l}\Delta_\bD+
2lx^{j_1}...x^{j_{l-1}}\partial^{j_l}\right\}\check p_{\Lambda,l}=\Ps^l{}^{i_1...i_l}_{j_1...j_l}x^{j_1}...x^{j_l}
\check M_{\Lambda,l+2} \quad\Rightarrow\nn
&& 0 =\check M_{\Lambda,l+2} :=
\Delta_\bD\,\check p_{\Lambda,l}+2l\left[2\left(x^\bD\right)^{\Lambda-l-2}\, b_{\Lambda,l+2}+4\left(x^\bD\right)^{\Lambda-l-4}r_\bD^2\, b_{\Lambda,l+4}+...\right]\nn
&& =\left(x^\bD\right)^{\Lambda-l-2} \left\{(\Lambda\!-\!l)(\Lambda\!-\!l\!-\!1)+
2b_{\Lambda,l+2}\left[2l\!+\!2(\Lambda\!-\!l\!-\!2)\!+\! \bD\right]\right\}\nn
&& +\left(x^\bD\right)^{\Lambda-l-4}r_\bD^2 \left\{b_{\Lambda,l+2}(\Lambda\!-\!l\!-\!2)(\Lambda\!-\!l\!-\!3)+4b_{\Lambda,l+4}\left[2l\!+\!2(\Lambda\!-\!l\!-\!4)\!+\! \bD\!+\!2\right]\right\}+... \nn
&& =\left(x^\bD\right)^{\Lambda-l-2} \left\{(\Lambda\!-\!l)(\Lambda\!-\!l\!-\!1)+
2b_{\Lambda,l+2}\left[2\Lambda\!-\!4\!+\! \bD\right]\right\}\nn
&& +\left(x^\bD\right)^{\Lambda-l-4}r_\bD^2 \left\{b_{\Lambda,l+2}(\Lambda\!-\!l\!-\!2)(\Lambda\!-\!l\!-\!3)+4b_{\Lambda,l+4}\left[2\Lambda\!-\!6\!+\! \bD\right]\right\}+...;\nonumber
\eea
the vanishing of the coefficient of each monomial $\left(x^\bD\right)^{\Lambda-l-2-2h}r_\bD^{2h}$ implies
\bea
&& b_{\Lambda,l+2}=-\frac{(\Lambda\!-\!l)(\Lambda\!-\!l\!-\!1)}{2(2\Lambda\!-\!4\!+\! \bD)}, \nn
&& b_{\Lambda,l+4}=-\frac{(\Lambda\!-\!l\!-\!2)(\Lambda\!-\!l\!-\!3)}{4(2\Lambda\!-\!6\!+\! \bD)}b_{\Lambda,l+2}=\frac{(\Lambda\!-\!l)(\Lambda\!-\!l\!-\!1)(\Lambda\!-\!l\!-\!2)(\Lambda\!-\!l\!-\!3)}{2\cdot 4(2\Lambda\!-\!4\!+\! \bD)(2\Lambda\!-\!6\!+\! \bD)},\nn[2pt]
&& ...; \nonumber
\eea
namely, more compactly, we obtain (\ref{bcoeff}).

The  $ \check F^{i_1...i_l}_{\bD,\Lambda}$ transform as the $X^{i_1...i_l}_{l}$
under the action of the $L_{hk}$, because the latter commute
with $\check p_{\Lambda,l}$. Using (\ref{ccr}), (\ref{cFDeco}), and the fact that $\partial^\bD$ annihilates all polynomials in the $x^i$, we find
\bea
&& iL_{h\bD}\check F^{i_1...i_l}_{\bD,\Lambda}=(x^h\partial^\bD-
x^\bD\partial^h)\check p_{\Lambda,l}X_{l}^{i_1...i_l}\nn
&& =\left\{x^h\left(x^\bD\right)^{\Lambda-l-1}(\Lambda\!-\!l)+x^h\left(x^\bD\right)^{\Lambda-l}\cancel{\partial^\bD}-
\left(x^\bD\right)^{\Lambda-l+1}\partial^h \right.\nn
&& +b_{\Lambda,l+2}r_\bD^2\left[x^h\left(x^\bD\right)^{\Lambda-l-3}(\Lambda\!-\!l\!-\!2)+x^h\left(x^\bD\right)^{\Lambda-l-2}\cancel{\partial^\bD}-\left(x^\bD\right)^{\Lambda-l-1}\partial^h\right]\nn
&& \left. +b_{\Lambda,l+4}r_\bD^4\left[x^h\left(x^\bD\right)^{\Lambda-l-5}(\Lambda\!-\!l\!-\!4)+x^h\left(x^\bD\right)^{\Lambda-l-4}\cancel{\partial^\bD}-\left(x^\bD\right)^{\Lambda-l-3}\partial^h\right]+...\right\}X_{l}^{i_1...i_l}\nn
&& =x^hX_{l}^{i_1...i_l}\check N_{\Lambda,l+1}- l\,x^\bD \check p_{\Lambda,l}\Ps^l{}^{i_1i_2...i_l}_{hj_2...j_l}X_{l-1}^{j_2...j_l}=X_{l+1}^{hi_1...i_l}\check N_{\Lambda,l+1}-\check Q_{\Lambda,l-1}\,\Ps^l{}^{i_1i_2...i_l}_{hj_2...j_l}X_{l-1}^{j_2...j_l},\nonumber
\eea
where $\check N_{\Lambda,l+1}$ is the homogeneous polynomials  of degree $l'\!-\! 1$   in  $x^\bD,r_\bD$ 
\bea
&& \check N_{\Lambda,l+1}:=\left(x^\bD\right)^{l'-1}\, l'\,+
b_{\Lambda,l+2}r_\bD^2\left(x^\bD\right)^{l'-3}(l'\!-\!2)+b_{\Lambda,l+4}r_\bD^4\left(x^\bD\right)^{l'-5} (l'\!-\!4)+...\nn
&& =\, l'\,\left[\left(x^\bD\right)^{l'-1}\!\!-\frac{(l'\!-\!1)(l'\!-\!2)}{2(2\Lambda\!-\!4\!+\! \bD)}r_\bD^2\left(x^\bD\right)^{l'-3}\!\!+ 
\frac{(l'\!-\!1)...(l'\!-\!4)}{8(2\Lambda\!-\!4\!+\! \bD)(2\Lambda\!-\!6\!+\! \bD)} r_\bD^4\left(x^\bD\right)^{l'-5}-...\right]\nn
&& =\, ( \Lambda\!-\!l)\,\check p_{\Lambda,l+1} \label{inter1}
\eea
(here $l'\!\equiv\! \Lambda\!-\!l$) and $\check Q_{\Lambda,l-1}$ is the homogeneous polynomial  of degree $l'\!+\! 1$   in  $x^\bD,r_\bD$ 
\bea
&& \check Q_{\Lambda,l-1}:= l\, x^\bD \check p_{\Lambda,l}-\frac{\zeta_{l+1}}D\,r_D^2 \check N_{\Lambda,l+1}=
l\, x^\bD \check p_{\Lambda,l}-\frac l{D\!+\!2l\!-\!2}\left[r_\bD^2-\left(x^\bD\right)^2\right] \check N_{\Lambda,l+1}\nn
&& =\left(x^\bD\right)^{l'+1}\!\left[\frac {ll'}{D\!+\!2l\!-\!2}+l\right]  - r_\bD^2\left(x^\bD\right)^{l'-1} l l' \left\{\frac {1}{D\!+\!2l\!-\!2}\left[1\!+\!
\frac{(l'\!-\!1)(l'\!-\!2)}{2(2\Lambda\!-\!4\!+\! \bD)}\right]+
\frac{l'\!-\!1}{2(2\Lambda\!-\!4\!+\! \bD)}\right\}\nonumber
\eea
\bea
&& +r_\bD^4\left(x^\bD\right)^{l'-3}\frac{l l'\,(l'\!-\!1)(l'\!-\!2)}{2(2\Lambda\!-\!4\!+\! \bD)}\left\{\frac {1}{D\!+\!2l\!-\!2}\left[1\!+\!
\frac{(l'\!-\!3)(l'\!-\!4)}{2(2\Lambda\!-\!6\!+\! \bD)}\right]+
\frac{l'\!-\!3}{2(2\Lambda\!-\!6\!+\! \bD)}\right\}+...\nn
&& =\frac {l(\Lambda\!+\!l\!+\!D\!-\!2)}{D\!+\!2l\!-\!2}\check p_{\Lambda,l-1}; \label{inter2}
\eea
(equalities (\ref{inter1}-\ref{inter2})  are proved by direct calculations), whence (\ref{L-su-cF}).

\subsection{Evaluating a  class of radial integrals, and proof of (\ref{xonpsi})}
\label{scpD}

Given a smooth $h(r)$ not depending on $k$, formula (98) of \cite{FioPis18}
gives the  (asymptotic expansion of)  the radial integral of its product with $g_l(r)g_L(r)$ ($l,L\in\NN_0$) 
at lowest order in $1/k$: 
\bea
\int_0^{\infty}\!\!\! g_L(r) g_l(r)h(r)dr= e^{-\frac{\sqrt{k_lk_L}\left(\widetilde{r}_l-\widetilde{r}_L\right)^2}{2\left(\sqrt{k_l}+\sqrt{k_L}\right)}}\:\sum_{n=0}^{+\infty}{ \frac{h^{(2n)}\left(\widehat{r}_{l,L}\right)}{(2n)!!\left(\sqrt{k_l}+\sqrt{k_L}\right)^n}}.                        \label{general}
\eea
$\widetilde{r}_l,k_l$ were defined in (\ref{definizioni1}), while \ $\widehat{r}_{l,L}\!:=\!\frac{\sqrt{k_l}\widetilde{r}_l+\sqrt{k_L}\widetilde{r}_L}{\sqrt{k_l}+\sqrt{k_L}}$, $h^{(n)}(r)\equiv d^nh/dr^n$.  
Up to $O\left(k^{-3/2} \right)$ the exponential in (\ref{general}) is 1, because by explicit computation 
\bea
&&\sqrt{k_l}=\sqrt{2k}\left(\!1\!+\!\frac{3b(l,\!D)}{4k}\right)+O\!\left(k^{-\frac{3}{2}} \right)\!,
 \quad \widetilde{r}_l-\widetilde{r}_L=\frac{b(l,\!D)}{3b(l,\!D)\!+\!2k}-\frac{b(L,\!D)}{3b(L,\!D)\!+\!2k}= O\!\left(k^{-1} \right)\!,\nn[6pt]
&&\sqrt{k_l}\widetilde{r}_l=\sqrt{2k}\left(1+\frac{3b(l,\!D)}{4k}\right)\!
\left(1+\frac{b(l,\!D)}{2k}\right)+O\!\left(k^{-\frac{5}{2}}\right)=\sqrt{2k}\left(1+\frac{5b(l,\!D)}{4k}\right)+O\left(k^{-\frac{3}{2}} \right),\nn[6pt]
&&\widehat{r}_{l,L}=\frac{2+\frac{5b(l,\!D)}{4k}+\frac{5b(L,\!D)}{4k}}{2+\frac{3b(l,\!D)}{4k}+\frac{3b(L,\!D)}{4k}}+O\!\left(k^{-2}\right)=
1\!+\!\frac{b(l,\!D)\!+\!b(L,\!D)}{4k}+O\!\left(k^{-2}\right), \quad \frac{\sqrt{k_lk_L}\left(\widetilde{r}_l-\widetilde{r}_L\right)^2}{2\left(\sqrt{k_l}+\sqrt{k_L}\right)}=O\!\left(k^{-\frac{3}{2}}\right).\qquad \nonumber
\eea

 By (\ref{Proj_l}),  (\ref{tTdeco}),
applying $P_{\Lambda}=\sum_{l=0}^\Lambda P^l_{\Lambda}$ 
to $x^i\bpsi
^{i_1i_2...i_l}=t^{i}\,T_{l}^{i_1i_2...i_l} \, r f_{l}(r)$ we find
\bea
P_{\Lambda}\left(x^i\bpsi
_l^{i_1i_2...i_l}\right) &\stackrel{(\ref{tTdeco})}{=}&
\widetilde{P}_{l+1}\left[T_{l+1}^{ii_1...i_l}\, r f_{l}(r)\right]+
\widetilde{P}_{l-1}\left[\frac{\zeta_{l+1}}D\,g^{ih}\Ps^{l}{}^{i_1i_2...i_l}_{hj_2...j_l}  T_{l-1}^{j_2...j_l}\, r f_{l}(r)\right]\nn 
&\stackrel{ (\ref{Proj_l})}{=}&c_{l+1}\, T_{l+1}^{ii_1...i_l}f_{l+1}(r)+c_{l}\, \frac{\zeta_{l+1}}D\,\Ps^{l}{}^{i_1i_2...i_l}_{ij_2...j_l} T_{l-1}^{j_2...j_l} f_{l-1}(r)\nn 
&\stackrel{ (\ref{Defpsi})}{=}&c_{l+1}\,\bpsi
_{l+1}^{ii_1...i_l}+
c_{l}\, \frac{\zeta_{l+1}}D\,\Ps^{l}{}^{i_1i_2...i_l}_{ij_2...j_l}\bpsi
_{l-1}^{j_2...j_l}, \nn[4pt]
\mbox{where }\quad c_{l} &:=& \left\{\!\!\ba{ll}
0 \:&\mbox{if }l=0,\Lambda, \\[6pt]
\displaystyle \rho_{l-1,l}:=\int^\infty_0\!\!\!\!r^d dr\, f_{l-1}^*(r) rf_l(r)=\int^\infty_0\!\!\!\! dr\, g_{l-1}(r) g_l(r)\,r\:&\mbox{if }0<l<\Lambda.
\ea\right. \quad
\nonumber
\eea
The last integral is of the type (\ref{general}) with \ $L=l\!-\!1$ and 
$h(r)=r$. \ By explicit computation,
\bea
\rho_{l-1,l} &=& 1+\frac{D^2\!-\!4D\!+\!3}{k}+\frac{l(l\!+\!D\!-\!2)\!+\!(l\!+\!1)(l\!-\!1\!+\!D)}{2k} \,+\,O\left(k^{-\frac{3}{2}} \right)\nn
 &=& \sqrt{1+\frac{(2D\!-\!5)(D\!-1\!)}{2k}+\frac{(l\!-\!1)(l\!+\!D\!-\!2)}{k}}
\,+\,O\left(k^{-\frac{3}{2}} \right). \label{general''}
\eea
By setting  the $O\left(k^{-3/2} \right)$ term equal to zero we finally arrive at
(\ref{defc}-\ref{xonpsi}).

\subsection{Proof of Proposition \ref{genDpropo}}

The equations in (\ref{linea1}), (\ref{linea4}),  (\ref{linea3}) follow from (\ref{Lvcr}),  (\ref{LLcr}),  (\ref{Lijrel}) and \ $\left[L_{ij},P_{\Lambda}\right]=0$.

By (\ref{HighestWeights}-\ref{HighestWeights'}), $\bpsi_{\Lambda,\pm}^{hk}:=T_{\Lambda,\pm}^{hk}f_\Lambda(r)\in \Hi_\Lambda^\Lambda$ are eigenvectors  with the highest and lowest eigenvalues, $\pm\Lambda$.
Given a basis $\left\{\bpsi_{m}^{\alpha}\right\}$
of eigenvectors of $L_{hk}$, $L_{hk}\bpsi_{m}^{\alpha}=m \,\bpsi_{m}^{\alpha}$ with
$m\in\{-\Lambda,1\!-\!\Lambda,...,\Lambda\}$ ($\alpha$ are some extra labels),  by (\ref{Lxhkcr}) we find that for all $m,\alpha$
$\left(\overline{x}^h\!+\!i\overline{x}^k\right)^j\bpsi_{m}^{\alpha}$ is either zero or an eigenvector of $L_{hk}$ with eigenvalue $j\!+\!m$, which must be $\le\Lambda$. 
Similarly, $\left(L_{hj}\!\pm\!iL_{kj}\right)^j\bpsi_{m}^{\alpha}$ is either zero or an eigenvector of $L_{hk}$ with eigenvalue $j\!+\!m$, which must be $\le\Lambda$. 
Therefore for $j>2\Lambda$ such vectors must be zero, and we obtain   (\ref{linea5}).

Applying (\ref{xonpsi}) twice we find
\bea
\overline{x}^i\,\overline{x}^j\bpsi
_l^{i_1i_2...i_l} &\stackrel{(\ref{xonpsi})}{=}&  \overline{x}^i\left(c_{l+1}\,\bpsi
_{l+1}^{ji_1...i_l}+
c_{l}\, \frac{\zeta_{l+1}}D\,g^{jh}\Ps^{l}{}^{i_1i_2...i_l}_{hj_2...j_l} \bpsi
_{l-1}^{j_2...j_l}\right)\nn
 &\stackrel{(\ref{xonpsi})}{=}& c_{l}\,\frac{\zeta_{l+1}}D\,g^{jh}\Ps^{l}{}^{i_1i_2...i_l}_{hj_2...j_l}
 \left(c_{l}\,\bpsi
_{l}^{ij_2...j_l}+
c_{l-1}\, \frac{\zeta_{l}}D\,g^{ik}\Ps^{l-1}{}^{j_2...j_l}_{kk_3...k_l} \bpsi
_{l-2}^{k_3...k_l}\right)\nn
&&+ c_{l+1}\,\left(c_{l+2}\bpsi
_{l+2}^{iji_1...i_l}+
c_{l+1}\, \frac{\zeta_{l+2}}D\,g^{ih}\Ps^{l+1}{}^{ji_1...i_l}_{hj_1...j_l} \bpsi
_{l}^{j_1...j_l}\right)\nn
 &=& c_{l+1}\,c_{l+2}\bpsi
_{l+2}^{iji_1...i_l}+ c_{l}\,c_{l-1}\,\frac{\zeta_{l+1}}D\, \frac{\zeta_{l}}D\, g^{jh}g^{ik} \Ps^{l}{}^{i_1i_2i_3...i_l}_{h\,k\,k_3...k_l} \bpsi
_{l-2}^{k_3...k_l}\nn
&&+(c_{l+1})^2\, \frac{\zeta_{l+2}}D\,g^{ih}\Ps^{l+1}{}^{ji_1...i_l}_{hj_1...j_l} \bpsi
_{l}^{j_1...j_l}+(c_{l})^2\,\frac{\zeta_{l+1}}D\,g^{jh}\Ps^{l}{}^{i_1i_2...i_l}_{hj_2...j_l}
\,\bpsi
_{l}^{ij_2...j_l}  \label{xxonX}
\eea
Taking the difference of (\ref{xxonX}) and  (\ref{xxonX})  with $i,j$ exchanged we find
\bea
[\overline{x}^i,\overline{x}^j]\bpsi
_l^{i_1i_2...i_l} &=&  \overline{x}^i\,\overline{x}^j\bpsi
_l^{i_1i_2...i_l}- \left(i\leftrightarrow j\right)\nn
 &\stackrel{(\ref{xxonX})}{=}& (c_{l+1})^2\, \frac{\zeta_{l+2}}D\left(g^{ih}\Ps^{l+1}{}^{ji_1...i_l}_{hj_1...j_l}
-g^{jh}\Ps^{l+1}{}^{ii_1...i_l}_{hj_1...j_l}\right) \bpsi
_{l}^{j_1...j_l}\nn
&&+(c_{l})^2\,\frac{\zeta_{l+1}}D\left(g^{jh}\Ps^{l}{}^{i_1i_2...i_l}_{hj_2...j_l}\bpsi
_{l}^{ij_2...j_l}
-g^{ih}\Ps^{l}{}^{i_1i_2...i_l}_{hj_2...j_l}\bpsi
_{l}^{jj_2...j_l}\right)\nn
&\stackrel{(\ref{LonX})}{=} & \left [(c_{l})^2\,\frac{\zeta_{l+1}}{Dl}-(c_{l+1})^2\frac{\zeta_{l+2}}{Dl}
\frac{\zeta_{l+1}}{\zeta_{l+2}}\right] i\,L^{ij}\bpsi
_l^{i_1...i_l} \nn
 &\stackrel{(\ref{difference-l})}{=}&    \frac{(c_{l+1})^2-(c_{l})^2}{D\!+\!2l\!- \!2}\, i\,L^{ji}\bpsi
_l^{i_1...i_l} \nn
 &\stackrel{(\ref{general''})}{=}&  i\, L^{ij}\bpsi
_l^{i_1...i_l}\times \left\{\!\!\ba{ll}-\frac 1k\:&\mbox{if }l<\Lambda\\[4pt]
\frac{(\rho_{\Lambda-1,\Lambda})^2}{D\!+\!2\Lambda\!- \!2} \:&\mbox{if }l=\Lambda, \ea\right.
\nonumber
\eea
up to order $O\left(k^{-3/2}\right)$, whence (\ref{linea2}), using  (\ref{general''}).
\ Contracting the indices $ij$ of (\ref{xxonX}) we find how  $\overline{\bx}^2:=\overline{x}^i\,\overline{x}_i$ acts on the $\Hi_\Lambda$ basis elements:
\bea
\overline{\bx}^2\,\bpsi
_l^{i_1i_2...i_l} &=& g_{ij}\overline{x}^i\,\overline{x}^j\bpsi
_l^{i_1i_2...i_l} \nn
&\stackrel{(\ref{xxonX})}{=}& c_{l+1}\,c_{l+2} 
\underset{(\ref{Plprojg})}{\cancel{g_{ij}\bpsi
_{l+2}^{iji_1...i_l}}}+ c_{l}\,c_{l-1}\,\frac{\zeta_{l+1}}D\, \frac{\zeta_{l}}D\, \underset{(\ref{Plprojg})}{\cancel{g^{hk} \Ps^{l}{}^{i_1i_2i_3...i_l}_{h\,k\,k_3...k_l}}}
\bpsi
_{l-2}^{k_3...k_l}\nn
&&+(c_{l+1})^2\, \frac{\zeta_{l+2}}D\, \Ps^{l+1}{}^{hi_1...i_l}_{hj_1...j_l} \bpsi
_{l}^{j_1...j_l}+(c_{l})^2\,\frac{\zeta_{l+1}}D\,\Ps^{l}{}^{i_1i_2...i_l}_{hj_2...j_l}
\,\bpsi
_{l}^{hj_2...j_l}\nonumber
\eea
\bea
&\stackrel{(\ref{contractPs})}{=}& (c_{l+1})^2\, \frac{\zeta_{l+2}}D\, \frac 1{l\!+\!1} \!\left[D+ l -
\frac{2l}{D\!+\!2l\!-\!2}\right]  \bpsi
_{l}^{i_1i_2...i_l}+(c_{l})^2\,\frac{\zeta_{l+1}}D\,\bpsi
_{l}^{i_1i_2...i_l}\nn
&\stackrel{(\ref{difference-l})}{=}& 
\left\{(c_{l+1})^2 \left[1 -
\frac{l}{D\!+\!2l\!-\!2}\right] +(c_{l})^2\,\frac{l}{D\!+\!2l\!-\!2}\right\}\,\bpsi
_{l}^{i_1i_2...i_l} \label{interm}
 \eea 
up to order $O\left(k^{-3/2}\right)$. We find (\ref{bx^2}), noting that up to order $O\left(k^{-3/2}\right)$ (\ref{interm}) becomes
\bea
\overline{\bx}^2\,\bpsi
_\Lambda^{i_1i_2...i_\Lambda}  &\stackrel{(\ref{defc})}{=}&\frac{(c_{\Lambda})^2\,\Lambda}{D\!+\!2\Lambda\!-\!2}\,\bpsi
_{\Lambda}^{i_1i_2...i_\Lambda}\nn
&\stackrel{(\ref{general''})}{=}& \frac{\Lambda}{D\!+\!2\Lambda\!-\!2}\,
 \left\{1+\frac{(2D\!-\!5)(D\!-1\!)}{2k}+
\frac{(\Lambda\!-\!1)(\Lambda\!+\!D\!-\!2)}{k}\right\}\,\bpsi
_{l}^{i_1i_2...i_l}\nn
\overline{\bx}^2\,\bpsi
_l^{i_1i_2...i_l}  &=& \left\{(c_{l+1})^2+\frac{l\left[(c_{l})^2-(c_{l+1})^2\right]}{D\!+\!2l\!-\!2}\right\}\,\bpsi
_{l}^{i_1i_2...i_l}\nn
&\stackrel{(\ref{general''})}{=}&  \left\{1+\frac{(2D\!-\!5)(D\!-1\!)}{2k}+\frac{E_l}{k}\right\}\,\bpsi
_{l}^{i_1i_2...i_l},\qquad \mbox{if } l<\Lambda.\nonumber
\eea

\subsection{Proof of Proposition \ref{Algebra-isomorphism'}}
\label{ProofAlgebra-isomorphism'}

We first show that indeed
$\A_\Lambda$ is generated by the 
$\overline{L}_{hi},\overline{x}^i\cdot$.
Since the action of $so(D)$ (which is spanned by the $\overline{L}_{hi}$) is transitive on each irreducible component
$\Hi_\Lambda^l\simeq V_D^l$ ($l\in\{0,1,...,\Lambda\}$) contained in $\Hi_\Lambda$,
it remains to show that   some  $\bpsi
_l\in \Hi_\Lambda^l$ can be mapped into some element
of $\Hi_\Lambda^m$  for all $m\neq l$ by applying polynomials in $\overline{x}^i\cdot$.
For all $\bpsi
_l\in \Hi_\Lambda^l$ we have $\overline{x}^i\bpsi
_l\in \Hi_\Lambda^{l+1}\oplus\Hi_\Lambda^{l-1}$; if
$l=\Lambda$ then $\overline{x}^i\bpsi
_l\in\Hi_\Lambda^{\Lambda-1}$ automatically; 
if $l<\Lambda$ one can map $\Hi_\Lambda^l$ into $\Hi_\Lambda^{l-1}$ by the contracted
multiplication of (\ref{tTcontraction}), while applying  $(\overline{x}^1\!+\!i\overline{x}^2)\cdot$ to  $(t^1\!+\!it^2)^lf_l(r)\in\Hi_\Lambda^l$  one obtains a vector proportional to 
$(t^1\!+\!it^2)^{l+1}f_{l+1}(r)\in\Hi_\Lambda^{l+1}$.

The Ansatz (\ref{Tcorr'}) with generic coefficients $a_{\Lambda,l}$ is 
manifestly  $Uso(D)$-equivariant, i.e. fulfills (\ref{compatibilityCond'}) for  all
$a= L_{hi}$; it is also invariant under permutations of $(i_1...i_l)$ and fulfills 
relations (\ref{gpsi=0})
(both sides give zero when contracted with any $\delta_{i_ai_b}$). More explicitly, (\ref{Tcorr'}) read
\bea
\ba{lll}
f_{0} & \stackrel{\varkappa_\Lambda}{\mapsto} & a_{\Lambda,0}F_{\bD,\Lambda}=a_{\Lambda,0}p_{\Lambda,0}=a_{\Lambda,0}\left(t^{\bD}\right)^\Lambda, \\[6pt]
t^if_1=T_{1}^if_1  & \stackrel{\varkappa_\Lambda}{\mapsto} & 
a_{\Lambda,1}F^{i}_{\bD,\Lambda}=a_{\Lambda,1} p_{\Lambda,1}\,T_{1}^i=a_{\Lambda,1}\left(t^{\bD}\right)^{\Lambda-1}t^i, \\[6pt]
T_{2}^{ij}f_2=\left(t^it^j-\frac{\delta^{ij}}D\right)f_2\quad & \stackrel{\varkappa_\Lambda}{\mapsto} \quad &   a_{\Lambda,2}F^{ij}_{\bD,\Lambda}=a_{\Lambda,2}p_{\Lambda,2}\,T_{2}^{ij}, \\[6pt]
... & \stackrel{\varkappa_\Lambda}{\mapsto} &  ...
\ea \nonumber
\eea
Similarly,  the Ansatz (\ref{Opcorr'}) with a generic
 function $m_\Lambda(s)$ of a real nonnegative variable $s$ is manifestly 
$Uso(D)$-equivariant, and by (\ref{L^2_DF}), (\ref{L-su-F}) we find
\bea
&&  m_\Lambda^*\!(\lambda)L_{\bD h}    m_\Lambda\!(\lambda)\,F^{i_1...i_l}_{\bD,\Lambda}
=i\mu_\Lambda\!(l)  ( \Lambda\!-\!l)  F_{\bD,\Lambda}^{hi_1...i_l}-i\mu_\Lambda^*\!(l\!-\!1)\frac {l(\Lambda\!+\!l\!+\!D\!-\!2)}{D\!+\!2l\!-\!2}\,\Ps^l{}^{i_1i_2...i_l}_{hj_2...j_l}  F_{\bD,\Lambda}^{j_2...j_l}, \quad\quad\label{LonT*''''}
\eea
where we have abbreviated \ $\mu_\Lambda(l)\equiv m_\Lambda(l)\,m_\Lambda^*(l\!+\!1)$. We determine the unknown $m_\Lambda,a_{\Lambda,l}$ requiring (\ref{compatibilityCond'})
for $a=\overline{x}^{h}\cdot$.
Applying $\varkappa_\Lambda$ to eq. (\ref{xonpsi}) with $l< \Lambda$, 
and imposing (\ref{compatibilityCond'}) we obtain
\bea
c_{l+1}\,a_{\Lambda,l+1}\, F_{\bD,\Lambda}^{hi_1...i_l}
+c_{l}\,a_{\Lambda,l-1}\, \frac{l}{D\!+\!2l\!-\!2}\,
\Ps^{l}{}^{i_1i_2...i_l}_{hj_2...j_l}F^{j_2...j_l}_{\bD,\Lambda}=m_\Lambda^*(\lambda)L_{\bD h} m_\Lambda(\lambda)
 a_{\Lambda,l}F^{i_1...i_l}_{\bD,\Lambda}  \nn
\stackrel{(\ref{LonT*''''})}{=}
ia_{\Lambda,l}\left[\mu_\Lambda(l)\, ( \Lambda\!-\!l)\, F_{\bD,\Lambda}^{hi_1...i_l}-\mu_\Lambda^*(l\!-\!1)\frac {l(\Lambda\!+\!l\!+\!D\!-\!2)}{D\!+\!2l\!-\!2}\,\Ps^l{}^{i_1i_2...i_l}_{hj_2...j_l}\, F_{\bD,\Lambda}^{j_2...j_l}\right],\nonumber
\eea
which implies the recursion relations for the coefficients $a_{\Lambda,l}$
\bea
 c_{l+1}a_{\Lambda,l+1}=i(\Lambda\!-\!l)a_{\Lambda,l}\mu_\Lambda(l), \qquad
c_{l}a_{\Lambda,l-1}=-i(\Lambda\!+\!l\!+\!D\!-\!2)a_{\Lambda,l}\mu_\Lambda^*(l\!-\!1). \label{recurs-a'}
\eea
Multiplying (\ref{recurs-a'}a) by
(\ref{recurs-a'}b)$_{l\mapsto l+1}$, i.e. by $c_{l+1}a_{\Lambda,l}=-ia_{\Lambda,l+1}\mu_\Lambda^*(l)(\Lambda\!+\!l\!+\!D\!-\!1)$, we find
\bea
c_{l+1}^2\,a_{\Lambda,l} a_{\Lambda,l+1} =a_{\Lambda,l+1}\mu_\Lambda^*(l)a_{\Lambda,l}\mu_\Lambda(l)(\Lambda\!-\!l)(\Lambda\!+\!l\!+\!D\!-\!1)\nn[6pt]
\Rightarrow\qquad|\mu_\Lambda(l)|^2\,(\Lambda\!-\!l)(\Lambda\!+\!l\!+\!D\!-\!1)=c_{l+1}^2,\nonumber
\eea
implying
\bea
m_\Lambda^*(l\!+\!1)m_\Lambda(l)=\mu_\Lambda(l)=\frac{\epsilon_{\Lambda,l}\,c_{l+1}}{\sqrt{(\Lambda\!-\!l)(\Lambda\!+\!l\!+\!D\!-\!1)}},\quad
 a_{\Lambda,l+1}=ia_{\Lambda,l}\epsilon_{\Lambda,l}\sqrt{\frac{\Lambda\!-\!l}{
(\Lambda\!+\!l\!+\!D\!-\!1)}}
\label{recurs-a'}
\eea
where $\epsilon_{\Lambda,l}$ are arbitrary phase factors. Choosing $\epsilon_{\Lambda,l}=1$, $m_\Lambda(s)$ positive-definite on $s\in \RR^+$, and using the property  $\Gamma(z+1)=z \Gamma(z)$ of the  Euler gamma function we find that (\ref{recurs-a'}) are solved by (\ref{mO(D)'}).
Applying $\varkappa_\Lambda$ to \ to eq. (\ref{xonpsi}) with $l= \Lambda$, 
and imposing (\ref{compatibilityCond'}) we obtain 
\bea
c_{\Lambda}\frac{\Lambda \,a_{\Lambda,\Lambda -1}}{D\!+\!2\Lambda \!-\!2}\,
\Ps^{\Lambda }{}^{i_1i_2...i_\Lambda }_{hj_2...j_\Lambda }F^{j_2...j_\Lambda }_{\bD,\Lambda}=m_\Lambda^*(\lambda)L_{\bD h} m_\Lambda(\lambda)
 a_{\Lambda,\Lambda }F^{i_1...i_l}_{\bD,\Lambda}  \nn
\stackrel{(\ref{LonT*''''})}{=}
-i \,a_{\Lambda,\Lambda }\mu_\Lambda^*(\Lambda \!-\!1)\frac {\Lambda(2\Lambda\!+\!D\!-\!2)}{D\!+\!2\Lambda \!-\!2}\,\Ps^\Lambda {}^{i_1i_2...i_\Lambda }_{hj_2...j_\Lambda }\, F_{\bD,\Lambda}^{j_2...j_\Lambda },\nonumber
\eea
which also is satisfied by (\ref{mO(D)'}).

\subsection{Proof of Theorem \ref{prophTTdeco}}
\label{proof-of-prophTTdeco}

\noindent
The proof is recursive. By (\ref{defhatT}) $\widehat{T}_{l+1}^{h_1...h_{l+1}}\!\!=\!\Ps^{l+1}{}_{i_1...i_lb }^{h_1...h_{l+1}} \overline{x}^{b }\widehat{T}_l^{i_1...i_l}$,
which applied to $T_m^{j_1...j_m}$ gives 
\bea
\widehat{T}_{l+1}^{h_1...h_{l+1}}T_m^{j_1...j_m}=\Ps^{l+1}{}_{i_1...i_lb }^{h_1...h_{l+1}} \overline{x}^{b }\widehat{T}_l^{i_1...i_l}T_m^{j_1...j_m}\stackrel{(\ref{hTTdeco})}{=}
\sum\limits_{n\in L^{lm}} \Ps^{l+1}{}_{i_1...i_lb }^{h_1...h_{l+1}}\,
\widehat{N}^{lm}_n \,V^{i_1...i_l,j_1...j_m}_{b_1...b_n} \overline{x}^{b }T_n^{b_1...b_n}\nn
\stackrel{(\ref{xonT})}{=}\sum\limits_{n\in L^{lm}} \widehat{N}^{lm}_n \,
\Ps^{l+1}{}_{i_1...i_lb }^{h_1...h_{l+1}} \Ps^l{}_{a_1...a_rb_{1}...b_{l-r}}^{i_1...i_ri_{r+1}...i_l}\Ps^m{}_{a_1...a_r b_{l-r+1}...b_n}^{j_1...j_rj_{r+1}...j_m}
\left(c_{n+1}\,T_{n+1}^{b_1...b_nb }+d_n\,c_n\,
\Ps^{n}{}^{b_1...b_n}_{b k_2...k_n} T_{n-1}^{k_2...k_n} \right)\nn
=\sum\limits_{n\in L^{lm}} \widehat{N}^{lm}_n \left[
\Ps^{l+1}{}_{a_1...a_rb_{1}...b_{l-r}b }^{h_1...h_rh_{r+1}..h_{l+1}} \Ps^m{}_{a_1...a_r b_{l-r+1}...b_n}^{j_1...j_rj_{r+1}...j_m}\left(c_{n+1}\,T_{n+1}^{b_1...b_nb }+d_n\,c_n\,
\Ps^{n}{}^{b_1...b_n}_{b k_2...k_n} T_{n-1}^{k_2...k_n} \right)\right]\qquad \label{interm10}
\eea
Now we note that 
\bea
\Ps^{l+1}{}_{a_1....a_rb_{1}....b_{l-r}b }^{h_1...h_rh_{r+1}..h_{l+1}}\,\Ps^{n}{}^{b_1b_2...b_n}_{b\, k_2...k_n} \, T_{n-1}^{k_2...k_n}=\nn
\Ps^{l+1}{}_{a_1...a_rb_{1}...b_{l-r}b }^{h_1...h_rh_{r+1}..h_{l+1}}
\Ps^{n-1}{}^{b_2b_3...b_n}_{hk_3...k_n} \frac 1{n}\left[\underset{(\ref{Plprojg})}{\cancel{
\delta^{b_1}_{b }}}\delta^h_{k_2}
\!+\!(n\!-\!1)\delta^{b_1}_{k_2}\delta^h_{b }\!-\!\frac {2(n\!-\!1)}{D\!+\!2n\!- \!4}\delta^{b_1h}\delta_{b k_2}\right]T_{n-1}^{k_2...k_n}\nn
=\frac {n\!-\!1}{n}\Ps^{l+1}{}_{a_1...a_rb_{1}...b_{l-r}b }^{h_1...h_rh_{r+1}..h_{l+1}} \left[
\Ps^{n-1}{}^{b_2b_3...b_n}_{b k_3...k_n} T_{n-1}^{b_1k_3...k_n}\!-\!
\frac {2}{D\!+\!2n\!- \!4}\Ps^{n-1}{}^{b_2b_3...b_n}_{b_1k_3...k_n}
T_{n-1}^{b k_3...k_n}\right]\nn
=\frac {(n\!-\!1)(D\!+\!2n\!- \!6)}{n(D\!+\!2n\!- \!4)}\Ps^{l+1}{}_{a_1...a_rb_{1}...b_{l-r}b }^{h_1...h_rh_{r+1}..h_{l+1}}\Ps^{n-1}{}^{b_2b_3...b_n}_{b k_3...k_n} T_{n-1}^{b_1k_3...k_n};\nonumber
\eea
for  the last equality we have used the symmetry of $\Ps^{l+1}{}_{a_1...a_rb_{1}...b_{l-r}b }^{h_1...h_rh_{r+1}..h_{l+1}}$ under the exchange $b\leftrightarrow b_1$.
Iterating the procedure we find
\bea
\Ps^{l+1}{}_{a_1....a_rb_{1}....b_{l-r}b }^{h_1...h_rh_{r+1}..h_{l+1}}\,\Ps^{n}{}^{b_1b_2...b_n}_{b\, k_2...k_n} \, T_{n-1}^{k_2...k_n}=\frac {(n\!-\!2)(D\!+\!2n\!- \!8)}{n(D\!+\!2n\!- \!4)}\Ps^{l+1}{}_{a_1...a_rb_{1}...b_{l-r}b }^{h_1...h_rh_{r+1}..h_{l+1}}\Ps^{n-2}{}^{b_3b_4...b_n}_{b k_4...k_n} 
T_{n-1}^{b_1b_2k_4...k_n}\nn
=...=\frac {(n\!-\!l\!+\!r)[D\!+\!2(n\!-\!l\!+\!r)\!- \!4]}{n(D\!+\!2n\!- \!4)}\Ps^{l+1}{}_{a_1...a_rb_{1}...b_{l-r}b }^{h_1...h_rh_{r+1}..h_{l+1}}\Ps^{n-l+r}{}^{b_{l-r+1}...b_n}_{b k_{l-r+2}...k_n} 
T_{n-1}^{b_1...b_{l-r}k_{l-r+2}...k_n},
\nonumber\label{useful1}
\eea
which replaced in (\ref{interm10}) gives
\bea
\widehat{T}_{l+1}^{h_1...h_{l+1}}T_m^{j_1...j_m}=
\sum\limits_{n\in L^{lm}} \widehat{N}^{lm}_n \left[
\Ps^{l+1}{}_{a_1...a_rb_{1}...b_{l-r}b}^{h_1...h_rh_{r+1}..h_{l+1}} \Ps^m{}_{a_1...a_r b_{l-r+1}...b_n}^{j_1...j_rj_{r+1}...j_m}\left(c_{n+1}\,T_{n+1}^{b_1...b_nb }
 \right.\right.\nn \left.\left. 
+ \frac {(n\!-\!l\!+\!r)[D\!+\!2(n\!-\!l\!+\!r)\!- \!4]}{n(D\!+\!2n\!-\!4)}\frac {n\,c_n}{D\!+\!2n\!- \!2}\,\Ps^{n-l+r}{}^{b_{l-r+1}...b_n}_{b k_{l-r+2}...k_n} 
T_{n-1}^{b_1...b_{l-r}k_{l-r+2}...k_n}\right)\right]\nn
=\sum\limits_{n\in L^{lm}} \widehat{N}^{lm}_n \left[c_{n+1}\,
\Ps^{l+1}{}_{a_1...a_rb_{1}...b_{l-r}b }^{h_1...h_rh_{r+1}..h_{l+1}} \Ps^m{}_{a_1...a_r b_{l-r+1}...b_n}^{j_1...j_rj_{r+1}...j_m}T_{n+1}^{b_1...b_nb }
\right.\nn \left. 
+ \frac {(n\!-\!l\!+\!m)[D\!+\!n\!-\!l\!+\!m\!- \!4]}{2(D\!+\!2n\!-\!2)(D\!+\!2n\!-\!4)}c_n\,\Ps^{l+1}{}_{a_1...a_ra_{r+1}b_{1}...b_{l-r}}^{h_1...h_rh_{r+1}..h_{l+1}} \Ps^m{}_{a_1...a_r a_{r+1} k_{l-r+2}...k_n}^{j_1...j_rj_{r+1}...j_m}
T_{n-1}^{b_1...b_{l-r}k_{l-r+2}...k_n}\right];\nonumber
\eea
renaming $n'\equiv n\!-\!1$ in the first sum, 
$n'\equiv n\!+\!1$ in the second, this becomes
\bea
\widehat{T}_{l+1}^{h_1...h_{l+1}}T_m^{j_1...j_m}=\sum\limits_{n'\in L^{lm}+1} \widehat{N}^{lm}_{n'-1} c_{n'}\,
\Ps^{l+1}{}_{a_1...a_{r'}bb_{1}...b_{l-r'}}^{h_1...h_rh_{r'+1}..h_{l+1}} \Ps^m{}_{a_1...a_{r'} b_{l-r'+1}...b_{n'-1}}^{j_1...j_{r'}j_{r'+1}...j_m} T_{n'}^{bb_1...b_{n'-1}} \nn 
+\sum\limits_{n'\in L^{lm}-1}\!\!  \widehat{N}^{lm}_{n'+1}\, c_{n'+1}\,\frac {(n'\!-\!l\!+\!m\!+\!1)[D\!+\!n'\!-\!l\!+\!m\!- \!3]}{2(D\!+\!2n')(D\!+\!2n'\!-\!2)}\nn
\Ps^{l+1}{}_{a_1...a_{r'}b_{1}...b_{l-r'+1}}^{h_1...h_{r'}..h_{l+1}} \Ps^m{}_{a_1...a_{r'} k_{l-r'+3}...k_{n'+1}}^{j_1...j_{r'}...j_m}
T_{n'}^{b_1...b_{l-r'+1}k_{l-r'+3}...k_{n'+1}}\nn
=\sum\limits_{n'\in L^{(l+1)m}} \widehat{N}^{(l+1)m}_{n'}
\Ps^{l+1}{}_{a_1...a_{r'}b_{1}...b_{l-{r'}+1}}^{h_1...h_{r'}...h_{l+1}}\Ps^m{}_{a_1...a_{r'} b_{l-r'+2}...b_{n'}}^{j_1...j_{r'}j_{{r'}+1}...j_m} T_{n'}^{b_1...b_{n'}}\nn\nonumber
\eea
where \ $\displaystyle r'\!:=\!\frac{l\!+\!1\!+\!m\!-\!n' }2$, \ we have used that $2(n\!-\!l\!+\!r)=n\!-\!l\!+\!m$ and we have set
\bea
\widehat{N}^{(l+1)m}_{n} = c_{n}\widehat{N}^{lm}_{n-1}+c_{n+1}\widehat{N}^{lm}_{n+1}
\frac {(n\!-\!l\!+\!m\!+\!1)[D\!+\!n\!-\!l\!+\!m\!- \!3]}{2(D\!+\!2n)(D\!+\!2n\!-\!2)}
\label{Recursive-hN}
\eea
for all $n\in  L^{(l+1)m}$. 
The relation among $N^{(l+1)m}_{n},N^{lm}_{n-1},N^{lm}_{n+1}$ is obtained  replacing $c_h\!\mapsto\! 1$ and removing \ $\widehat{}$. \
Since $c_h\le c_{\Lambda}$, by the induction hypotesis (\ref{hN})$_l$ we find
(\ref{hN})$_{l\!+\!1}$, as claimed:
\bea
&& \widehat{N}^{(l+1)m}_{n} \le c_{\Lambda}\left[\widehat{N}^{lm}_{n-1}+\widehat{N}^{lm}_{n +1}
\frac {(n \!-\!l\!+\!m\!+\!1)[D\!+\!n \!-\!l\!+\!m\!- \!3]}{2(D\!+\!2n )(D\!+\!2n \!-\!2)}\right]\nn
&& \le (c_{\Lambda})^{l+1}\left[N^{lm}_{n -1}+N^{lm}_{n +1}
\frac {(n \!-\!l\!+\!m\!+\!1)[D\!+\!n \!-\!l\!+\!m\!- \!3]}{2(D\!+\!2n )(D\!+\!2n \!-\!2)}\right]= (c_{\Lambda})^{l+1} N^{(l+1)m}_{n}.\nonumber
\eea

\subsection{Proof of Theorem \ref{convergD}}
\label{ProofconvergD}

Setting \ 
$\bphi_\Lambda^\perp:=\bphi\!-\!\bphi_\Lambda$ \ we find
\bea
(f -\hat f_{2\Lambda})\bphi=(f - f_{2\Lambda})\bphi+(f_{2\Lambda} -\hat f_{2\Lambda})\bphi
\nn[6pt]
=(f - f_{2\Lambda})\bphi+ (f_{2\Lambda} -\hat f_{2\Lambda})\bphi_\Lambda^\perp+(f_{2\Lambda} -\hat f_{2\Lambda})\bphi_\Lambda,\nonumber
\eea
\bea
 \left\Vert(f -\hat f_{2\Lambda})\bphi\right\Vert & \le & 
\left\Vert(f - f_{2\Lambda})\bphi\right\Vert+\left\Vert(f_{2\Lambda} -\hat f_{2\Lambda})\bphi_\Lambda^\perp\right\Vert+\left\Vert(f_{2\Lambda} -\hat f_{2\Lambda})\bphi_\Lambda\right\Vert
\nn[2pt]  & \le & 
\left\Vert f - f_{2\Lambda}\right\Vert_{op}\,\left\Vert\bphi\right\Vert+\left\Vert f_{2\Lambda} -\hat f_{2\Lambda}\right\Vert_{op}\,\left\Vert\bphi_\Lambda^\perp\right\Vert+\left\Vert(f_{2\Lambda} -\hat f_{2\Lambda})\bphi_\Lambda\right\Vert \label{TriangIneq}
\eea
Since $f\in C(S^d)$, then
$\Vert f_{2\Lambda}\!-\!f\Vert_{op}=\Vert f_{2\Lambda}\!-\!f\Vert_\infty\equiv \sup_{p\in S^d}|f_{2\Lambda}(p)\!-\!  f(p)|$, which goes to zero because $f_{2\Lambda}\to f$ {\it uniformly} over $S^d$ (in fact, $f_{2\Lambda}\!-\!  f$ is  a continuous function  over the compact
manifold $S^d$, and therefore by Heine-Cantor theorem is also
uniformly continuous); 
therefore the first term at the rhs(\ref{TriangIneq}) goes to zero as $\Lambda\to\infty$.
As $\Lambda\to\infty$ also the second term at the rhs(\ref{TriangIneq}) goes to zero because so does $\Vert\bphi_\Lambda^\perp\Vert=\Vert\bphi\!-\!\bphi_\Lambda\Vert$,
while $\Vert f_{2\Lambda} -\hat f_{2\Lambda}\Vert_{op}$ certainly is bounded   (actually one can easily show that this goes to zero as well).

To show that the lhs(\ref{TriangIneq}) goes to zero as $\Lambda\to\infty$ 
we now show that the last term at the rhs does, as well. 
Using the decompositions (\ref{decophi}-(\ref{phi-coeff-prop}) for $\bphi_\Lambda,\hat f_{2\Lambda},f_{2\Lambda}$ we can write
\bea
(\hat f_{2\Lambda}-f_{2\Lambda})\bphi_\Lambda=
\sum_{l=0}^{2\Lambda} \! f_{i_1...i_l}^l\big(\widehat{T}_l^{i_1...i_l}
-T_l^{i_1...i_l}\big)\,\sum_{m=0}^{\Lambda}\phi_{j_1...j_m}^m T_m^{j_1...j_m}\nn
\stackrel{(\ref{hTTdeco}),(\ref{TTdeco})}{=}\sum_{l=0}^{2\Lambda} \sum_{m=0}^{\Lambda}\! f_{i_1...i_l}^l\phi_{j_1...j_m}^m
\sum\limits_{n\in L^{lm}}\,\big(\widehat{N}^{lm}_n- N^{lm}_n\big)\,\Ps^l{}_{a_1...a_rc_{1}...c_{l-r}}^{i_1...i_l}\Ps^m{}_{a_1...a_r c_{l-r+1}...c_n}^{j_1...j_rj_{r+1}...j_m} \Ps^n{}_{c_1...c_n} ^{k_1...k_n}\, T_n^{k_1...k_n}\nn
=\sum_{l=0}^{2\Lambda} \sum_{m=0}^{\Lambda}\sum\limits_{n\in L^{lm}}\,
 f_{a_1...a_rc_{1}...c_{l-r}}^l\phi_{a_1...a_r c_{l-r+1}...c_n}^m
\big(\widehat{N}^{lm}_n- N^{lm}_n\big) \, T_n^{c_1...c_n}
\nonumber
\eea
and taking the square norm in $\Hi_s$,
\bea
\left\Vert (\hat f_{2\Lambda}-f_{2\Lambda})\bphi_\Lambda\right\Vert^2 
=\sum_{l,l'=0}^{2\Lambda} \,\sum_{m,m'=0}^{\Lambda}\, \sum\limits_{n\in L^{lm}}
\sum\limits_{n'\in L_{l'}^{m'}}\!  \big(\widehat{N}^{lm}_n- N^{lm}_n\big) 
\big(\widehat{N}^{l'm'}_{n'}- N^{l'm'}_{n'}\big) \nn
\big(f_{a_1...a_rc_{1}...c_{l-r}}^l\phi_{a_1...a_r c_{l-r+1}...c_n}^m\big)^*
\,f_{b_1...b_{r'}d_{1}...d_{l'-r'}}^{l'}\phi_{b_1...b_{r'} d_{l'-r'+1}...d_{n'}}^{m'}
\,\langle T_n^{c_1...c_n} ,T_{n'}^{d_1...d_n}\rangle\nonumber
\eea
\bea
=\sum_{l,l'=0}^{2\Lambda} \,\sum_{m,m'=0}^{\Lambda}\, \sum\limits_{n\in L^{lm}}
\sum\limits_{n'\in L_{l'}^{m'}}\!  \big(\widehat{N}^{lm}_n- N^{lm}_n\big) 
\big(\widehat{N}^{l'm'}_{n'}- N^{l'm'}_{n'}\big) \nn
\big(f_{a_1...a_rc_{1}...c_{l-r}}^l\phi_{a_1...a_r c_{l-r+1}...c_n}^m\big)^*
\,f_{b_1...b_{r'}d_{1}...d_{l'-r'}}^{l'}\phi_{b_1...b_{r'} d_{l'-r'+1}...d_{n'}}^{m'}
\,Q_n \delta_{nn'}\Ps^n{}_{c_1...c_n} ^{d_1...d_n}\nn
=\sum_{l,l'=0}^{2\Lambda} \,\sum_{m,m'=0}^{\Lambda}\, \sum\limits_{n\in L^{lm}\cap L_{l'}^{m'}}\!  \big(\widehat{N}^{lm}_n- N^{lm}_n\big) 
\big(\widehat{N}^{l'm'}_{n}- N^{l'm'}_{n}\big) Q_n\nn
\big(f_{a_1...a_rc_{1}...c_{l-r}}^l\phi_{a_1...a_r c_{l-r+1}...c_n}^m\big)^*
\,f_{b_1...b_{r'}c_{1}...c_{l'-r'}}^{l'}\phi_{b_1...b_{r'} c_{l'-r'+1}...c_{n}}^{m'}
\label{interm9}
\eea
By (\ref{defc}),  (\ref{Norm-bx^2}) it is $(c_\Lambda)^l\le(1+\epsilon)^{l/2}<e^{l\epsilon/2}\le e^{\Lambda\epsilon}$, because  we are using $l\le 2\Lambda$; hence by
Theorem \ref{prophTTdeco} we find 
$$
0\le \big(\widehat{N}^{lm}_n- N^{lm}_n\big) \le N^{lm}_n [(c_\Lambda)^l-1]
< N^{lm}_n \big(e^{\Lambda\epsilon}\!-\!1\big).
$$
Replacing these inequalities  in the result above we obtain
\bea
\left\Vert (\hat f_{2\Lambda}-f_{2\Lambda})\bphi_\Lambda\right\Vert^2 & < &
\big(e^{\Lambda\epsilon}\!-\!1\big)^2\sum_{l,l'=0}^{2\Lambda} \,\sum_{m,m'=0}^{\Lambda}\, \sum\limits_{n\in L^{lm}\cap L_{l'}^{m'}}\!   N^{lm}_n
 N^{l'm'}_{n} Q_n\nn
&&\big(f_{a_1...a_rc_{1}...c_{l-r}}^l\phi_{a_1...a_r c_{l-r+1}...c_n}^m\big)^*
\,f_{b_1...b_{r'}c_{1}...c_{l'-r'}}^{l'}\phi_{b_1...b_{r'} c_{l'-r'+1}...c_{n}}^{m'}\nn[6pt]
&=& \big(e^{\Lambda\epsilon}\!-\!1\big)^2\left\Vert f_{2\Lambda}\bphi_\Lambda\right\Vert^2.
 \label{smart-ineq}
\eea
The last equality holds because the sum   is nothing but $\left\Vert f_{2\Lambda}\bphi_\Lambda\right\Vert^2$, i.e.
what we would obtain from (\ref{interm9}) replacing $\widehat{N}^{lm}_n\mapsto 0$, and therefore $\hat f_{2\Lambda}\mapsto 0$. Moreover,
$\Vert f_{2\Lambda}\bphi_\Lambda \Vert
\le\Vert f_{2\Lambda}\Vert_{op}\,\Vert\bphi_\Lambda\Vert$; but 
both factors have $\Lambda$-independent bounds: 
$\Vert\bphi_\Lambda \Vert\le \Vert\bphi\Vert$, while by the triangular inequality, $\Vert f_{2\Lambda}\Vert_{op}\le
\Vert f\Vert_{op}+\Vert f_{2\Lambda}-f\Vert_{op}$, and the second term goes to zero
as $\Lambda\to\infty$, hence is bounded by some $\eta\ge 0$.
So we end up with
\bea
\left\Vert (\hat f_{2\Lambda}-f_{2\Lambda})\bphi_\Lambda\right\Vert<
\big(e^{\Lambda\epsilon(\Lambda)}\!-\!1\big)\,\left(\Vert f\Vert_{op}+\eta\right)\,\left\Vert\bphi\right\Vert\stackrel{\Lambda\to\infty}{\longrightarrow}0; \label{strong-convergence'}
\eea
the limit is zero because by (\ref{consistencyD}) $k(\Lambda)$ diverges at least as $\Lambda^4$, 
then  by  (\ref{Norm-bx^2}) $\Lambda\epsilon(\Lambda)$ goes to zero as $\Lambda^{-1}$.
Replaced in (\ref{TriangIneq}) this yields for all $f\in C(S^d)$
\bea
 \left\Vert(f -\hat f_{2\Lambda})\bphi\right\Vert\stackrel{\Lambda\to\infty}{\longrightarrow}0 \label{strong-convergence}
\eea
i.e. $\widehat{f}_{2\Lambda}\to f\cdot$ strongly, as claimed. 
Replacing $f\mapsto fg$,
 we find also that $\widehat{(fg)}_{2\Lambda}\to (fg)\cdot$ strongly for all $f,g\in C(S^d)$, as claimed.
On the other hand, since $\Vert\bphi_\Lambda^\perp\Vert\le\Vert\bphi\Vert$ 
(because $\bphi=\bphi_\Lambda+\bphi_\Lambda^\perp$ with $\la\bphi_\Lambda,\bphi_\Lambda^\perp\ra=0$), 
relations (\ref{TriangIneq})  and  (\ref{TriangIneq})   imply also 
\bea
 \left\Vert(f -\hat f_{2\Lambda})\bphi\right\Vert  \le
\left[\left\Vert f - f_{2\Lambda}\right\Vert_{op}+\left\Vert f_{2\Lambda} -\hat f_{2\Lambda}\right\Vert_{op}+\big(e^{\Lambda\epsilon(\Lambda)}\!-\!1\big)\,\left(\Vert f\Vert_{op}+\eta\right)\right] \Vert\bphi\Vert \le F \Vert\bphi\Vert,
\nonumber 
\eea
where  $F>0$ is an upper bound for the expression in the square bracket; hence
\bea
\Vert \hat f_{2\Lambda}\bphi\Vert 
\le \Vert (\hat f_{2\Lambda}\!-\! f)\bphi\Vert +\Vert f\bphi\Vert 
\le F \Vert\bphi\Vert+\Vert f\Vert_\infty \Vert\bphi\Vert\le \left(F+\Vert f\Vert_\infty\right) \Vert\bphi\Vert \nonumber
\eea
i.e. the operator norms $\Vert \hat f_{2\Lambda}\Vert_{op}$ of the $ \hat f_{2\Lambda}$ are uniformly bounded: \
$\Vert \hat f_{2\Lambda}\Vert_{op}\le  \left(F+\Vert f\Vert_\infty\right) \Vert\bphi\Vert$.
Therefore  (\ref{strong-convergence}) implies the last claim of the theorem
\bea
\Vert(fg -\hat f_{2\Lambda}\hat g_{2\Lambda})\bphi\Vert
&\le & \Vert (f -\hat f_{2\Lambda})g\bphi\Vert+\Vert \hat f_{2\Lambda}(g-\hat g_{2\Lambda})\bphi\Vert \nn[6pt]
&\le & \Vert (f -\hat f_{2\Lambda})(g\bphi)\Vert+\Vert \hat f_{2\Lambda}\Vert_{op} \: \: \Vert(g -\hat g_{2\Lambda})\bphi\Vert
\stackrel{\Lambda\to\infty}{\longrightarrow}0.      \label{strong-convergence''}
\eea

\end{document}